\documentclass[12pt,a4paper]{article}
\pdfoutput=1

\usepackage[utf8]{inputenc}
\usepackage{uniinput}
\usepackage{jheppubmod}
\usepackage{amsmath}
\usepackage{url}
\usepackage{bm}
\usepackage{mathtools}
\usepackage{relsize}
\usepackage{cancel}
\usepackage{lipsum,cite,mathrsfs}
\usepackage{hyperref}
\usepackage[plain]{fancyref}
\usepackage{xcolor}
\usepackage{float}
\usepackage{subcaption}
\usepackage{physics}

\definecolor{newred}{HTML}{DC3220}

\newcommand{\bea}{\begin{eqnarray}}
\newcommand{\eea}{\end{eqnarray}}
\newcommand{\be}{\begin{equation}}
\newcommand{\ee}{\end{equation}}

\usepackage{microtype}

\def\({\left (}
\def\){\right )}

\renewcommand{\i}{\text{i}}

\def\Label#1{\label{#1}%
  \smash{\hbox to0pt{\raise1ex\hbox{\tiny[#1]}\hss}}}
\def\noLabels{\let\Label=\label}
\def\nobbibitem{\let\bbibitem=\bibitem}

\title{Blackish Holes}
\author[a]{Souvik Banerjee,}
\emailAdd{souvik.banerjee@uni-wuerzburg.de}
\affiliation[a]{Institut für Theoretische Physik und Astrophysik, Julius-Maximilians-Universität Würzburg,\\ Am Hubland, 97074 Würzburg, Germany}

\author[b]{Suman Das,}
\emailAdd{sumand250496@gmail.com}
\affiliation[b]{Mandelstam Institute for Theoretical Physics, School of Physics, University of the Witwatersrand, Johannesburg, WITS 2050, South Africa.}

\affiliation[c]{Saha Institute of Nuclear Physics, 1/AF, Salt Lake, Kolkata 700064, India.}

\affiliation[d]{Homi Bhabha National Institute, Training School Complex, Anushaktinagar, Mumbai 400094,
India.}

\author[a,c,d]{Arnab Kundu,}
\emailAdd{arnab.kundu@saha.ac.in}

\author[a]{Michael Sittinger}
\emailAdd{michael.sittinger@stud-mail.uni-wuerzburg.de}

\abstract{Based on previous works, in this article we systematically analyze the implications of the explicit normal modes of a probe scalar sector in a BTZ background with a Dirichlet wall, in an asymptotically AdS-background. This is a Fuzzball-inspired geometric model, at least in an effective sense. We demonstrate explicitly that in the limit when the Dirichlet wall approaches the event horizon, the normal modes condense fast to yield an effective branch cut along the real line in the complex frequency plane. In turn, in this approximation, quasi-normal modes associated to the BTZ black hole emerge and the corresponding two-point function is described by a thermal correlator, associated with the Hawking temperature in the general case and with the right-moving temperature in the extremal limit. We further show, analytically, that the presence of a non-vanishing angular momentum non-perturbatively enhances this condensation. The consequences are manifold: {\it e.g.}~there is an emergent {\it strong thermalization} due to these modes, adding further support to a quantum chaotic nature associated to the spectral form factor. We explicitly demonstrate, by considering a classical collapsing geometry, that the one-loop scalar determinant naturally inherits a Dirichlet boundary condition, as the shell approaches the scale of the event horizon. This provides a plausible dynamical mechanism in the dual CFT through a global quench, that can create an emergent Dirichlet boundary close to the horizon-scale. We offer comments on how this simple model can describe salient features of Fuzzball-geometries, as well as of extremely compact objects. This also provides an explicit realization of how an effective thermal physics emerges from a non-thermal microscopic description, within a semi-classical account of gravity, augmented with an appropriate boundary condition.}

\begin{document}
\maketitle


\section{Introduction}
\noindent 

Black hole thermodynamics has shaped much of modern theoretical research in quantum aspects of gravity, since the early results of Bekenstein and Hawking \cite{Bekenstein1972, Hawking:1975vcx}. Several aspects of this can be understood within the framework of Euclidean Quantum Gravity, which, however, is a mathematically ill-defined quantity. The working principle here is similar to the equivalence between Euclidean QFT and Statistical Field Theory. Unlike Euclidean QFT though, the Euclidean gravitational path integral is unbounded from below. Moreover, the formulation does not shed any light on how it can capture physical answers that are expected to be sensitive to the UV-description of the system.

Nevertheless, there is plenty of circumstantial evidence that Euclidean Quantum Gravity does capture salient physical features of black holes, ranging from thermodynamic properties of black holes to much fine-grained entropic measures, {\it e.g.}~Renyi entropies, of the system. See {\it e.g.}~\cite{Almheiri:2012rt, Almheiri:2013hfa, Almheiri:2019hni, Almheiri:2020cfm, Lewkowycz:2013nqa} for (only a representative of the) recent works on this. The latter has also a vast literature on various aspects of realizing the Page curve in the context of black hole information paradox. See {\it e.g.}~\cite{Penington:2019npb, Saad:2019lba, Penington:2019kki, Saad:2021rcu},  for a partial list on some of these progress, and \cite{Bousso:2022ntt, Harlow:2022qsq} for recent reviews including these advances.

While it inherits the issues of Euclidean quantum gravity,\footnote{In particular, the Euclidean path integral is not bounded from below and therefore the corresponding path integral is mathematically ill-defined.} the fine-grained entropic discussions are mostly also confined to low-dimensional gravitational systems, in particular in two dimensions. Despite being extremely interesting and tractable models, it remains unclear to what extent they capture universal and generic mechanism for quantum black holes in higher dimensions. Nonetheless, a robust feature of this formulation is the so-called Euclidean regularity at the origin, which physically translates to a smooth event horizon of a black hole.

Alternatively, the ``Fuzzball paradigm" attempts to construct explicit microscopic descriptions of a quantum black hole, starting from a manifestly UV-complete framework of string theory. See {\it e.g.}~\cite{Bena:2022ldq, Bena:2022rna} for a recent review on this. Both perturbative and non-perturbative degrees of freedom in string theory (such as D-branes) are required to construct the explicit description. In the massless limit of string theory, these degrees of freedom can be as large as the scale of the event horizon and therefore in this paradigm, the event horizon is not smooth. Instead, it is replaced by a stringy star-like object whose dimension can be arbitrarily close to the Schwarzschild radius.\footnote{Note that, in order to achieve this, one needs a mechanism to avoid the Buchdahl limit\cite{Buchdahl:1959zz}. In the Fuzzball-paradigm, the Buchdahl theorem can be violated since the extra-dimensions and fluxes turned on them violate the underlying assumptions of the theorem. See {\it e.g.}~\cite{Gibbons:2013tqa, Mathur:2016ffb} for more discussions on this.}

By now, an impressive family of such solutions have been explicitly constructed within supergravity, which is the massless limit of string theory. However, most of these geometries carry a large number of symmetries and, in particular, supersymmetry. The latter makes it unlikely to realize quantum-thermal aspects of black holes, {\it e.g.}~scrambling, thermalization and quantum chaos. These features, on the other hand, can be realized rather generically from semi-classical analyses of a smooth horizon.

It is therefore an interesting and important question whether an ``effective" description of the fuzzball paradigm exists in which salient features of black holes can be captured, without making explicit reference to the precise details of the UV degrees of freedom. At the outset, this is analogous to the existence of Euclidean quantum gravity path integral itself, which can be viewed as an effective tool to capture certain physical phenomena; however, it is not known to be derivable from a first principle.

In \cite{tHooft:1984kcu} 't Hooft introduced an effective model of quantum black holes, by replacing the purely ingoing classical boundary condition (in the Lorentzian picture) by a Dirichlet boundary condition slightly outside the event horizon of the black hole. It was dubbed as a ``brickwall", which completely reflects all modes that are incident on it. By tuning the position of the brickwall, it was argued that the entropy of a probe scalar sector becomes dominated by a black hole area term at the leading order. This can be viewed as a ``mechanism" of generating area-worth thermal entropy for an otherwise ordinary system. 
\begin{figure}[H]
    \centering
    \includegraphics[width=.77\textwidth]{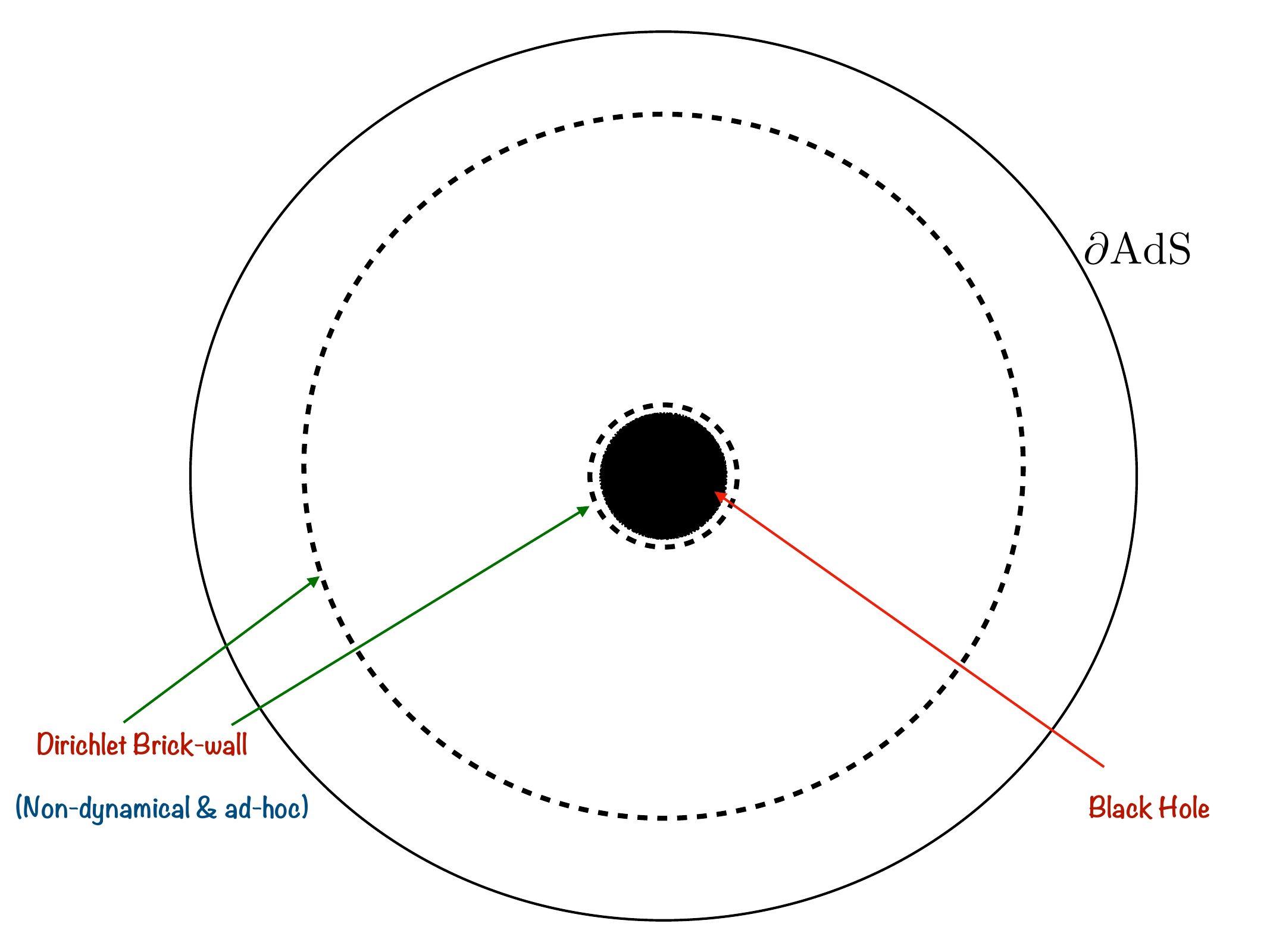}
   \caption{A pictorial representation of the Brickwall model. The dashed circles denote the location of the brickwall, which can be placed at an arbitrary location.}
    \label{model}
\end{figure}

In figure \ref{model}, we have schematically represented the model, in an asymptotically AdS-background.\footnote{We emphasize that while AdS asymptotics make the framework better defined, the physics does not depend on it. The relevant features emerge from the near horizon geometry.} Clearly, the crucial input here is the brickwall itself, which is {\it a priori} a non-dynamical object, but resembles to D-branes in string theory. Recently, in \cite{Das:2022evy, Das:2023ulz, Das:2023xjr, Das:2023yfj}, this model has been explored further and it was demonstrated that the probe scalar sector contains signatures of quantum chaos, even though it is not as strong as the Random Matrix Theory universality. Furthermore, in \cite{Krishnan:2023jqn, Burman:2023kko}, a more precise ``entropy-matching" was performed based on the explicit modes found earlier. It was also shown in \cite{Banerjee:2024dpl} that the black hole quasi-normal modes can be obtained from the analytic properties of the Green's function which {\it a priori} contains the information about the normal modes. See also \cite{Krishnan:2024kzf, Burman:2024egy, Krishnan:2024sle} for further recent progress based on this model.

Taken seriously, the above circumstantial evidences perhaps hint that an ``effective description", which can be described geometrically, exists for quantum black holes. This, however, does not shed any light on how this could be ``derived" from a UV-complete description, nor does it guarantee which physical questions this is able to address. It is important to nonetheless emphasize that it incorporates a crucial UV-data of the Fuzzball paradigm: there is structure at the horizon, which is now described by a Dirichlet boundary condition. One primary goal of this program is to understand these aspects systematically.  

It is important to note at this point that a possible resolution of the black hole information paradox is also proposed from a completely different technical and mathematical perspective. In terms of the so-called mirror operator construction, which necessitates the existence of a black hole interior. See {\it e.g.}~\cite{Papadodimas:2012aq, Papadodimas:2013jku, Papadodimas:2013wnh, Papadodimas:2015jra, Papadodimas:2015xma} as well as a recent review in \cite{Raju:2020smc}. Furthermore, the resolution based on the Island-rule has been criticized in \cite{Geng:2021hlu, Raju:2021lwh}, while the Fuzzball-approached has been criticized in \cite{Raju:2018xue}. It is also important to note that our ``effective geometry" is explicitly at loggerheads with the mirror operator construction. At this point, however, we will not attempt to offer any resolution between any of these tensions.

Coming back to our work, this program can be rephrased and motivated from a different perspective. Given the interesting aspects of the quantum black holes, we can ask: Is there a class of boundary conditions that can yield a qualitatively similar physics? This question will essentially revolve around the aspects of area-worth thermal entropy, maximal chaos as seen by out-of-time-correlators, quantum chaos in the spectral form factor and so on. On one hand, a similar class of questions have been addressed in quantum quench in QFTs, especially CFTs. In critical quenches, the boundary state yields a kinematic thermal description for low-point correlators, and can display a maximally chaotic dynamics for a large-$c$ CFT in two-dimensions\cite{Calabrese:2009qy, Cardy:2014rqa, Calabrese:2016xau, Das:2021qsd}. A very similar physics is also realized in the ``moving-mirror" models in which the dynamics of eternal or evaporating black holes can be realized by choosing an appropriate mirror profile\cite{Akal:2021foz, Akal:2022qei, Cotler:2022weg, Biswas:2024mlq}. For the CFT degrees of freedom, these are essentially certain boundary conditions. This class of questions can now be posed for an arbitrary QFT in a curved geometry, augmented by a class of boundary conditions.

Returning back to the geometric picture, there is a rather pragmatic and perhaps observational perspective on which this program can also shed light. This is to distinguish between black holes and black hole mimickers. While it is certainly true that recent observations in the Event Horizon Telescope or the LIGO black hole mergers strongly support the existence of black holes in our observed Universe, we cannot rule out the possibility of (quantum) structures at the horizon scale. Therefore, it becomes important to understand and catalogue possibilities in which such signatures may be detected, albeit in future. Such a program already exists within the Fuzzball paradigm itself where the intricate detailed geometric structures are currently explored to extract observational signatures in {\it e.g.}~mesoscopic physics\cite{Mayerson:2023wck}.

The idea of distinguishing black holes from black hole mimickers have received a wide attention in \cite{Banerjee:2023uto, Banerjee:2021qei, Giri:2024cks, Danielsson:2023onu}, beginning with the very early work of Israel-Mukhoyama in \cite{Mukohyama:1998rf} where it was already advocated that a low-energy observer is unable to distinguish the Boulware vacuum from the Hartle-Hawking vacuum and therefore for a compact enough star, the observed physical thermodynamic relations are indistinguishable from that of a black hole. In a very precise sense, our program is to revisit this aspect in an asymptotically AdS spacetime, with the added understanding of gauge-gravity duality. Furthermore, our focus is on finer-grained physical quantities, such as the spectral form factor, quasi-normal modes and the related.

In this article, we first expand on the key observations in \cite{Banerjee:2024dpl} and explore its consequences in various directions. In particular, we identify the presence of a {\it strong thermalization} due to the angular momenta modes, which adds strength to the non-trivial spectral rigidity in the corresponding scalar spectrum. Secondly, by considering a collapsing shell model and quantizing a scalar determinant in this dynamical geometry, we explicitly demonstrate that a Dirichlet boundary condition emerges when the collapsing shell approaches the event horizon. In the dual CFT, this provides a plausible mechanism by which a global quench can result into an effective Dirichlet boundary condition for the scalars modes in the bulk. Subsequently, the quantization yields results that we have already observed in the static case.

This article is divided into the following parts: We begin with setting up the framework in the next two sections, in which we also summarize and review many of the earlier results. In section $2$, among several physical properties, we extensively study the analytic structure of the Green's function focussing on the particular role of the angular momenta modes. In particular, we argue that there are two emergent notions of thermalization when the Dirichlet wall approaches the event horizon: ``weak thermalization" and ``strong thermalization". In the next section, we offer comments on the case of rotating BTZ geometry, focussing on similar aspects. The next section is devoted to a detailed account of the one-loop scalar sector in a collapsing geometry, where we also demonstrate that a Dirichlet boundary condition for the scalar sector emerges dynamically at the horizon scale, in this framework. Finally we review how a type III von Neumann algebra emerges from a type I algebra in the limit when the brickwall approaches the horizon. We then conclude with several open and possible future directions. Various technical details, relevant reviews (including a discussion on the Boulware vacuum) and technical results are summarized in seven Appendices.  

A note on terminology: In this article, we will use various terms to mean the same physical degree of freedom. These terms are ``brickwall", ``Dirichlet hypersurface", ``stretched horizon", ``cut-off surface".

\section{Probe Scalar Quantization in a BTZ-Geometry}
\label{s.scalaronbtz}
Let us begin with a discussion on quantization of a probe scalar field in a given background geometry, in particular a black hole geometry. For simplicity, we will consider only a free scalar theory. In a semi-classical quantization, the gravitational Euclidean path integral takes the following qualitative form\cite{Denef:2009kn}:
\begin{eqnarray}
    Z_{\rm Euclidean} = \sum_{g_*} {\rm det} \left( - \nabla_{g_*}^2 \right)^{\pm 1} e^{- S_{\rm E}[g_*]} \ ,
\end{eqnarray}
where $g_*$ denotes the Euclidean saddles and ${\rm det} \left( - \nabla_{g_*}^2 \right)$, the product of all operators that characterize fluctuations around the Euclidean saddles. This determinant can be explicitly evaluated imposing the Euclidean regularity condition at the horizon \cite{Denef:2009kn} and, as a result, in terms of the quasi-normal modes of the particular fluctuation. We will perform a similar analyses, however, we will manifestly sacrifice Euclidean regularity. In the Lorentzian picture, this implies that we allow both incoming and outgoing modes at the/close to the event horizon. In our case, the corresponding determinant is therefore more directly related to the resulting normal modes that for the scalar fluctuations have been previously encountered in \cite{Das:2022evy}.


To set the stage, 
let us first review the solution of a free scalar theory in three-dimensional BTZ-geometry \cite{Banados:1992wn} and discuss its associated quantization. The background geometry is given by
\begin{equation} \label{metric}
   ds^2_{\rm out}= - \left (r^2-r_{\rm H}^2 \right)dt^2+\frac{dr^2}{ \left (r^2-r_{\rm H}^2 \right)} + r^2 d\psi^2 \ ,
\end{equation}
where $r= r_{\rm H}$ is the position of the horizon and we have set the radius of curvature to unity. The dynamics of a probe scalar field of mass $\mu$ in this background satisfies the Klein-Gordon equation:
\begin{equation}\label{eom1}
    \Box \Phi\equiv \frac{1}{\sqrt{|g|}}\partial_{\nu}\left(\sqrt{|g|}\partial^{\nu}\Phi\right) = \mu^2 \Phi \ .
\end{equation}
The general solution of the Klein-Gordon equation can be written as:
\begin{eqnarray}
\label{eqn:mode-functions}
\Phi = \sum_{n, m} \left( a_{n, m} f_{n,m}( t, r \psi ) + a_{n,m}^\dagger f_{n,m}^*(t, r, \psi)\right)  \ , \quad f(t,r,\psi) = \frac{1}{\sqrt{r}} e^{- i \omega_{n,m} t} e^{i m \psi} f(r) \ .  \nonumber \\ 
\end{eqnarray}
Note that, in the general solution above, we have kept both incoming and outgoing modes explicitly. A particular ratio of these modes ensures a Dirichlet boundary condition at a constant radius hypersurface, while infalling boundary condition is picked up by simply throwing away the outgoing piece on that hypersurface.  The wavefunctions $f_{n, m}$ are normalized with respect to the Klein-Gordon inner product \cite{Birrell:1982ix}. Now, a canonical quantization can be carried out by identifying $a_{n,m}^\dagger$ and $a_{n,m}$ as creation and annihilation operators. If the cut-off is put exactly at the horizon the corresponding Hilbert space can be realized as constructed out of the Boulware vacuum. The latter is defined in terms of the annihilation operator: $a_{n,m} |0\rangle_{\cal B} = 0$. Similar to the free theory in flat space, the corresponding Fock space is now built on acting by creation operators on the vacuum:
\begin{eqnarray}
\left| N_{n, m}\right \rangle = \prod_{n,m} \left( a_{n, m}^\dagger \right)^{N_{n,m}} | 0 \rangle_{\cal B} \,
\end{eqnarray}
The Boulware Hilbert space is related to the Minkowski and the Rindler Hilbert space in the following manner: A Minkowski vacuum can be written as the maximally entangled state of two complementary Rindler Hilbert spaces. Each state in this product Rindler Hilbert space can be thought of as excitations on the product of two Boulware vacua. To make this statement precise, let us consider {\it e.g}~the Minkowski vacuum:
\begin{eqnarray}
\left| 0 \right \rangle_{\rm M} = \frac{1}{\sqrt{Z}} \sum_i e^{- \beta E_i/2} \left| E_i\right \rangle_{\rm L}  \left| E_i\right \rangle_{\rm R}  \ , 
\end{eqnarray}
where $\left| E_i\right \rangle_{\rm L,R}$ are eigenstates of the left and right Rindler Hamiltonians. The state $\left| 0 \right \rangle_{\rm M}$ can be thought of as a thermofield double state which is built from excited state at a temperature $T = 1/(2\pi \beta)$, starting from the Boulware vacuum $\left| 0 \right \rangle_{\rm L} \left| 0 \right \rangle_{\rm R}$. See {\it e.g.}~for a recent work with quantum disentanglement in the Boulware vacuum in \cite{Leutheusser:2021frk, Emparan:2023ypa}\footnote{As we will discuss towards the end of this paper, this is also intimately connected to the emergence of type III von Neumann algebra in the corresponding quantum description.}.
Evidently, in an asymptotically flat geometry, the Boulware vacuum approaches the Minkwoski vacuum in the large radii limit and therefore the yields a vanishing stress-tensor. At smaller radii, there is a non-trivial stress-tensor flux and it yields a thermal expectation value for the same with a negative stress-tensor. Close to the event horizon, this stress-tensor expectation value diverges towards a negative infinity. This happens both 
at the past and the future event horizons of an eternal black hole geometry.

As already chalked out in the introduction, our aim here is to keep the cut-off initially at a small but finite distance away from the horizon and move this cut-off gradually towards the horizon, eventually enabling us to analyze the properties of the spectrum when the cut-off is at an infinitesimal distance away from the event horizon. Therefore, in this case, the stress-tensor expectation value will be always finite in the vacuum state defined as $a_{n,m} |0\rangle_{r} = 0$. Here the subscript $r$ is to remind ourselves that this way of choosing the vacuum allows for the quantization of the scalar sector (for instance) 
as a function of the distance of the Dirichlet hypersurface from the event horizon. In this sense, there is a richer class of quantization available at our disposal.
Clearly, this cut-off vacuum\footnote{In \cite{Burman:2023kko}, this vacuum was referred to as a stretched horizon vacuum.} coincides with the Boulware vacuum when the Dirichlet hypersurface coincides with the event horizon. 

It is straightforward to check that the Klein-Gordon field introduced in \eqref{eom1} indeed yields a diverging stress-tensor once the Dirichlet brickwall is placed at the event horizon. If we place the Dirichlet wall infinitesimally close to the horizon, it is still very large, thus, seemingly, raising the issue of a non-trivial backreaction by the scalar sector on the classical geometry. However, if we combine this scalar field stress-tensor with the large negative stress-tensor possessed by the cut-off vacuum, the divergences cancel and one obtains an order one answer. Thus, the probe limit analysis still holds. It was first noted in in \cite{Mukohyama:1998rf} in the context of Boulware vacuum which involves cancellation of actual divergences. For the convenience of the readers we will provide further details of this argument in appendix \ref{s.bwm}.

With this lightning introduction to the choice of vacuum, and leaving the rest of the details and formal definitions to the appendix \ref{s.boulwarehartlehawking} of this paper, let us now focus on the solution of the mode functions in \ref{eqn:mode-functions}. These are obtained by solving the radial Klein-Gordon equation:
\begin{align}
\label{rstar} 
& \left( r^2 - r_{\rm H}^2 \right) \frac{d^2\Phi}{dr^2} + 2 r \left( r^2 - r_{\rm H}^2 \right) \frac{d\Phi}{dr} + \left( \omega^2 - V \right) \Phi = 0, \\
& {\rm yielding} \quad \frac{d^2}{dr_*^2} f(r) - V(r) f(r) + \omega^2 f(r) = 0\, , \\
& {\rm with } \quad V(r) = \left( r^ 2 - r_{\rm H}^2\right)  \left( \frac{1}{r^2} \left( m^2 + \frac{1}{4} \right)  + \nu^2 - \frac{1}{4}\right) \,.  
\end{align}
Here $\nu^2 = 1 + m^2$ and we have introduced the tortoise coordinate:
\begin{eqnarray}
dr_* = \frac{dr}{r^2 - r_{\rm H}^2} \quad \implies \quad r_* = - \frac{1}{r_{\rm H}} \tanh^{-1} \left( \frac{r}{r_{\rm H}} \right)  \equiv \frac{1}{2 r_{\rm H}} \log \left| \frac{r - r_{\rm H}}{r + r_{\rm H}}\right|  \ . 
\end{eqnarray}
Thus, the range of the Schwarzschild radial coordinate $r \in [r_{\rm H}, \infty]$ maps to $r_* \in [- \infty, 0]$. Evidently, near the horizon, $V(r) \to 0$ and therefore the solution of the Klein-Gordon equation is simply: $f(r) \sim e^{\pm i \omega r_*}$. We can now easily estimate the number of oscillations of the classical solution in the near-horizon region \cite{Mathur:2024mvo}.

Towards that, let us consider that the Dirichlet brane is located at $r = r_{\rm H} + \epsilon$, where $\epsilon \ll r_{\rm H}$. In this regime:
\begin{eqnarray}
r_* = \frac{1}{2 r_{\rm H}} \log \left| \frac{\epsilon}{2  r_{\rm H}} \right| + {\cal O}(\epsilon)  \equiv r_*^{(\epsilon)} + {\cal O}(\epsilon)\ ,
\end{eqnarray}
which is obtained by carrying out a straightforward expansion of (\ref{rstar}) in the small $\epsilon$-regime. The number of oscillations of the classical solution is given by
\begin{eqnarray}
n = \omega \left|  r_*^{(\epsilon)} \right |= \frac{\omega}{2 r_{\rm H}} \log \left( \frac{2 r_{\rm H}}{\epsilon} \right)  =  - \frac{\omega}{2 r_{\rm H}} \log(\delta) \ , \label{nosc}
\end{eqnarray}
where $\delta = \int_{r_{\rm H}}^{r_{\rm H}+ \epsilon} dr \sqrt{g_{rr}} \sim \epsilon^{1/2}$ is the proper length between the horizon and the Dirichlet brane.\footnote{Note that we are measuring the proper length in units of the AdS curvature scale, which has been set to unity here. Restoring it, one will obtain: $n \sim \frac{\omega\ell^2}{r_{\rm H}} \log (\delta/ \ell)$, where $\ell$ is the curvature scale.}

Note that, given $r_{\rm H}$, the black hole BTZ geometry is assigned a Hawking temperature: $T_{\rm BH} = \frac{r_{\rm H}}{\pi\ell^2}$, where $\ell$ is the AdS-curvature. Thus, general modes for which $\omega/T_{\rm BH} \sim {\cal O}(1)$ implies $n \gg 1$ (using (\ref{nosc})) and diverges logarithmically as we take the Dirichlet brane closes and closer to the horizon, {\it i.e.}~$\delta \to 0$.\footnote{Note that, hydrodynamic modes typically satisfy $\omega/T_{\rm BH} \ll 1$ limit and therefore for these low energy modes, the number of oscillations can be of order unity. Nonetheless, given a small but fixed value of $\omega/T_{\rm BH}$, we can choose an $\epsilon$ such that $n \gg 1$. This raises an intriguing possibility for the brickwall model: The associated low-energy (hydrodynamic) modes will also have an IR cut-off, {\it i.e.}~the hydrodynamic description will be valid in a regime: $\pi (\log\frac{\delta}{\ell})^{-1} \ll \omega / T_{\rm BH} \ll 1$. } In this limit, the stress-tensor originating from quantizing the scalar field in the BTZ geometry with the Dirichlet brane geometry is identical to that of the black hole. This yields an ideal fluid stress-tensor, see {\it e.g.}~\cite{Mathur:2024mtf} for a more elaborate argument.

Let us now be more explicit with the solution of equation (\ref{eom1}). Consider an ansatz: $ \Phi=\sum_{\omega, m}e^{-i\omega t}e^{i m\psi}{\tilde\phi}(r)$. Introducing a new coordinate, $z = 1- {r_{\rm H}^2}/{r^2}$ such that $z \in [0,1]$, we obtain two linearly independent solutions in terms of hypergeometric functions. The general solution is thus a linear combination:
\begin{align}\label{sol1}
    {\tilde\phi}\left(r(z)\right) & \equiv \phi\left(z\right)  =(1-z)^{\beta}\left[ C_1\, z^{-i \alpha} {}_2F_1(a, b; c; z) \right. \nonumber \\
    & \left. + \,  C_2\, z^{i \alpha} \,{}_2F_1(1+a-c, 1+b-c; 2-c; z)\right] \ ,
\end{align} 
where
\begin{align}\label{ab}
  &  a=\beta-\frac{i}{2r_{\rm H}}(\omega + m) \ , \ b=\beta-\frac{i}{2r_{\rm H}}(\omega - m) \ , \ c=1-2i\alpha \ , \nonumber \\
  & \text{with}  \ \ \
   \alpha=\frac{\omega}{2r_{\rm H}} \ , \ \ \beta=\frac{1}{2}(1-\sqrt{1+\mu^2}) \ .
\end{align}
Near the horizon $z=0$ the first first term of \eqref{sol1} behaves as $e^{-i \omega r_{*}}$, where $r_*$ is the tortoise coordinate $r_* = \frac{1}{2 r_{\rm H}}\log{\frac{r-r_{\rm H}}{r+r_{\rm H}}}$, the horizon being at $r_{\rm H} \rightarrow -\infty$. Substituting this term into the full solution $\Phi$ yields the solution ingoing towards the horizon. Analogously, the second term of \eqref{sol1} yields the outgoing solution.

Consider the position of the Dirichlet brickwall to be $z=z_0 > 0$. This system has been recently studied in detail in \cite{Das:2022evy, Das:2023ulz, Das:2023xjr}. The scalar field now satisfies a Dirichlet boundary condition at $z_0$ and a normalizable condition at the asymptotic conformal boundary. This boundary condition fixes the ratio of the undetermined constants $C_1$ and $C_2$ of the general solution as:
\begin{equation}\label{rat}
    R_{C_2C_1}=\frac{C_2}{C_1}=-z_0^{-2i\alpha} \frac{{}_2F_1(a, b; c; z_0)}{{}_2F_1(1+a-c, 1+b-c; 2-c; z_0)} \ .
\end{equation}
Accordingly, \eqref{sol1} takes the form:
\begin{align}
\label{eqn:sol-dir}
    \phi(z) & = C_1(1-z)^{\beta}\left( z^{-i \alpha} {}_2F_1(a, b; c; z)  +  R_{C_2C_1} z^{i \alpha} {}_2F_1(1+a-c, 1+b-c; 2-c; z)\right) \  .
\end{align}
We can also expand \eqref{eqn:sol-dir} near the boundary at $z = 1$ to obtain:
\begin{equation}
    \phi(z)_{\rm bdry}\sim R_1 (1-z)^{\frac{1}{2}(1-\sqrt{1+\mu^2})} +R_2 (1-z)^{\frac{1}{2}(1+\sqrt{1+\mu^2})} \ ,
\end{equation} 
where
\begin{align}
    R_1 &=\frac{\Gamma(c)\Gamma(c-a-b)}{\Gamma(c-a)\Gamma(c-b)}+R_{C_2C_1}\frac{\Gamma(2-c)\Gamma(c-a-b)}{\Gamma(1-a)\Gamma(1-b)}  \ , \\
    R_2 &= \frac{\Gamma(c)\Gamma(a+b-c)}{\Gamma(a)\Gamma(b)}+R_{C_2C_1} \frac{\Gamma(2-c)\Gamma(a+b-c)}{\Gamma(1+a-c)\Gamma(1+b-c)} \ . 
\end{align}
Normalizability at the conformal boundary imposes $R_1=0$. Therefore, the quantized spectrum can now be obtained as solutions of (\ref{rat}) and the normalizability condition. This spectrum has been analyzed in detail in \cite{Das:2022evy, Das:2023ulz, Das:2023xjr}, which is accessible numerically and we will not review them here. We will, however, note that an analytic formula can be obtained for the spectrum with a WKB-approximation\cite{Das:2023ulz}, when the Dirichlet brickwall is placed very close to the event horizon. This is given by
\begin{eqnarray}
\omega(n,m) = \pi \left( 8 n + 3\right) \left[ 4 W \left( \frac{8\sqrt{2} \pi n + 3 \sqrt{2} \pi}{4 m \sqrt{\epsilon}} \right) \right]^{-1} \ , \label{modeWKB}
\end{eqnarray}
where $W$ is the product log function,\footnote{The product log function $W(z)$ yields the principal solution for $x$ that satisfies: $z = x e^x $.} and $n$ and $m$ are integer-valued. As is explained in \cite{Das:2023ulz}, the above formula can be used to access up to the high end of the $m$-spectrum, provided $\epsilon$ is suitably chosen.

Given the analytical expression in (\ref{modeWKB}), several instructive and qualitative lessons can be drawn. Evidently, we obtain a spectrum $\omega(n,m)$ as a function of the location of the Dirichlet brane which is characterized by $\epsilon$. As summarized in \cite{Mathur:2024mvo}, we can estimate the number of modes up to a given energy-scale $E_{\rm cut}$ for a given choice of $\epsilon$. This number then can be analyzed as a function of the proximity of the brane to the event horizon.

Note that, the product log function admits a standard expansion for an infinitely large argument, which, for us, corresponds to setting $\epsilon \to 0$ keeping $\{n, m\}$ fixed. This is best expressed as follows:
\begin{eqnarray}
W(x) =  \frac{(\log (\log (x))-2) \log (\log (x))}{2 \log
   ^2(x)}+\log (x)-\log (\log (x))+\frac{\log (\log
   (x))}{\log (x)}  + {\cal O}\left(\frac{1}{x} \right) \ ,\nonumber\\
\end{eqnarray}
in the limit $x\to\infty$. A straightforward check further reveals that ${\rm Abs}(W(x))$, in the limit $x\to \infty$, is dominated by the ${\rm Abs}(\log(x))$-term which dominates all the other nested logarithmic terms in the expansion. Thus, to a first approximation, we can simply use $W(x) \sim \log(x)$ in this limit. Furthermore, note that 
\begin{eqnarray}
\frac{\partial}{\partial x} W(x) = \frac{W(x)}{x + x W(x)} \ . 
\end{eqnarray}

Given the analytical formula in (\ref{modeWKB}), we can estimate how fast the modes condense by estimating the density of the modes along each of the $n$ and $m$-directions separately. This is simply determined by the ``inverse density of states" obtained as:
\begin{eqnarray}
&& \left. \frac{\partial \omega}{\partial n} \right|_{m = {\rm fixed}} = \frac{8 \pi}{W(x)} - \frac{\pi (8n+3)}{W(x) x (1+ W(x))} \frac{\partial x}{\partial n} \ , \quad x = \frac{8\sqrt{2}\pi n + 3\sqrt{2} \pi}{4 m \sqrt{\epsilon}} \ , \\
&& \left. \frac{\partial \omega}{\partial m} \right|_{n = {\rm fixed}} = - \pi (8 n+3)\frac{1}{W(x) x (1+ W(x))} \frac{\partial x}{\partial m} \ . 
\end{eqnarray}
Note that, the term $x^{-1} \partial x/\partial n$ is $\epsilon$-independent and the leading order contribution to $\partial \omega/\partial n$ comes from the $W(x)^{-2}\sim \log(x)^{-2}$ term. The above expressions can be expressed as follows:
\begin{eqnarray}
&& \left. \frac{\partial \omega}{\partial n}\right|_{m = {\rm fixed}} = \frac{2\pi} {1 + (8n + 3) \frac{\pi}{4\omega}} \ , \\
&& \left. \frac{\partial \omega}{\partial m} \right|_{n = {\rm fixed}} = \sqrt{\frac{\epsilon}{2}} e^{\frac{(8n+3)\pi}{4\omega}} \frac{1}{1 + \frac{(8n+3)\pi}{4\omega}} \ .
\end{eqnarray}

Therefore, using the leading order behaviour of the product log function, we obtain:
\begin{eqnarray}
&& \left. \frac{\partial n} {\partial \omega} \right|_{m = {\rm fixed}}\equiv \left( \left. \frac{\partial \omega} {\partial n} \right|_{m = {\rm fixed}}\right)^{-1} \approx  a + b |\log \epsilon | \ , \label{omegan}\\
&& \left. \frac{\partial m} {\partial \omega} \right|_{n = {\rm fixed}}\equiv \left( \left. \frac{\partial \omega} {\partial m} \right|_{n = {\rm fixed}}\right)^{-1} \approx p + q |\log \epsilon |^2 \ . \label{omegam}
\end{eqnarray}
Here, $\{a,b\}$ and $\{p,q\}$ are order-one numbers. There is already a visible hierarchy of scales in the corresponding densities. Even though the density of states along both quantum numbers are roughly determined by powers of $\log\epsilon$, the total number along the angular direction will be enhanced by a factor of $\sqrt{\epsilon}$, which can already be intuited from the exact WKB result in (\ref{modeWKB}).

This is most easily seen by the following argument. Let us consider a stripped-down version of the analytical WKB-formula in (\ref{modeWKB}) and equate it with the cut-off energy-scale:
\begin{eqnarray}
    \omega(n,m) = \frac{n}{\log\left( \frac{n}{m\sqrt{\epsilon}}\right) } = E_{\rm cut} \ . 
\end{eqnarray}
Suppose we keep $m$ fixed and keep $n_{\rm cut}$ number of states to reach the energy cut-off. This implies:
\begin{eqnarray}
    && E_{\rm cut} = \frac{n_{\rm cut}}{\log(n_{\rm cut}) - \log(\sqrt{\epsilon})} \quad \implies \quad n_{\rm cut} = \sqrt{\epsilon}\left( 1 + \delta_1 \right)\ , \\
    && \delta_1 \ll 1 \ , \quad E_{\rm cut} \sim \frac{\sqrt{\epsilon}}{\delta_1} \ .
\end{eqnarray}
In order to obtain $E_{\rm cut} \gg 1$, we must set $\sqrt{\epsilon} \gg \delta_1$. Therefore, it is possible to reach a cut-off energy scale which is parametrically larger than order one scales, by keeping only very low-lying modes in $n$.

On the other hand, keeping $n$ fixed, we obtain:
\begin{eqnarray}
    && E_{\rm cut} = \frac{1}{\log(m_{\rm cut}) + \log(\sqrt{\epsilon})} \quad \implies \quad m_{\rm cut} = \frac{1}{\sqrt{\epsilon}}\left( 1 + \delta_2 \right) \ ,\\
    && E_{\rm cut} \sim \frac{1}{\delta_2} \ .
\end{eqnarray}
Contrary to the previous case, we only require $\delta_2 \ll 1$, but it is independent of how $\epsilon$ scales. The upshot is that in order to reach a parametrically large energy cut-off, we need to keep a parametrically large number of $m$ states in the counting. The scaling of the total number of states can now be determined as:
\begin{eqnarray}
N_{m={\rm fixed}} = \alpha_1 + \alpha_2 \left|\log(\sqrt{\epsilon}) \right| 
 + {\cal O}(\epsilon)\ , \label{Nalongn}
\end{eqnarray}
where $\alpha_{1,2}$ are $\epsilon$-independent numbers, but can depend on the quantum numbers $\{m,n\}$. A similar estimate can be done with $n = {\rm fixed}$, which yields:
\begin{eqnarray}
&& N_{n={\rm fixed}} = \beta_1 + \beta_2 \frac{\left|\log(\sqrt{\epsilon}) \right|}{\sqrt{\epsilon}}  + {\cal O}(\epsilon) \ , \label{Nalongm}
\end{eqnarray}
where $\beta_{1,2}$ are $\epsilon$-independent numbers, as before. It is evident from (\ref{Nalongn}) and (\ref{Nalongm}) that there is a non-perturbative enhancement of pole-condensation along the angular momentum direction.

\subsection{A Word on Entropy Matching}

In \cite{Burman:2023kko}, the Authors described a matching of the scalar-sector entropy with the BTZ entropy which fixes the maximum angular momenta quantum number in terms of other parameters in the description, in particular in terms of the cut-off $\epsilon$. Interestingly, this matching fixed the temperature parameter to be the Hawking temperature. This matching is motivated by the brickwall analyses of 't Hooft in \cite{tHooft:1984kcu} where the temperature was set to be the Hawking temperature by hand. 

The analyses in \cite{Burman:2023kko} is based on the multi-particle sector canonical partition function, which can be obtained in terms of the single-particle partition function, since the scalar fields are non-interacting. In this section we will essentially re-confirm the conclusions of \cite{Burman:2023kko} working with a single-particle partition function. Our motivation behind focusing on the single-particle sector is two-fold: First, we know that only the single-particle spectral form factor displays chaotic features in this model.\footnote{See, {\it e.g.}~our earlier works in \cite{Das:2022evy,Das:2023xjr, Das:2023yfj, Das:2023ulz} for a detailed discussion on this. Also note that, in the absence of any interaction, the multi-particle spectral form factor is not expected to display any level-repulsion. This is simply because there is no mechanism for this repulsion when the interaction vanishes.} Secondly, while \cite{Burman:2023kko} used an analytic fitting formula for the spectrum, we will use a WKB formula obtained in \cite{Das:2023xjr}. While there are specific differences between the fitting formula of \cite{Burman:2023kko} and the WKB result of \cite{Das:2023xjr}, we will reach a similar conclusion. This demonstrates the robustness of the result which is determined by a $(\log\epsilon)$ term in both cases.

Given the WKB-modes in (\ref{modeWKB}), let us estimate the corresponding single-particle partition function and subsequently estimate the ``entropy".\footnote{Since the multi-particle partition function is simply a product of the single-particle partition function, the key features of entropy is already encoded in an entropic measure that can be defined using the single-particle partition function.} The partition function is given by
\begin{eqnarray}
Z = \sum_{n,m} e^{-\beta \omega(n,m)} \approx \int dm dn e^{-\beta \omega(n,m)} \ ,
\end{eqnarray}
where in the second line we have replaced the discrete sum by a continuum integral. This is evidently an approximation, which yields errors of order $\Delta\omega$ along the discrete quantum numbers. The integral over $n$ can be performed exactly, which yields:
\begin{eqnarray}
&&  Z = - \frac{(8n+3)\pi}{2\sqrt{2\epsilon}}\int \frac{dy}{y^2} \left[ \frac{\text{Ei}\left(-\frac{(8 n+3) \pi  \beta }{4
   W(y)}\right)-(W(y)+1) e^{-\frac{\pi  \beta  (8 n+3)}{4
   W(y)}}}{2 \pi  \beta } \right]  \ , \\
   && y = \frac{\pi}{m} \frac{8n+3}{2\sqrt{2\epsilon}} \ ,
\end{eqnarray}
which can not be further integrated analytically. However, in the limit $\epsilon\to 0$, we can expand the integrand and estimate an $\epsilon$-dependence of the partition function. At the leading order, this yields:
\begin{eqnarray}
Z \approx \frac{1}{2\pi\beta} \left( c_1 \frac{m\sqrt{\epsilon}}{8n+3} + c_2 \left| \log(m\sqrt{\epsilon}) \right|  \right) + {\cal O}(\epsilon)\ .
\end{eqnarray}
With this, one obtains:
\begin{eqnarray}
S & = & \log Z - \beta \partial_\beta \log Z \\
& = & \log\left( m \frac{\sqrt{\epsilon}}{\sqrt{r_{\rm H}}}\right) + {\rm sub-leading} \ , \label{entone}
\end{eqnarray}
where we have re-instated the factor of $r_{\rm H}$. Recall that the standard BTZ-entropy is given by $S_{\rm BTZ} = (\pi r_{\rm H})/(2G_N)$, where $G_N = \ell_P$ is the Newton's constant and $\ell_P$ is the Plack length. Equating this BTZ entropy with the entropic measure derived above yields a maximum angular quantum number 
\begin{eqnarray}
m_{\rm max} & = & r_{\rm H}^{1/2} \epsilon^{-1/2} {\rm exp} \left( \frac{\pi r_{\rm H}}{2 G_N} \right)  \\
& = & \left( \frac{r_{\rm H}}{\epsilon}\right)^{1/2} {\rm exp} \left( \frac{\pi r_{\rm H}^{3/2}}{2\sqrt{2}} \epsilon^{1/2}\right) \approx \epsilon^{-1/2}+ {\rm sub-leading}\ , 
\end{eqnarray}
where we have used $\epsilon = \frac{r_{\rm H}}{2} \ell_P^2$. Note that, the WKB result is valid up to $m_{\rm cut-off} \sim \epsilon^{-2/3}$\cite{Das:2023xjr} and therfore the matching can be done well within the regime of its validity. This clearly demonstrates a power-law hierarchy in $\epsilon$ between the maximum quantum number required to reproduce the Hawking entropy and the regime of the WKB-result.

Note further that, by comparing (\ref{entone}) with (\ref{Nalongn}), it appears that $S\sim N_{m={\rm fixed}}\equiv N_m$. This is expected in the single-particle sector. Consider {\it e.g.}~the two particle sector. Given $S\sim N_{m}$, the total number of states, which estimates the dimension of the Hilbert space, will scale as $2^{N_{m}}=e^{N_{m} \log 2}$. This recovers the expected exponential scaling.

Before leaving this section, let us summarize. We have demonstrated here that in the brickwall model, the probe scalar sector entropy can already be matched parametrically to the Bekenstein-Hawking entropy. This can be done within a single-particle sector since the multi-particle partition function is simply a product of the single-particle partition functions. Furthermore, the parametric matching yields a power-law hierarchy between the maximum number of modes needed for the matching and the reliable regime of the WKB-formula. This adds strength to the claims already made in \cite{Burman:2023kko}.

\subsection{Aspects of the Green's Function}

In this section we will discuss in detail various aspects of two-point correlation function in the scalar sector. Several salient and important features are already summarized in our earlier work in \cite{Banerjee:2024dpl}. Our goal here is to present and explicit and detailed account of our analyses that lead to the conclusions in \cite{Banerjee:2024dpl}.

As we have mentioned earlier, let us impose Dirichlet boundary condition $\phi(z=z_0)=0$, instead of ingoing boundary condition on the horizon. Physically it implies there is a hard (brick) wall at $z=z_0$. In terms of the solution of the massless Klein-Gordon equation discussed earlier, this boundary condition fixes the ratio of  the undetermined constants $C_1$ and $C_2$ to:
\begin{equation}\label{rat}
    R_{C_2C_1}=\frac{C_2}{C_1}=-z_0^{-2i\alpha} \frac{{}_2F_1(a, b; c; z_0)}{{}_2F_1(1+a-c, 1+b-c; 2-c; z_0)} \ , 
\end{equation}
and the explicit solution in \eqref{sol1} becomes:
\begin{equation}
    \phi(z)=C_1(1-z)^{\beta}\left( z^{-i \alpha} {}_2F_1(a, b; c; z) +  R_{C_2C_1} z^{i \alpha} {}_2F_1(1+a-c, 1+b-c; 2-c; z)\right).
\end{equation}
Using identities of hypergeometric functions this explicit solution can be expanded near the conformal boundary of AdS, as follows:
\begin{equation}
    \phi(z)_{\rm bdry}\sim R_1 (1-z)^{\frac{1}{2}\left(1-\sqrt{1+\mu^2} \right)} +R_2 (1-z)^{\frac{1}{2}\left(1+\sqrt{1+\mu^2} \right)} \ ,
\end{equation}
where
\begin{align}
    R_1 &=\frac{\Gamma(c)\Gamma(c-a-b)}{\Gamma(c-a)\Gamma(c-b)}+R_{C_2C_1}\frac{\Gamma(2-c)\Gamma(c-a-b)}{\Gamma(1-a)\Gamma(1-b)} \ , \\
    R_2 &= \frac{\Gamma(c)\Gamma(a+b-c)}{\Gamma(a)\Gamma(b)}+R_{C_2C_1} \frac{\Gamma(2-c)\Gamma(a+b-c)}{\Gamma(1+a-c)\Gamma(1+b-c)} \ . 
\end{align}
The corresponding two-point function can be computed by taking the ratio of the normalizable mode and the non-normalizable mode and is given by
\begin{equation}
    G(\omega,m)=\frac{R_2}{R_1}=\frac{ \frac{\Gamma(c)\Gamma(a+b-c)}{\Gamma(a)\Gamma(b)}+R_{C_2C_1} \frac{\Gamma(2-c)\Gamma(a+b-c)}{\Gamma(1+a-c)\Gamma(1+b-c)}}{\frac{\Gamma(c)\Gamma(c-a-b)}{\Gamma(c-a)\Gamma(c-b)}+R_{C_2C_1}\frac{\Gamma(2-c)\Gamma(c-a-b)}{\Gamma(1-a)\Gamma(1-b)}} \ . \label{greenR2R1}
\end{equation}
Given a two-point correlator, one can read off the corresponding spectrum by analyzing the analytic structure. In a generic QFT, this data is encoded in the so-called spectral function which is defined using a Kallen-Lehman representation of the two-point correlator.\footnote{Note that, since we are considering a non-interacting scalar field theory, albeit in a curved geometry, the Kallen-Lehman representation is automatically guaranteed.} 
\begin{figure}[H]
\begin{subfigure}{0.47\textwidth}
    \centering
    \includegraphics[width=\textwidth]{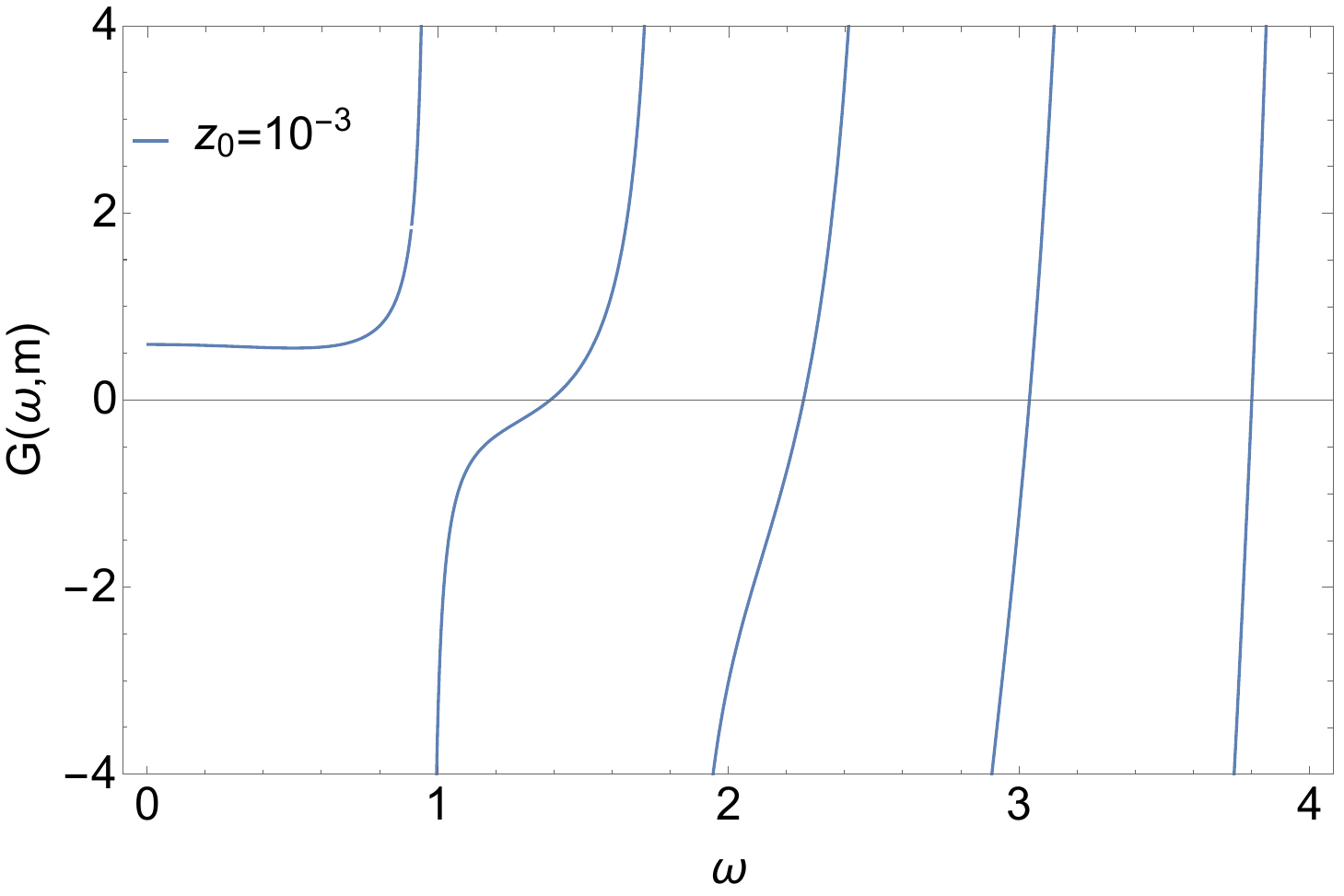}
    \end{subfigure}
    \hfill
    \begin{subfigure}{0.47\textwidth}
    \includegraphics[width=\textwidth]{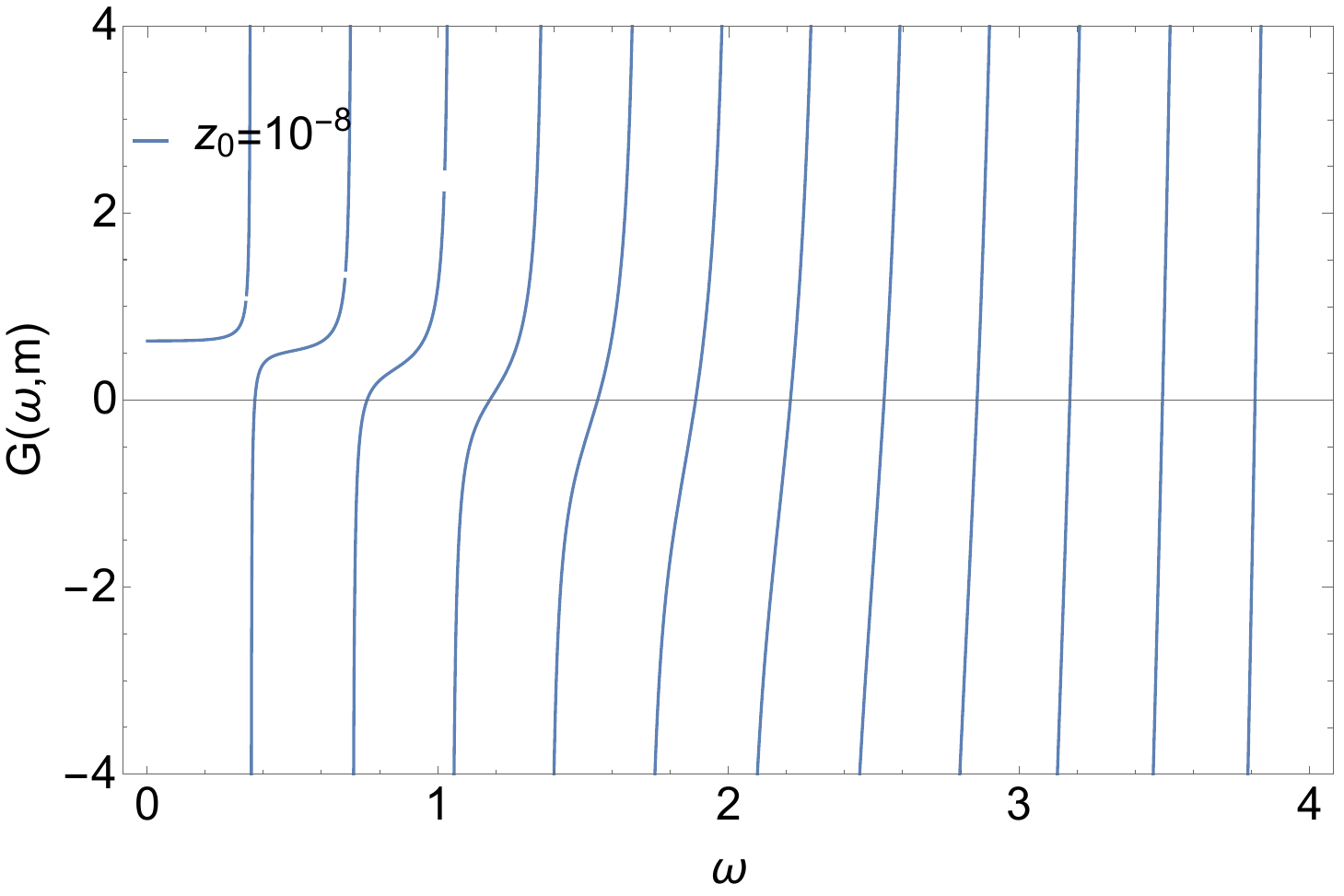}
    \end{subfigure}
    \caption{Pole Structure of $G_{\omega}(n, m)$ for fixed $m=1$. Poles are coming closer and closer as we move the position of the stretched horizon towards the event horizon. $\mu=1$ here for the both figures.}
    \label{green1}
\end{figure}
We are interested in the non-trivial \footnote{Equation \eqref{greenR2R1} also has poles for $a+b-c=-n$ where $n \in \mathcal{Z}_{\geq}0$ which implies $\mu=\sqrt{n^2-1}=0, \sqrt{3}, \sqrt{8}, ...$. We are not interested in those because poles are in the mass axis.} pole structure of \eqref{greenR2R1} and poles appear when the denominator vanishes, i.e.
\begin{equation}\label{deno}
    \frac{\Gamma(c)}{\Gamma(c-a)\Gamma(c-b)}+R_{C_2C_1}\frac{\Gamma(2-c)}{\Gamma(1-a)\Gamma(1-b)}=0 \ ,
\end{equation}
which is identical to the normal mode equation of \cite{Das:2022evy, Das:2023ulz, Das:2023xjr}. The corresponding normal modes can be obtained by numerically solving the algebraic equation in (\ref{deno}), or in an analytic regime using the WKB-approximation. For visual purposes, we have shown how the poles of the Green's function are distributed in figure~\ref{green1}. For more details on the numerical solution and the WKB-analyses, we refer the reader to \cite{Das:2022evy, Das:2023xjr}.

\subsection{Limiting Case: When the Brickwall is Very Close to the Horizon}

Recall that in our framework the position of the brickwall, denoted by $z_0$ (as well as $\epsilon$ before) is a free kinematic parameter. Two extreme limits, namely when the brickwall is very close to the BTZ horizon and when it is close to the conformal boundary of AdS, are of particular importance. As is already demonstrated in earlier works, a non-trivial spectrum ({\it i.e.}~poles of the correlator) emerges when the brickwall approaches the BTZ-horizon. In the limit when the brickwall is close to the conformal boundary of AdS, the spectrum is a linear one. The latter is similar to a harmonic oscillator and is not therefore interesting from the point of view of quantum chaos.  

\begin{figure}[H]
\begin{subfigure}{0.47\textwidth}
    \centering
    \includegraphics[width=\textwidth]{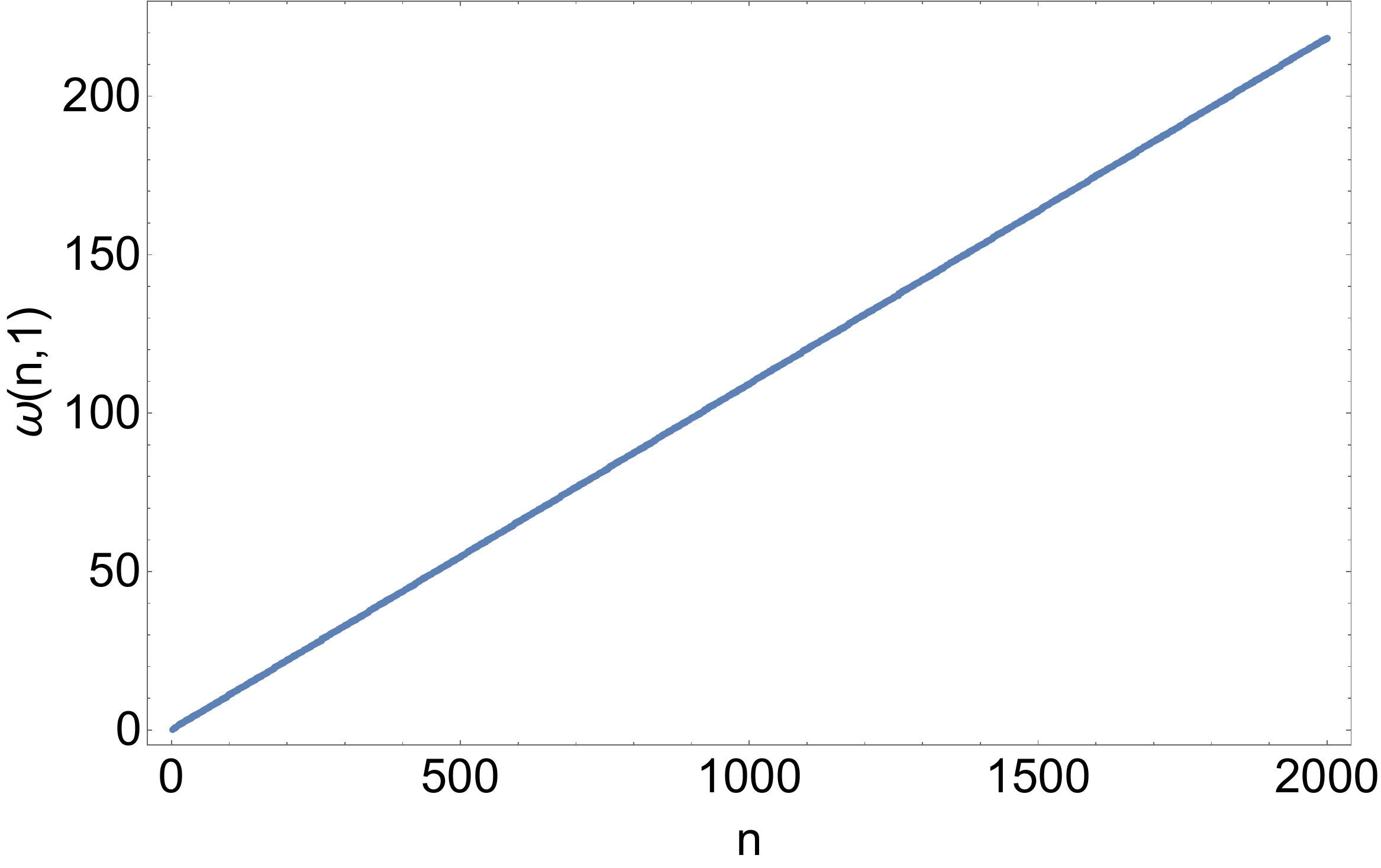}
    \end{subfigure}
    \hfill
    \begin{subfigure}{0.47\textwidth}
    \includegraphics[width=\textwidth]{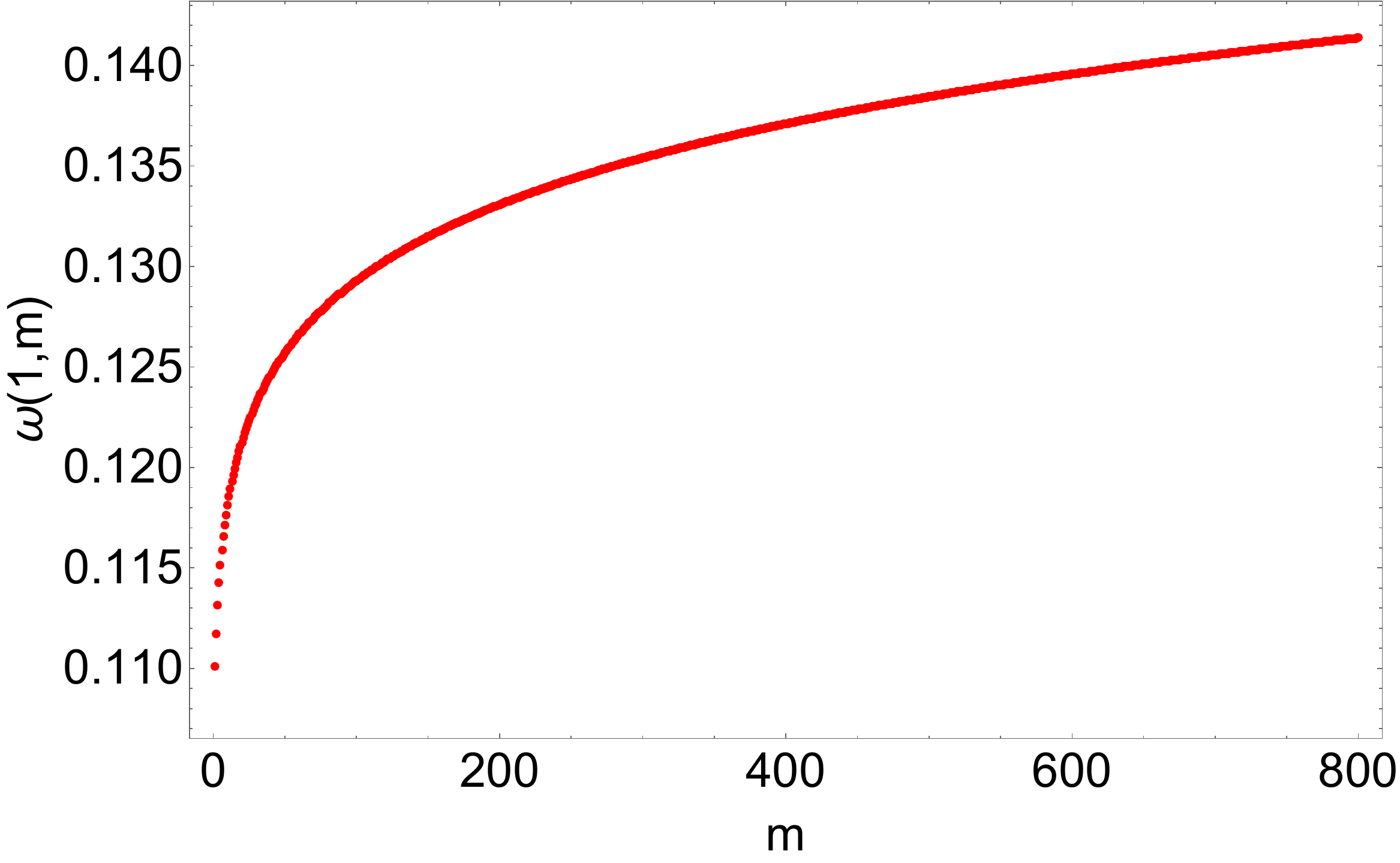}
    \end{subfigure}
    \caption{Normal modes along $n$ (fixed $m=1$) and $m$ (fixed $n=1$) direction are plotted in left and right panel respectively. Here $z_0=10^{-25}$ and $\mu=0$. }
    \label{normal_modes}
\end{figure}

\begin{figure}[H]
\begin{subfigure}{0.47\textwidth}
    \centering
    \includegraphics[width=\textwidth]{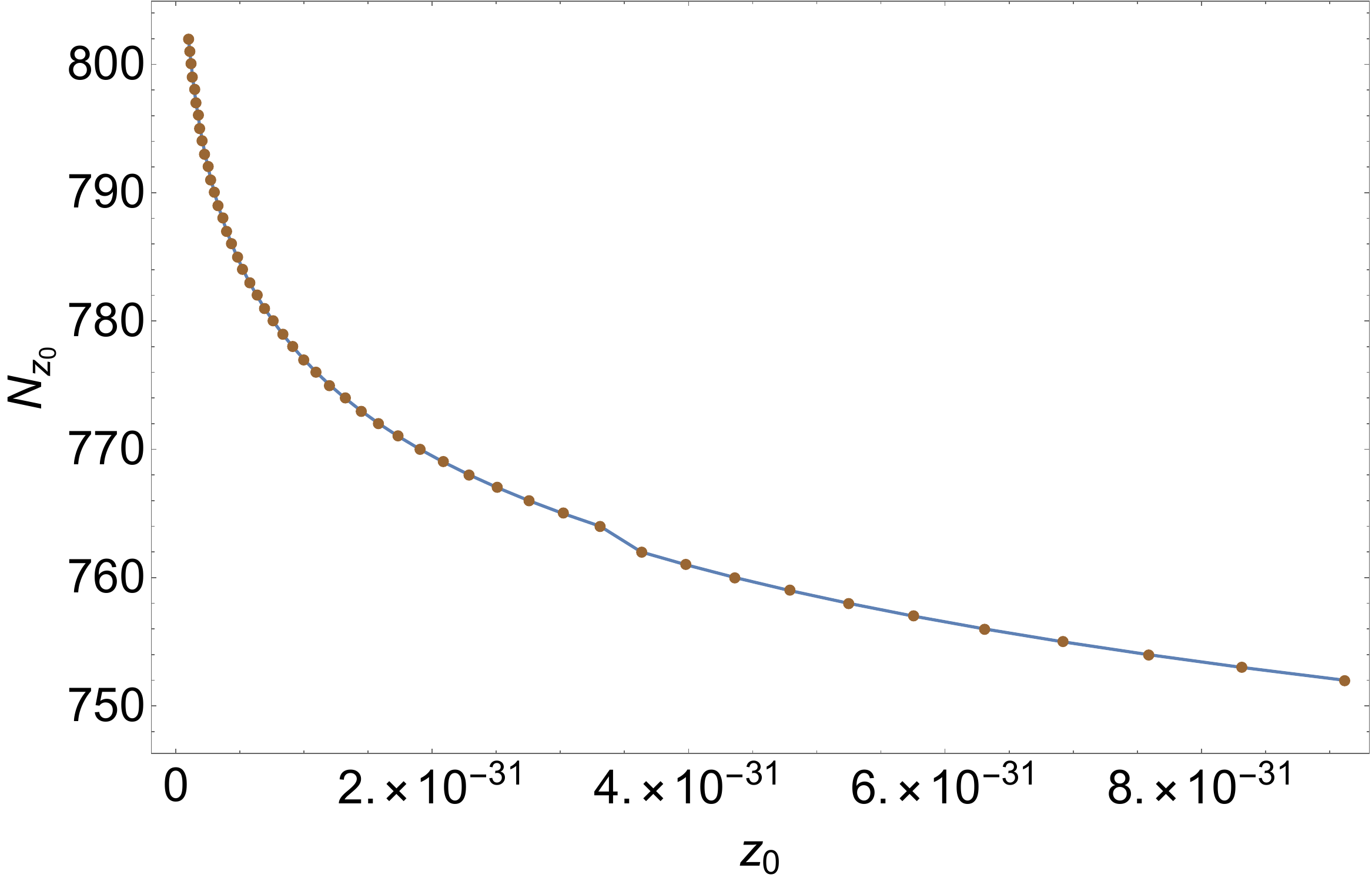}
    \end{subfigure}
    \hfill
    \begin{subfigure}{0.47\textwidth}
    \includegraphics[width=\textwidth]{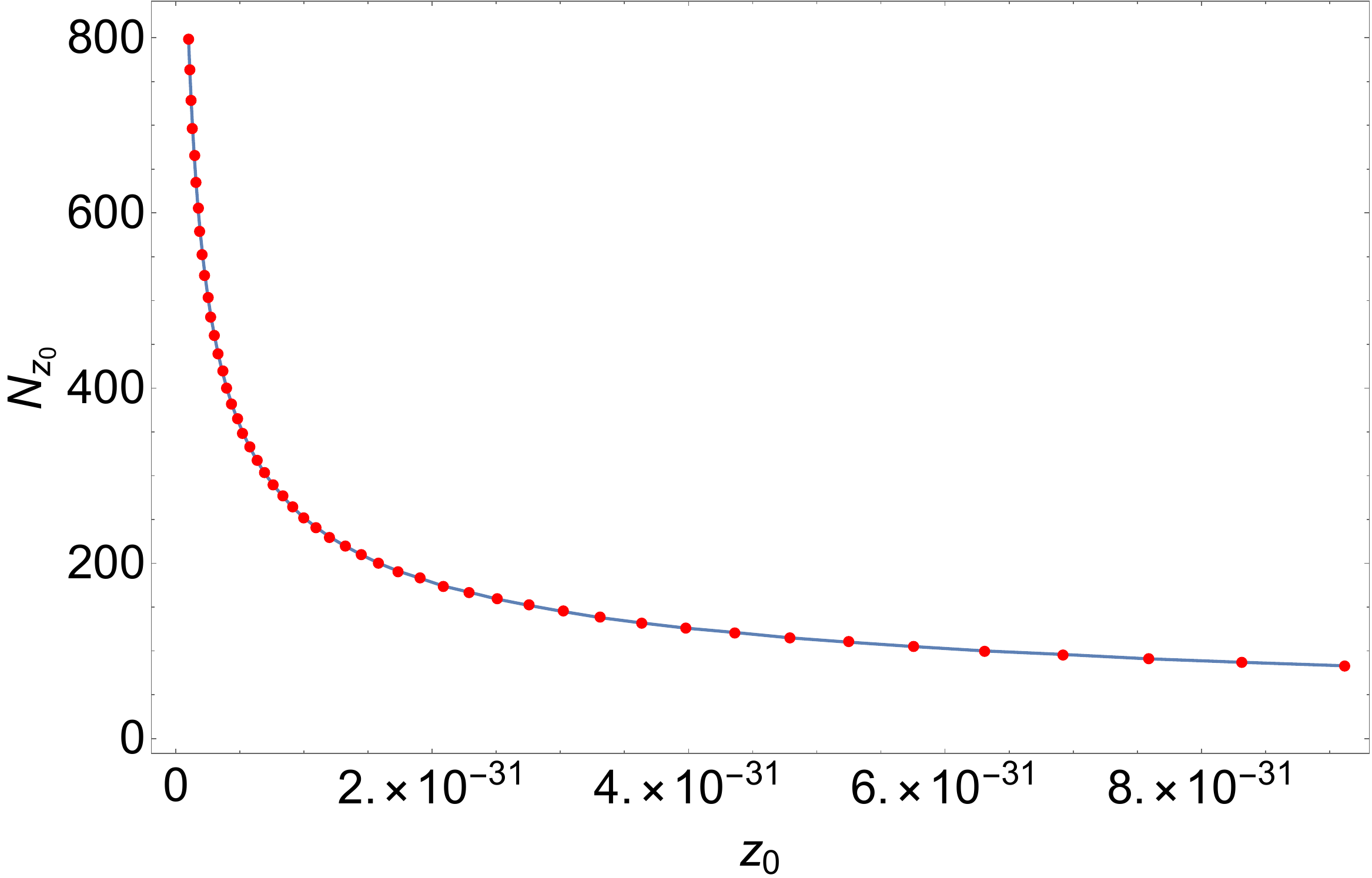}
    \end{subfigure}
    \caption{ $N_{z_0}$ vs. $z_0$ for fixed $m$ and fixed $n$ respectively. $\mu=0$ here. }
    \label{poles1}
\end{figure}

\begin{figure}[H]
    \centering
    \includegraphics[width=.47\textwidth]{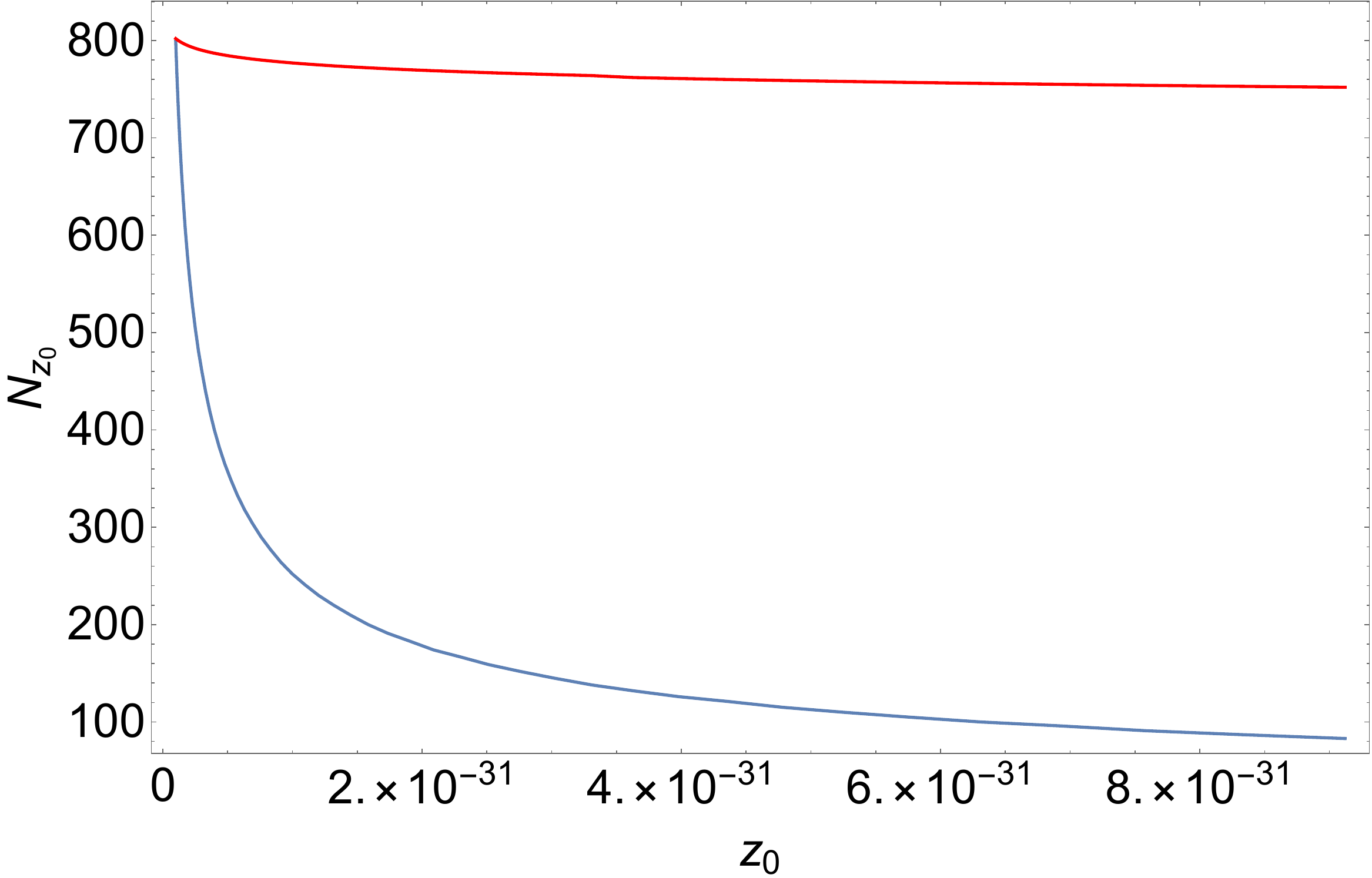}
   \caption{ Showing $N_{z_0}$ vs. $z_0$ plots for fixed $m$ and fixed $n$ in a single picture. $\mu=0$ here.}
    \label{poles2}
\end{figure}

Let us therefore consider the scenario when position of the stretched horizon is very close to the event horizon, so that we can write \eqref{rat} as $R_{C_1C_2}\approx -z_0^{-2i\alpha}$. With this approximation \eqref{deno} becomes:
\begin{eqnarray}\label{quant1}
     & z_0^{2i\alpha} \frac{\Gamma(c)}{\Gamma(c-a)\Gamma(c-b)}\frac{\Gamma(1-a)\Gamma(1-b)}{\Gamma(2-c)}\equiv e^{i \theta(\omega, m)} =1 \ , \nonumber \\
     \nonumber \\ 
     & \Rightarrow \quad \theta(\omega_n, p)= 2n \pi \hspace{1cm}\text{where} \quad  n\in \mathbf{Z} \ . 
\end{eqnarray}
This leads to the following quantization condition:
\begin{equation}\label{quant2}
    \alpha \log z_0+\text{Arg}\left[\frac{\Gamma(c)}{\Gamma(c-a)\Gamma(c-b)} \right]=n \pi \ , \hspace{1.5cm}\text{with} \hspace{0.2cm}  n\in \mathbf{Z} \ . 
\end{equation}
It is important to emphasize that the solution of this equation, $\omega_{n,m}$'s, are the normal modes as appeared in \cite{Das:2022evy, Das:2023ulz, Das:2023xjr}. Figure \ref{normal_modes} shows the behaviour of normal modes $w(n, m)$ with $n$ and $m$ which are the two quantum numbers in this framework.
At these values of $\omega$, the denominator of \eqref{greenR2R1} vanishes and therefore these correspond to the poles of the Green's function. We have already observed in Figure \ref{green1}, the density of poles along $n$-direction increases as we move the position of stretched horizon towards the BTZ-horizon. To get an intuition of how fast that density increases, here in Figure \ref{poles1} we have plotted the behaviour of $N_{z_0}$ with $z_0$. Where $N_{z_0}$ counts the  number of modes within a given $\omega_{\rm max}$ for fixed $m$ or $n$. It is clear from the figures that the rate of growth is faster for poles along $m$-direction. In Figure \ref{poles2} we have shown both panels of Figure \ref{poles1} in a single picture.
A few comments are in order. First, note that based on our analytical result in the WKB-approximation, we already observed this hierarchy in the pole-condensation rates, see equations (\ref{Nalongn}) and (\ref{Nalongm}). Figure~\ref{poles2} already demonstrate this hierarchy explicitly. Along both $m$ and $n$-directions, poles are getting closer and closer but as we have seen in {\it e.g.}~\cite{Das:2022evy}, the corresponding spectral form factor along $n$-direction do not have a linear ramp as opposed to $m$-direction which have clear slope one ramp.\footnote{To be more precise, one can take a projection of the full partition function along a constant quantum number, and correspondingly compute the spectral form factor.} So the rate at which this accumulation of poles occur plays an important role in the strength of thermality, i.e. there is already a hint of a ``strong thermalization" and a ``weak thermalization". We will elaborate on these ideas in subsequent sections. Before that, we will now discuss analytic structure of the correlator.

\subsection{Analytic Structure of Green's Function}
\label{s.structureofgreen}

We have already discussed the pole structures in details. In this section, we will now explore the analytic structure of the Green's function further, with a particular focus on the residues at the poles.  

In order to compute the residues of \eqref{greenR2R1} at the poles $\omega_i$,\footnote{Here we introduce a collective notation for the spectrum: $\omega_i$ implies the normal modes which in our case depends on two independent quantum numbers.} note that, \eqref{greenR2R1} can be written as:
\begin{eqnarray}\label{gre0}
    G(\omega, m)=\frac{\frac{\Gamma(a+b-c)\Gamma(c-a)\Gamma(c-b)}{\Gamma(a)\Gamma(b)\Gamma(c-a-b)}+R_{C_2C_1}\frac{\Gamma(2-c)\Gamma(a+b-c)}{\Gamma(1+a-c)\Gamma(1+b-c)}\frac{\Gamma(c-a)\Gamma(c-b)}{\Gamma(c)\Gamma(c-a-b)}}{1+R_{C_2C_1} \frac{\Gamma(2-c)}{\Gamma(1-a)\Gamma(1-b)}\frac{\Gamma(c-a)\Gamma(c-b)}{\Gamma(c)}} \ ,
\end{eqnarray}
where we have rewritten the numerator according to our convenience. Recall that, a similar calculation for the standard BTZ geometry with an ingoing boundary condition at the horizon yields a retarded Green's function of the following form:
\begin{eqnarray}
G_{\rm bh} = \frac{\Gamma(a+b-c)\Gamma(c-a)\Gamma(c-b)}{\Gamma(a)\Gamma(b)\Gamma(c-a-b)} \ ,
\end{eqnarray}
which has poles in the lower half of the complex frequency plane. These poles correspond to the well-known quasi-normal modes for the BTZ-geometry:
\begin{eqnarray}
\omega_i \equiv \omega_{n,m} = \pm m - i \left(n + \frac{\Delta}{2} \right) \ ,  \quad \Delta = \sqrt{1+\mu^2} \ , \quad n \in {\mathbb Z} \ .
\end{eqnarray}
In our case, with a Dirichlet boundary condition at the stretched horizon, the Green's function can now be written as:
\begin{eqnarray}
  G(\omega, m)  =\frac{G_{\text{bh}}(\omega)+R_{C_2C_1}\frac{\Gamma(2-c)\Gamma(a+b-c)}{\Gamma(1+a-c)\Gamma(1+b-c)}\frac{\Gamma(c-a)\Gamma(c-b)}{\Gamma(c)\Gamma(c-a-b)}}{1+R_{C_2C_1} \frac{\Gamma(2-c)}{\Gamma(1-a)\Gamma(1-b)}\frac{\Gamma(c-a)\Gamma(c-b)}{\Gamma(c)}} \ . 
\end{eqnarray}
The residues corresponding to the simple poles of $G(\omega, m)$ can be obtained to be:
\begin{equation}\label{gre1}
    \text{Res}(G, \omega_i)=\frac{\left(G_{\text{bh}}(\omega)+R_{C_2C_1}\frac{\Gamma(2-c)\Gamma(a+b-c)}{\Gamma(1+a-c)\Gamma(1+b-c)}\frac{\Gamma(c-a)\Gamma(c-b)}{\Gamma(c)\Gamma(c-a-b)} \right)\bigg\rvert_{\omega=\omega_i}}{\partial_{\omega} \left(1+R_{C_2C_1} \frac{\Gamma(2-c)}{\Gamma(1-a)\Gamma(1-b)}\frac{\Gamma(c-a)\Gamma(c-b)}{\Gamma(c)} \right)\bigg\rvert_{\omega=\omega_i}} \ ,
\end{equation}
while, as already pointed out earlier, the poles are found as solutions of: 
\begin{equation*}
    R_{C_2C_1}=-\frac{\Gamma(1-a)\Gamma(1-b)\Gamma(c)}{\Gamma(2-c)\Gamma(c-a)\Gamma(c-b)} \ . 
\end{equation*}
Using this, the numerator of \eqref{gre1} can be written as:
\begin{eqnarray}
    \left(   G_{\text{bh}}-\frac{\Gamma(1-a)\Gamma(1-b)\Gamma(a+b-c)}{\Gamma(1+a-c)\Gamma(1+b-c)\Gamma(c-a-b)}  \right)\bigg\rvert_{\omega=\omega_i} =&( G_{\text{bh}}- G_{\text{bh}}^*)\big\rvert_{\omega=\omega_i}\nonumber \\
    =& 2i \text{Im}G_{\text{bh}}(\omega, m)\big\rvert_{\omega=\omega_i}  \ . \nonumber\\
\end{eqnarray}

On the other hand, the denominator can be abbreviated as:
\begin{equation}
    \partial_{\omega} \left(1+R_{C_2C_1} \frac{\Gamma(2-c)}{\Gamma(1-a)\Gamma(1-b)}\frac{\Gamma(c-a)\Gamma(c-b)}{\Gamma(c)} \right)\bigg\rvert_{\omega=\omega_i}=\partial_{\omega}e^{-i \theta(\omega, m)}\bigg\rvert_{\omega=\omega_i} \ .
\end{equation}
Finally we can write \eqref{gre1} in the following form:
\begin{equation}
\label{eq:res-ImG}
    \text{Res}(G, \omega_i)=2i \frac{\text{Im}G_{\text{bh}}(\omega, m)\bigg\rvert_{\omega=\omega_i}}{\partial_{\omega}e^{-i \theta(\omega, m)}\bigg\rvert_{\omega=\omega_i}} \ .
\end{equation}
where $\theta(\omega, m)$ is given in \eqref{quant1}.

As we have argued and demonstrated before, when the stretched horizon is very close to the event horizon, the poles condense and we can approximate the sum oven $i$ by integral over $\omega_i$:
\begin{equation}\label{greenid}
    G(\omega, m)=\sum_{i} \left(\frac{\omega^2}{\omega_i^2}  \right)^{\Delta}\frac{\text{Res}(G, \omega_i)}{\omega-\omega_i} \approx \int d\omega_i \frac{\rho_{\omega}(\omega_i)}{\omega-\omega_i} + {\cal O}\left( \Delta \omega_i \right) \ ,
\end{equation}
with
\begin{eqnarray}
\rho_\omega (\omega_i) = \left(\frac{\omega^2}{\omega_i^2} \right)^\Delta {\rm Res}\left(G, \omega_i \right) \frac{di}{d\omega_i} \ , 
\end{eqnarray}
where $\Delta \omega_i$ denotes the ``typical" gap in the spectrum, which we ignore when we go to the continuum description. In the first step in (\ref{greenid}) we have replaced the sum $\sum_i \to \int d\omega_i$ at the cost of introducing a measure $(di)/(d\omega_i)$. There is an intrinsic error is this approximation, which can be roughly estimated by the ${\cal O}\left( \Delta \omega_i \right)$ where $\Delta \omega_i$ will scale differently along the two different quantum numbers.

Now using \eqref{quant1} we can write:
\begin{equation}
    \frac{di}{d\omega_i}=\frac{1}{2 \pi} \frac{d\theta(\omega_i, m)}{d\omega_i}=-\frac{1}{2\pi i}e^{i \theta(\omega_i, m)}\partial_{\omega_i}e^{-i \theta(\omega_i, m)}=-\frac{1}{2\pi i}\partial_{\omega_i}e^{-i \theta(\omega_i, m)} \ . 
\end{equation}
Plugging this back to \eqref{greenid}, it is easy to write $\rho_{\omega}(\omega_i, m)$ as:
\begin{equation}
    \rho_{\omega}(\omega_i, m)=-\frac{1}{\pi}\left(\frac{\omega^2}{\omega_i^2}  \right)^{\Delta} \text{Im}G_{\text{bh}}(\omega, m)\bigg\rvert_{\omega=\omega_i}  \ . \label{specfn}
\end{equation}
Furthermore, using (\ref{greenid}), it is straightforward to show that
\begin{equation}
 2 i {\rm Im}G_R(\omega) \equiv  G(\omega+i\epsilon, m)- G(\omega-i\epsilon, m) \simeq \int d\omega_i  \rho_{\omega}(\omega_i, m) 2\pi i \delta(\omega-\omega_i)=2\pi i \rho_{\omega}(\omega, m) \ . \label{imGreen}
\end{equation}
So as the poles become dense, correlator develops a branch cut with the discontinuity given by \eqref{imGreen}. A comparison between equations (\ref{specfn}) and (\ref{imGreen}) now shows that
\begin{eqnarray}
{\rm Im}G_R(\omega) = - {\rm Im}G_{\rm bh}(\omega) \ . \label{nmqnm}
\end{eqnarray}
The relation above essentially demonstrates that the retarded correlator of a scalar field with an ingoing boundary condition in the BTZ geometry can be exactly captured by the Green's function of the same scalar field with a Dirichlet boundary condition, peovided the condition is imposed on a stretched horizon which is sufficiently close to the event horizon.

\subsection{Position space representation of the Green's function}
\label{s.positionspace}
Given the equality in (\ref{nmqnm}) above, it is now guranteed that the corresponding position-space Green's function will be given by a standard thermal correlation function. Nonetheless, for completeness, we demonstrate below the technical aspects of this derivation, explicitly. 

To begin with, for the simplicity of the calculation, we take an infinite radius limit of the compact space direction of the CFT. This implies that we can replace the discrete $m$-sum as an integral over $m$. Let us also introduce a new set of coordinates $p=\frac{m+\omega}{2}$ and $w=\frac{m-\omega}{2}$ with their conjugate lebeled as $v$ and $u$. Clearly, these are the null directions and the time-like direction is defined as $(u+v)$. The position space Green's function is given by
\begin{equation}
\begin{split}
\label{eq:GF-position}
{\cal F}(u, v) & = \mathcal{N} \int_{\mathbf{R}^2} dw \, dp \, e^{ipv-iwu} G(w, p) \\
 & = \mathcal{N} \int_{\mathbf{R}^2} dw \, dp \, e^{ipv-iwu} \sum_{j} \left(\frac{w^2}{w_j^2}  \right)^{\Delta}\frac{\text{Res}(G(w, p), w_j)}{w-w_j+i\epsilon} \ .
\end{split}
\end{equation}
Here in the denominator $i \epsilon$ is added to fix the contour in the lower half plane.  This means that we are interested in the $u> 0$ part of the correlator. Note that the expression above is exact, modulo the approximation in going from $\sum_m \to \int$. This approximation can introduce errors at order $\Delta m = 1$ and can therefore be ignored safely, since there is no parameter that can grow arbitrarily large in this approximation.  

When the stretched horizon is very close to the event horizon, the collection of discrete poles can be replaced by a branch cut i.e. sum over $j$ in the above equation can be approximated by an integral over $w_j$ which we will write as $w$. So we have:
\begin{equation}
\label{FImGbh}
 {\cal F}(u, v)= \theta(u) \, \mathcal{N} \int_{\mathbf{R}^2} dp \, dw \, e^{ipv-iwu}   \text{Im}G_{\text{bh}}(w, p) \ , 
\end{equation}
where
\begin{equation}\label{bhgreen}
    \text{Im}G_{\text{bh}}(w, p)=\frac{\Gamma(1-\Delta)}{\Gamma(\Delta-1)}  \left(  \frac{\Gamma(\frac{\Delta}{2}-\frac{i w}{r_H})  \Gamma(\frac{\Delta}{2}-\frac{i p}{r_H})}{\Gamma(1-\frac{\Delta}{2}-\frac{i w}{r_H})  \Gamma(1-\frac{\Delta}{2}-\frac{i p}{r_H})}   -\frac{\Gamma(\frac{\Delta}{2}+\frac{i w}{r_H})  \Gamma(\frac{\Delta}{2}+\frac{i p}{r_H})}{\Gamma(1-\frac{\Delta}{2}+\frac{i w}{r_H})  \Gamma(1-\frac{\Delta}{2}+\frac{i p}{r_H})}   \right) \ . 
\end{equation}
We can first carry out the $w$-integration. The first term in \eqref{bhgreen} has poles in the lower half plane at $w=-i r_H \left(n+\frac{\Delta}{2}   \right)$, whereas the second term has poles in the upper half of the complex $w$-plane. As $u>0$, we have to pick the contour in the LHP and the entire contribution comes only from the first term of \eqref{bhgreen}. The final $w$-integration gives the following:
\begin{equation}
\begin{split}
{\cal F}(u, v) & = \theta(u) \, \mathcal{N} \int_{\mathbf{R}} dp \, e^{ipv} \frac{\Gamma(1-\Delta)}{\Gamma(\Delta-1)}  \frac{ \Gamma(\frac{\Delta}{2}-\frac{i p}{r_H})}{\Gamma(1-\frac{\Delta}{2}-\frac{i p}{r_H})} \sum_{n=0}^\infty \frac{(-1)^n/n! e^{-u r_H \left(n+\frac{\Delta}{2}\right)}}{\Gamma(1-\Delta-n)}  \\
 & = \theta(u) \, \mathcal{N} \int_{\mathbf{R}} dp \, e^{ipv} \frac{1}{\Gamma(\Delta-1)}  \frac{ \Gamma(\frac{\Delta}{2}-\frac{i p}{r_H})}{\Gamma(1-\frac{\Delta}{2}-\frac{i p}{r_H})} \left(  2 \sinh{\frac{u r_H}{2}} \right)^{-\Delta} \ . 
\end{split}
\end{equation}
Without performing any further integral, we can already observe that the large $u$-behaviour of the correlator is determined by $e^{- \Delta u r_H/2}$, where $u$ is the null direction. Thus, it is suggestive that the resulting position space correlator looks like a thermal correlator, with an exponential fall-off behaviour in the time-like direction. In this case, it is possible to carry out the $p$-integration as well. The relevant poles are in the LHP at $p=-i r_H \left( n+\frac{\Delta}{2}  \right)$, which, when integrated, picks the $v<0$ part and yields:
\begin{equation}
    \begin{split}
        {\cal F}(u, v) & =\theta(u) \theta(-v)  \frac{\mathcal{N}}{\Gamma(\Delta-1)} \left(  2 \sinh{\frac{u r_H}{2}} \right)^{-\Delta}  \sum_{n=0}^\infty \frac{(-1)^n/n! e^{v r_H \left(n+\frac{\Delta}{2}\right)}}{\Gamma(1-\Delta-n)} \\
         & =  \frac{(1)^{-\Delta} \mathcal{N}}{\Gamma(\Delta-1)\Gamma(1-\Delta)} \, \theta(u)\theta(-v) 2^{-2 \Delta} \left(  \sinh{\frac{u r_H}{2}} \sinh{\frac{v r_H}{2}}\right)^{-\Delta}\\
         & = \widetilde{\mathcal{N}} \, \theta(u)\theta(-v) \, 2^{-2 \Delta} \left(  \sinh{\frac{u \pi}{\beta}} \sinh{\frac{v \pi}{\beta}}\right)^{-\Delta} \ .
    \end{split}
\end{equation}
It is now evident that the correlator above has the correct temporal behaviour in the large time limit.

It is worth noting that the identification of the temporal behaviour follows directly from
\eqref{FImGbh}. Recall, \eqref{FImGbh} is an approximation that results from replacing the dense set of poles with a branch cut along the real axis of the frequency plane. 
In this limit, when an observer no longer distinguishes discrete poles from a smooth distribution, or equivalently, cannot resolve the gap between poles, one can identify the residue of the poles with the imaginary part of the black hole retarted correlator through \eqref{eq:res-ImG}. This fact enters the computation of the position space Green's function leading to \eqref{FImGbh} through \eqref{eq:GF-position}. In other words, identifying the residues of the poles of Green's function with the imaginary part of the black hole Green's function in the continuum limit ensures the thermal identification of the position space Green's function. One very important point\footnote{We thank Shiraz Minwalla for raising this point and for useful discussions regarding the thermal identification of the Green's function.} here is that, in the computation of Green's function with Dirichlet condition on the brickwall, we need to keep both the positive-frequency incoming and positive-frequency outgoing modes near the wall. Therefore, naturally, this Green's function cannot be rendered as the retarded Green's function. It only behaves like a retarded Green's function in the small-radius continuum limit. In the dynamical scenario, for instance in a situation where the cutoff is a collapsing shell as we will elaborate in section \ref{sec:collapse}, the small radius limit can be identified naturally as a late-time limit. For this, the emergence of thermalilty becomes even more natural.
This is very similar to the version of the information paradox of \cite{Maldacena:2001kr}. This was observed already in \cite{Giusto:2023awo} in a very different model with a supergravity solution. Note that, this observation is very generic: It does not depend on the specific nature of the spectrum, as long as there exists a parameter in the problem tuning which the gap in the spectrum can be made parametrically smaller. For a generic quantum mechanical system, it appears plausible that a similar physics will generally hold, although we are not aware of a precise statement. It will be very interesting to explore and understand this aspect better.\footnote{Note that, the approximation of discrete energy levels by a smooth distribution function, holographically, becomes equivalent to imposing an ingoing boundary condition at the event horizon. The latter is also equivalent to a regularity condition in the Euclidean description. It is likely that the smooth distribution approximation can be mapped to a Euclidean regularity condition in a more general framework, {\it i.e.}~independent of the degrees of freedom, the geometric background or perhaps even without gravity. We would like to address this issue in near future.}

The explicit knowledge of the spectrum becomes relevant in the correction terms, which depart from an exponential decay at late times. As long as we consider a quantum mechanical system with a discrete spectrum, the two-point function cannot decay for an arbitrarily long time, simply because at some point we will begin recovering the information of the pure state.\footnote{This is sometimes dubbed as echoes in the two-point function. See {\it e.g.}~\cite{Bena:2019azk, Ikeda:2021uvc} for recent explicit computations of echoes in fuzzball geometries.} We will now estimate the role of the corrections for our case. In the absence of an analytical result in the entire regime, we will only be able to provide an estimate based on the WKB-result for the spectrum.

Let us recall that the WKB-answer is give by
\begin{eqnarray}
\omega(n,m) = \pi (8 n+3) (4 W(x))^{-1} \ , \quad x = \frac{\pi(8n+3)}{2\sqrt{2}m \sqrt{\epsilon}}\ ,
\end{eqnarray}
in the limit $\epsilon \to 0$, where $\epsilon$ is the coordinate distance of the stretched horizon from the event horizon. As we have already elucidated in (\ref{greenid}), the error in going from the discrete sum to a smooth integral is proportional to be gap in the spectrum, as $\epsilon \to 0$. Suppose, the observer has an access up to an energy-scale $E_{\rm cut}$. This defines the total number of states in the spectrum: $E_{\rm cut} = \sum_i^{N_{\rm cut}} \omega_i$. As we have demonstrated previously, we $\epsilon \to 0$, $N_{\rm cut} \to \infty$.\footnote{Very precisely, we have explicitly shown that as $\epsilon$ becomes smaller and smaller $N_{\rm cut}$ becomes larger and larger. A formal statement of an infinite limit cannot be made precisely, since we do not have the required control over the spectrum, nonetheless, this is enough for our purposes.}

The error in (\ref{greenid}) can therefore be estimated as:
\begin{eqnarray}
\Delta G(\omega_i) = \frac{1}{N_{\rm cut}} \Delta \omega_i \ ,
\end{eqnarray}
where $\Delta\omega_i$ denotes the gap along the $n$ or $m$ quantum numbers. This gap can be expressed as:
\begin{eqnarray}
 && \left.   \Delta \omega \right|_{m ={\rm fixed}} = \frac{2\pi}{1 + \frac{\pi(8n+3)}{4\omega}} \ , \\
 && \left. \Delta \omega  \right|_{n={\rm fixed}} = \sqrt{\frac{\epsilon}{2}} \, {\rm exp}\left[ \frac{\pi(8n+3)}{4\omega} \right] \frac{1}{1 + \frac{\pi(8n+3)}{4\omega}}  \ .
\end{eqnarray}
These can now be used to estimate the error in the position-space two-point function, respectively. We emphasize again that, given our WKB-spectrum, it is rather cumbersome to carry out the explicit Fourier transformation to the position space, if possible at all. For example, it is possible to carry out the $u$-integral of the error term for fixed $m$, which yields:
\begin{eqnarray}
    \int e^{- i w u} \Delta G(\omega_i) \sim \frac{1}{N_{\rm cut}} \left[ \alpha_1 \delta(u) + \alpha_2 e^{i \alpha_3 u} + \ldots \right] \ ,
\end{eqnarray}
where $\alpha_{1,2,3}$ are real constants. While the first term is simply a contact-term, the second term is oscillatory. We expect an oscillatory behaviour at late times and therefore it is an encouraging estimate.

Thus, on general grounds, the error term is expected to behave as:
\begin{eqnarray}
    \Delta {\cal F} \sim \frac{1}{N_{\rm cut}} e^{i \alpha_3 t} + \ldots \ .
\end{eqnarray}
Comparing ${\cal F}$ with $\Delta {\cal F}$, the time-scale at which the oscillatory terms appear is given by
\begin{eqnarray}
    t_*^{(1)} = \log N_{\rm cut} \ . 
\end{eqnarray}
The bottom-line is: At $t_*$ onwards the thermal correlator will display echoes. Following a similar argument when $n$ is fixed yields:
\begin{eqnarray}
    \Delta {\cal F} \sim \frac{1}{N_{\rm cut}} \sqrt{\epsilon} \, e^{i \alpha_4 t} + \ldots  \ ,
\end{eqnarray}
where $\alpha_4$ is a real-constant. The corresponding echo time-scale is now augmented further by an explicit factor of $\epsilon$, which yields: 
\begin{eqnarray}
    t_*^{(2)} = \log \left(\frac{N_{\rm cut}}{\sqrt{\epsilon}} \right) \ , 
\end{eqnarray}
and therefore $t_*^{(1)} \ll t_*^{(2)}$. This implies that information recovery takes a non-perturbatively longer time once the angular momentum quantum numbers are included.

The same statement can be recast in a slightly different manner. Consider the case when the angular momentum is kept fixed. Suppose we perform an ``ergodic averaging" of the position-space correlator, in particular of the error term, as follows:
\begin{eqnarray}
    \frac{1}{\Lambda} \lim_{T \to \Lambda} \int_0^T \Delta {\cal F} dt \sim {\cal O}(\Lambda^{-1}) \ .
\end{eqnarray}
Arranging $\Lambda \sim \epsilon^{-1/2}$ will tune the error term to be of the same order as the one in which different angular momenta modes are summed over. Thus, in one case the resulting position-spcae correlator approaches to its thermal value upon performing an ergodic averaging, while in the other case no such averaging is required. This feature has a qualitative similarity to the notion of ``weak thermalization" and ``strong thermalization" of \cite{Ba_uls_2011}.\footnote{We thank Mohsen Alishahiha for raising this point to us.} It is interesting to note that this notion of weak and strong thermalization appears to be directly tied to the absence and presence of quantum chaos in the corresponding part of the spectrum.\footnote{Recall that, earlier works in {\it e.g.}~\cite{Das:2022evy, Das:2023xjr} have demosntrated how a non-trivial ramp (of unit slope) structure emerges in the spectral form factor because of the non-trivial level correlations along the angular momenta spectrum. This does not happen along the $n$ quantum numbers which are appear linear to a good approximation.}

\subsection{Some Comments on Poincar\'{e} Recurrence}
\label{s.somecom}
As we have seen in previous sections, having an explicit analytic handle on the spectrum allows for an explicit estimate on various aspects of the associated physics. In this section, we will offer comments on Poincar\'{e} recurrence of the scalar-sector. In this section we will not present a new calculation but we will review the core arguments that were already discussed in \cite{Barbon:2003aq}\footnote{See also \cite{Barbon:2014rma} for related discussions on the nature of long-time quantum noise for black holes in an asymptotically AdS geometry.} Poincar\'{e} recurrence, which is a theorem in classical dynamics and results from a compact phase-space and the existence of phase-space volume preserving canonical flow, does not generalize trivially in the quantum regime. Nonetheless, it is reasonable to expect that this holds for a generic class of quantum systems with unitary time-evolution and a discrete energy spectrum.

To quantify the recurrence, one may consider auto-correlation functions of local operators in the quantum system: $G(t)= {\rm Tr}\left( \rho {\cal O}(t) {\cal O}(0)\right)$, for a general self-adjoint operator ${\cal O}$, given the specific state $\rho$. In our case, the density matrix can represent any energy eigenstate of the scalar sector as the pure state. As we have argued previously, the auto-correlation function is dominated by a thermal correlation function in which the temperature is fixed at the Hawking temperature of the black hole. Nonetheless, this thermal answer emerges as we approximate the collection of the poles in the corresponding Green's function by a smooth branch-cut. 

As explained in \cite{Barbon:2003aq}, the periodicity of the correlator can be quantified by defining the observable $ L(t) = \left| G(t)/G(0)\right|^2$, and subsequently performing an ergodic averaging: $\bar{L} = \lim_{T\to \infty} \int_0^T dt L(t)$. Based on general arguments, the ergodic average is bounded from below: ${\bar L} \sim e^{- S}$, where $S$ is the microcanonical entropy. As long as the spectrum is discrete, the ergodic average is exponentially small, but non-vanishing. By construction, in the scalar one-loop determinant sector the recurrence index is exponentially suppressed in entropy. The knowledge of the explicit modes now enable us to calculate the sub-leading terms in the ergodic averaged correlator. At present, however, we are unaware whether these sub-leading contributions carry any universal or interesting physics in them, and therefore we leave a systematic analyses of the same for future.

\section{Probe Scalar Quantization in a Rotating BTZ-Geometry}

In this section, we will collect the salient observations made in \cite{Das:2023xjr}. We will not present a new computation here, however, we include a general discussion to complement our discussions so far. Our special focus will be on the ramification of the non-trivial level-correlation in the angular momenta. The background metric is the well-known rotating BTZ geometry:
\begin{eqnarray}
ds^2 = && \left(M - \frac{r^2}{\ell^2} \right) dt^2  + r^2 d\psi^2 + 2 \left(- \frac{J}{2} \right) dt d\psi + \nonumber\\
&& \left( - M + \frac{J^2}{4r^2} + \frac{r^2}{\ell^2} \right)^{-1} dr^2 \ .
\end{eqnarray}
The corresponding Klein-Gordon equation 
\begin{eqnarray}
    \Box \Phi = \mu^2 \Phi \ , 
\end{eqnarray}
can be written as:
\begin{eqnarray}
    z (1-z) \frac{d^2F}{dz^2} + \left(c - (1+a+b)z \right) \frac{dF}{dz} - a b F(z) =0 \ ,
\end{eqnarray}
where
\begin{eqnarray}
    a = \beta - i \frac{\ell^2}{2(r_+ + r_-)} \left( \omega + \frac{m}{\ell}\right)  \ , \quad b = \beta - i \frac{\ell^2}{2(r_+ + r_-)} \left( \omega - \frac{m}{\ell}\right) \ , \quad c = 1 - 2 i \alpha \ , \nonumber\\
\end{eqnarray}
with
\begin{eqnarray}
\alpha = \frac{\ell^2 r_+}{2(r_+^2 - r_-^2)} \left( \omega - \Omega_H m  \right)  \ , \quad \beta = \frac{1}{2}\left(1 - \sqrt{1 + \mu^2} \right) \ , \quad \Omega_H = \frac{J}{2r_+^2} \ .
\end{eqnarray}
Also, $r_\pm$ are the two roots denoting the outer and the inner horizons: $r_\pm = \frac{M\ell^2}{2} \left(1 \pm \sqrt{1 - \frac{J^2}{M^2 \ell^2}}\right)$. The scalar field $\Phi$ has been expanded in the modes corresponding to the $t$ and $\psi$ translation symmetries: $\Phi = e^{-i \omega t} e^{i m \psi} z^{-i\alpha}(1-z)^{\beta} F(z)$, where the radial coordinate is redefined as: $z = (r^2 - r_+^2)/ (r^2 - r_-^2)$. 

The quantization of the scalar sector can now be carried out in a straightforward manner, as is also described in detail in \cite{Das:2023xjr}. The quatization condition now breaks the ${\mathbb Z}_2$ symmetry under $m \to - m$ and the resulting spectrum lacks the non-trivial level correlations, as long as $J$ is non-vanishing. However, since the scalar normal mode spectrum is not strictly positive, it is natural to define a {\it regulated} spectrum $\tilde{\omega} = \omega - \Omega_H m$, which turns out to be always positive.\footnote{The physical reason behind excluding the negative modes is the presence of a superradiance instability of the modes in the regime $\omega < \Omega_H m$. Essentially, in this limit, the incident scalar mode with a given angular momentum is able to extract energy from the scattering potential of the rotating black hole. See {\it e.g.}~for more detailed discussions in \cite{Cardoso:2004hs, Cardoso:2004nk}.} As described in \cite{Das:2023xjr}, the regulated modes $\tilde{\omega}$ define a grand-canonical partition functions for the scalar one-loop determinant:
\begin{eqnarray}
    Z\left[\beta \right] = \sum e^{-\beta \omega} \to Z\left[\beta, \Omega_h \right] = \sum e^{-\beta \left(\omega - \Omega_H m \right)} \ .
\end{eqnarray}

The modes $\tilde{\omega}$ exhibits the very same level-correlatations that is observed for the modes $\omega$, when $J$ vanishes. Therefore, the corresponding spectral form factor $Z[\beta + i t, \Omega_H]$ displays the Dip-Ramp-Plateau behaviour of \cite{Das:2022evy}. As a result, all local operators in the canonical ensemble will exhibit a ``weak thermalization" as the brickwall is placed close enough to the event horizon, in the sense we have described earlier. However, in the grand-canonical ensemble, the hierarchy of the pole condensation along the angular momentum will be restored and there will be both a ``weak thermalization" (when the operator is projected onto a fixed angular quantum number) as well as a ``strong thermalization" (when the operator excites all angular momenta, generically). Evidently, all observations that result from the non-trivial level correlations of the angular momenta will be present in the grand canonical ensemble. 

This is not particularly surprising and bears a resemblance to the Generalized Gibbs ensemble to describe thermalization of integrable systems in general. The take-away message here is simply that these features depend crucially on the respective ensemble.

The extremal limit, on the other hand, is much simpler and yields a linear spectrum. In this case, independent of the ensemble, there is only a ``weak thermalization", provided we do not identify the temperature of the scalar sector with the vanishing temperature of the extremal BTZ geometry. For more details on this case, we refer the interested Reader to appendix C and to \cite{Das:2023xjr}, where several relevant details are provided.

The important aspect to note is that in the limit when the brickwall approaches the event horizon, {\it i.e.}~when the poles are replaced by a branch cut, the corresponding position space correlator is given by equation (\ref{Cuvext}), which is essentially a Fourier transformation of the imaginary part of the extremal black hole correlator $G_{\rm ext}$ in the momentum space. Using equations (\ref{GextGextstarext}) and (\ref{resGext}), in the evaluation of this integral, only $G_{\rm ext}$ will contribute since the poles of $G_{\rm ext}^*$ lie in the upper half plane.

The computation proceeds in an identical manner as is performed in {\it e.g.}~\cite{Bena:2019azk} for a family of six-dimensional $\frac{1}{8}$-BPS geometries that are asymptotically AdS$_3\times S^3$. As a result, the position space correlator will be given by a thermal correlator where the temperature is fixed at $T_R =(r_- + r_+)/(2\pi)$, {\it i.e.}~the right-moving temperature. Note that, in the extremal limit both the Hawking temperature as well as the left-moving temperature vanishes. It may be confusing to note that, while the spectral form factor is agnostic about the temperature about the scalar sector, from the two-point correlation function, it is unambiguously fixed. Recall, however, that the temperature is fixed only when the Dirichlet wall approaches the event horizon. In this limit, the scalar one-loop determinant is also placed at the same temperature as that of the black hole, at least for the non-extremal case. This is clear in the Euclidean description, since the corresponding section almost touches the tip of the Euclidean cigar. While we are still free to choose an arbitrary periodicity for the scalar sector at the cost of inducing a conical defect in the geometry, Euclidean regularity is a natural choice.

\section{Probe Scalar Quantization: Collapsing Black Holes}
\label{sec:collapse}

In this section we will analyze the free scalar field theory in a geometry where a spherical shell is collapsing. The primary motivation is to factor in a dynamical geometric background, generalizing the Dirichlet brickwall model that we have discussed above. There is a further crucial difference. We will consider a geometry which is patch-wise glued at the location of the shell. The inside-shell geometry is global AdS while the outside is a BTZ-background. As a result of the smooth and consistent gluing, we can no longer impose a Dirichlet boundary condition for the probe scalar at the shell.\footnote{Physically, this implies that there are probe scalar degrees of freedom inside the shell as well as outside of it. Since this model is dual to a pure state in the dual CFT, there is no information paradox in this case till the shell collapses and forms a black hole. If we are to bring in the fuzzball-intuition, it is likely that before the collapse to a classical black hole happens, the end state of the collapse forms a fuzzball-like geometry. However, at this point, this is a plausible scenario at best, the details of which remain a challenge to understand. It will, nonetheless, be very interesting to capture the fuzzball-like state starting from a reasonable initial data and evolving in time.} Nevertheless, we will observe that the key features of pole-accumulation of the retarded correlator persist in this model as well.

Let us consider a collapsing shell in $(2+1)$-dimensions. Figure \ref{model} shows a snapshot of the collapse at some instant $t=t_0$. The position of the shell, denoted by $r=r_s$, is depicted in black. Outside this black circle, the geometry is given by the BTZ metric, and inside it is global AdS. Note that, similar analyses have previously appeared in \cite{Danielsson:1999zt, Danielsson:1999fa}, however, there are two crucial differences with our work here. First, we are using a smooth boundary condition at the surface of the shell and not an ingoing boundary condition. Secondly, we will focus on the analytic structure of the retarded correlator and draw conclusions about an emerging thermal physics from it. 
\begin{figure}[H]
    \centering
    \includegraphics[width=.87\textwidth]{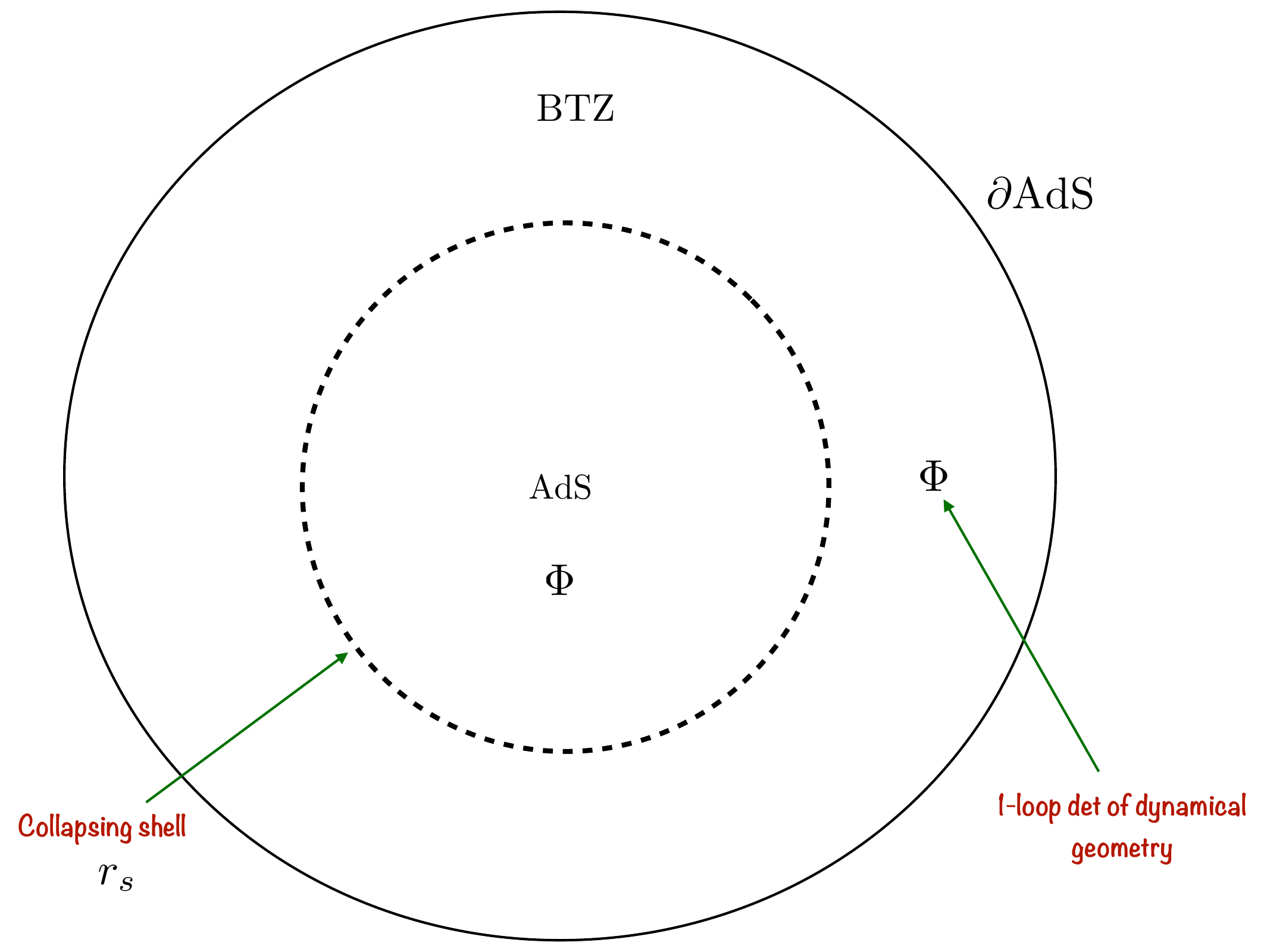}
   \caption{A toy model for a $2+1$-dimensional collapsing shell at some instant $t=t_0$. Outside the shell, the metric is BTZ, and inside it is global AdS.}
    \label{model}
\end{figure}
The metric is given by the following:
\begin{equation}\label{metric}
    ds^2=-f(r) dt^2+\frac{dr^2}{f(r)}+r^2 d\psi^2 \ ,
\end{equation}
where
\begin{equation}\label{metfn}
  f(r)=\begin{cases}
    f_1=1+r^2, & \text{for $r<r_s$} \ .\\
    f_2=r^2-r_H^2, & \text{for $r>r_s$}  \ .
  \end{cases}
\end{equation}
As before, we want to quantize a probe scalar of mass $\mu$ in this background {\it i.e.}~we want solve the following Klein-Gordon equation:
\begin{equation}\label{eom1}
    \Box \Phi\equiv \frac{1}{\sqrt{|g|}}\partial_{\nu}\left(\sqrt{|g|}\partial^{\nu}\Phi\right)=\mu^2 \Phi \ .
\end{equation}
Since the metric is invariant under the translation of $t$ and $\psi$, we can use the ansatz, $ \Phi = \sum_{\omega, J} e^{-i\omega t}  e^{i J\psi} \phi_{\omega, m}(r)$, and with this ansatz, the radial part of \eqref{eom1} satisfies:
\begin{equation}\label{eom1}
    \frac{1}{r} \frac{d}{dr} \left( r f(r) \frac{d \phi(r)}{dr}   \right) +\left( \frac{\omega^2}{f(r)}-\frac{m^2}{r^2}-\mu^2 \right) \phi(r)=0 \ .
\end{equation}
Where we have suppressed the $\omega, m$ indices of $\phi(r)$. Let $\phi_1$ and $\phi_2$ represent solutions inside and outside the shell, respectively. The matching conditions are:
\begin{align}\label{bdry_condn}
    \phi_1\vert_{r=r_s}\, &=\phi_2\vert_{r=rs} \ , \nonumber \\
    f_1(r) \partial_r \phi_1 \vert_{r=r_s}\, &= f_2(r) \partial_r \phi_2 \vert_{r=r_s} \ .
\end{align}
We are interested in the Green's function of the system. Following Son-Starinets\cite{Son:2002sd} prescription and using the conditions \eqref{bdry_condn}, we can write the Green's function, which is the ratio of the normalizable mode to the non-normalizable mode, as:
\begin{equation}\label{green1}
    G(\omega,m)=- \frac{\phi_1 f_2 \partial_r \phi_2^{(-)}-\phi_2^{(-)}f_1 \partial_r \phi_1 }{ \phi_1 f_2 \partial_r \phi_2^{(+)} - \phi_2^{(+)}f_1 \partial_r \phi_1  } \ ,
\end{equation}
where everything is evaluated at $r=r_s$. Here:
\begin{itemize}
\item $\phi_1$: solution of the Klein-Gordon (K-G) equation that is regular at the origin $r=0$.
\item $\phi_2^{(+)}$: normalizable part of the solution of the Klein-Gordon equation outside the shell which is regular at the conformal boundary, $r\rightarrow \infty$.
\item $\phi_2^{(-)}$: non-normalizable part of the solution of the Klein-Gordon equation outside the shell which blows up at the boundary, $r\rightarrow \infty$.
\end{itemize}
There are, of course, ambiguities up to multiplicative constants in the solutions. These constants simply appear as overall factors in \eqref{green1} and we can ignore them.

\subsection{Inside the shell: $r < r_s$}

Inside $r=r_s$, the geometry is that of an empty AdS in global coordinates and is given by
\begin{equation}
    ds^2_{\rm in}=-(1+r^2)dt^2+\frac{dr^2}{1+r^2}+r^2 d\psi^2 \ .
\end{equation}
The radial part of \eqref{eom1} becomes: 
\begin{equation}
    (1+r^2) \phi''(r)+ \left(3r+\frac{1}{r} \right) \phi'(r)+ \left( \frac{\omega ^2}{r^2+1}-\frac{m^2}{r^2} -\mu ^2 \right) \phi(r)=0 \ .
\end{equation}
The solution is given by
\begin{equation}\label{solin1}
    \phi_{in}(r)=(1+r^2)^{\omega/2} \bigg( C_1 \, r^m \, {}_2 F_1 (a, b, c; -r^2)+ C_2 \, (m\rightarrow -m) \bigg) \ , 
\end{equation}
where
\begin{align}
   a &=\frac{1}{2} (1+m-\sqrt{1+\mu^2}+\omega) \ , \nonumber \\
   b &=\frac{1}{2}(1+m+\sqrt{1+\mu^2}+\omega) \ . \nonumber \\
   c &=1+m  \ . \nonumber
\end{align}
Here in \eqref{solin1}, the second term blows up at the origin ($r\rightarrow 0$), so regularity at the origin (for positive $m$) sets $C_2=0$,\footnote{Note that this condition implies that there are both ingoing and outgoing modes at the location of the shell, unlike the Dirichlet boundary condition on the static brickwall.} thus the solution inside the shell, which is regular at $r=0$, is given by
\begin{equation}\label{solin2}
    \phi_1(r)=C_1 \, (1+r^2)^{\omega/2} \, {}_2F_1 (a, b, c; -r^2)  \ .
\end{equation}
%

\subsection{Outside the shell: $r>r_s$}

Outside $r=r_s$, the spacetime looks like a BTZ black hole and its metric is given by 
\begin{equation}
    ds^2_{\rm out}=-(r^2-r_H^2)dt^2+\frac{dr^2}{(r^2-r_H^2)}+r^2 d\psi^2 \ ,
\end{equation}
The radial part of the scalar field equation, using the same ansatz as before, is given by
\begin{equation}
    (r^2-r_H^2) \phi''(r)+ \left(3 r-\frac{r_H^2}{r} \right) \phi'(r)+ \left( \frac{\omega ^2}{r^2-r_H^2}-\frac{m^2}{r^2} -\mu ^2 \right) \phi(r)=0 \ ,
\end{equation}
whose solution is the following:
\begin{align}\label{solout1}
    \phi_{\text{out}}(r)= &\left(1-\frac{r_H^2}{r^2} \right)^{-\frac{i \omega}{2 r_H}}  \Bigg( D_1 \left(\frac{r_H}{r} \right)^{\Delta_-} {}_2F_1 \left(a, b, 1+a+b-c; \frac{r_H^2}{r^2} \right) \nonumber \\
    & \hspace{3.5cm}+D_2 \left(\frac{r_H}{r} \right)^{\Delta_+} {}_2F_1 \left (c-a, c-b, 1-a-b+c; \frac{r_H^2}{r^2} \right) \Bigg) \ ,
\end{align}
where
\begin{align}
    \Delta_{\pm} &=1\pm \sqrt{1+\mu^2} \ , \nonumber \\
    a,b &= \frac{\Delta_{-}}{2}-\frac{i}{2r_H} (\omega \pm m) \ , \nonumber \\
    c &=1-\frac{i \omega}{r_H} \ . \nonumber
\end{align}
In \eqref{solout1}, the first term is non-normalizable whereas the second term is normalizable. So, in terms of the notations of the previous section, we obtain:
\begin{align}
    \phi_2^{(+)} &=\left(1-\frac{r_H^2}{r^2} \right)^{-\frac{i \omega}{2 r_H}}  \left(\frac{r_H}{r} \right)^{\Delta_+} {}_2F_1  \left(c-a, c-b, 1-a-b+c; \frac{r_H^2}{r^2} \right) \ , \label{phi21}  \\
     \phi_2^{(-)} &=\left(1-\frac{r_H^2}{r^2} \right)^{-\frac{i \omega}{2 r_H}}  \left(\frac{r_H}{r} \right)^{\Delta_-} {}_2F_1 \left(a, b, 1+a+b-c; \frac{r_H^2}{r^2} \right) \ . \label{phi22}
\end{align}
%

\subsection{Green's function}

\begin{figure}
\begin{subfigure}{0.47\textwidth}
    \centering
    \includegraphics[width=\textwidth]{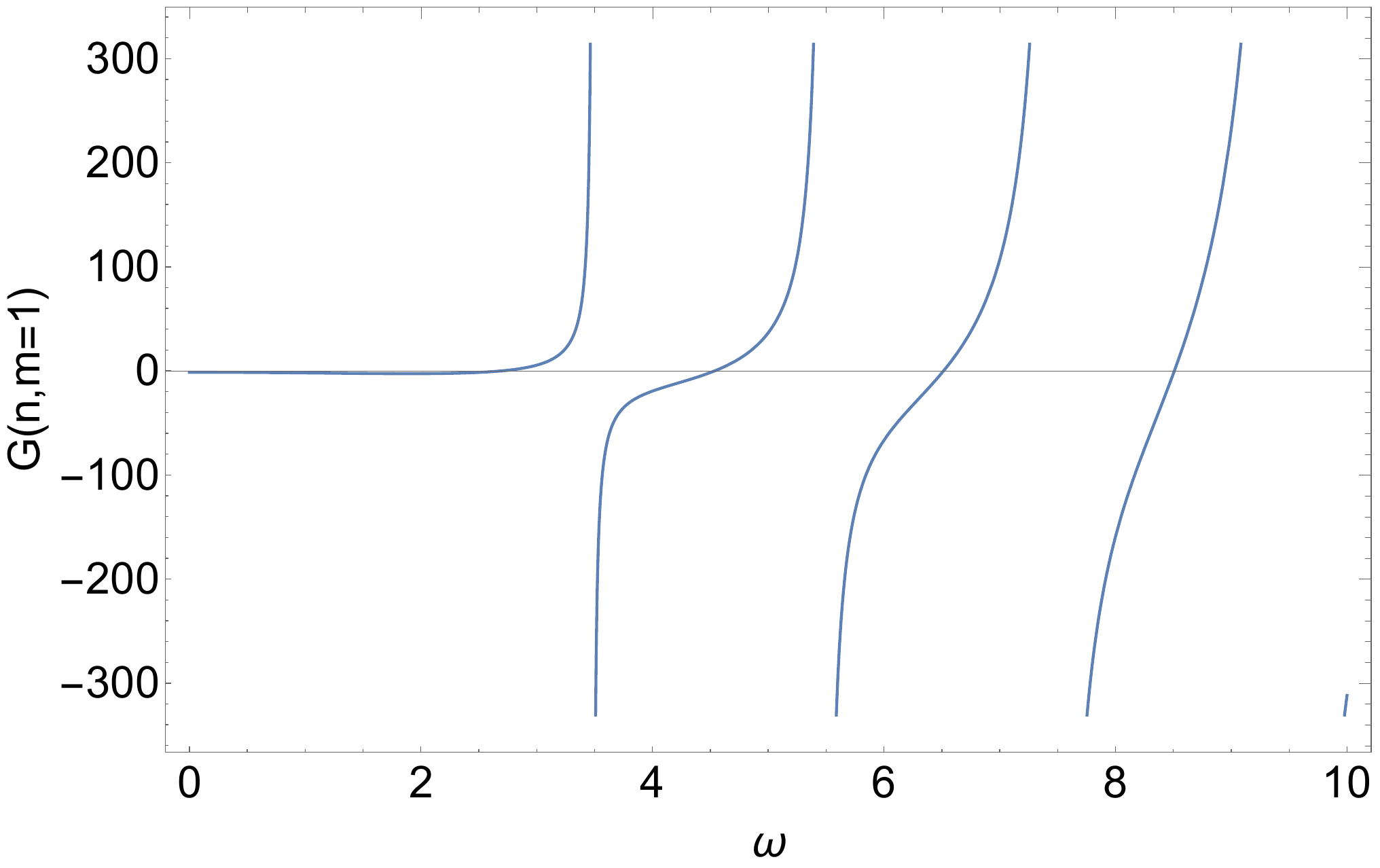}
    \end{subfigure}
    \hfill
    \begin{subfigure}{0.47\textwidth}
    \includegraphics[width=\textwidth]{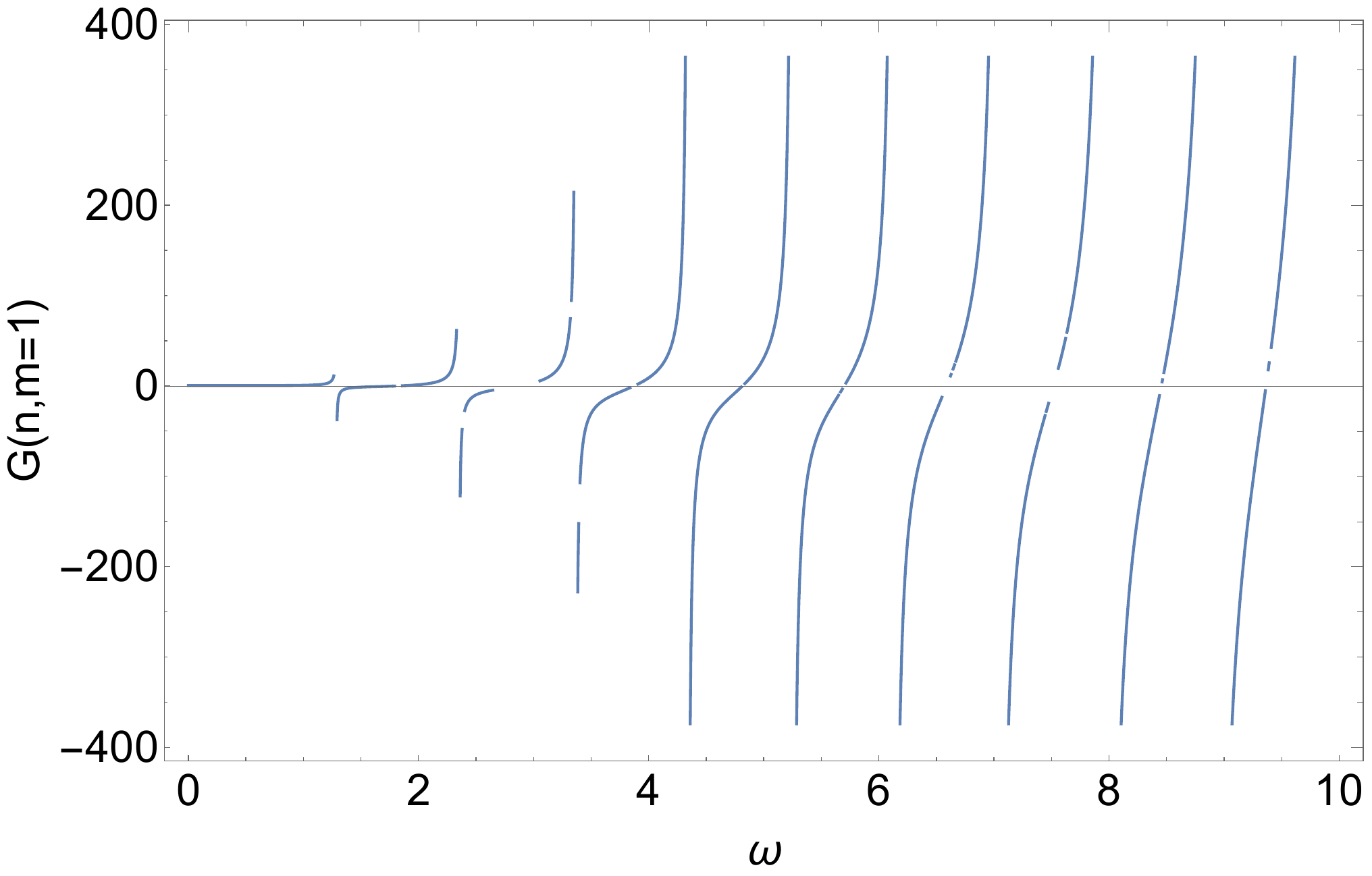}
    \end{subfigure}
    \caption{Plot of the Green's function for fixed $m=1$. $r_s=10$ for the left, whereas $r_s=1.01$ for the right. This shows the poles are coming closer and closer as the shell starts moving from the boundary towards the horizon. Other parameters are $\mu=1.1$ and $r_H=1$.}
    \label{green_plot}
\end{figure}

\begin{figure}[H]
    \centering
    \includegraphics[width=.47\textwidth]{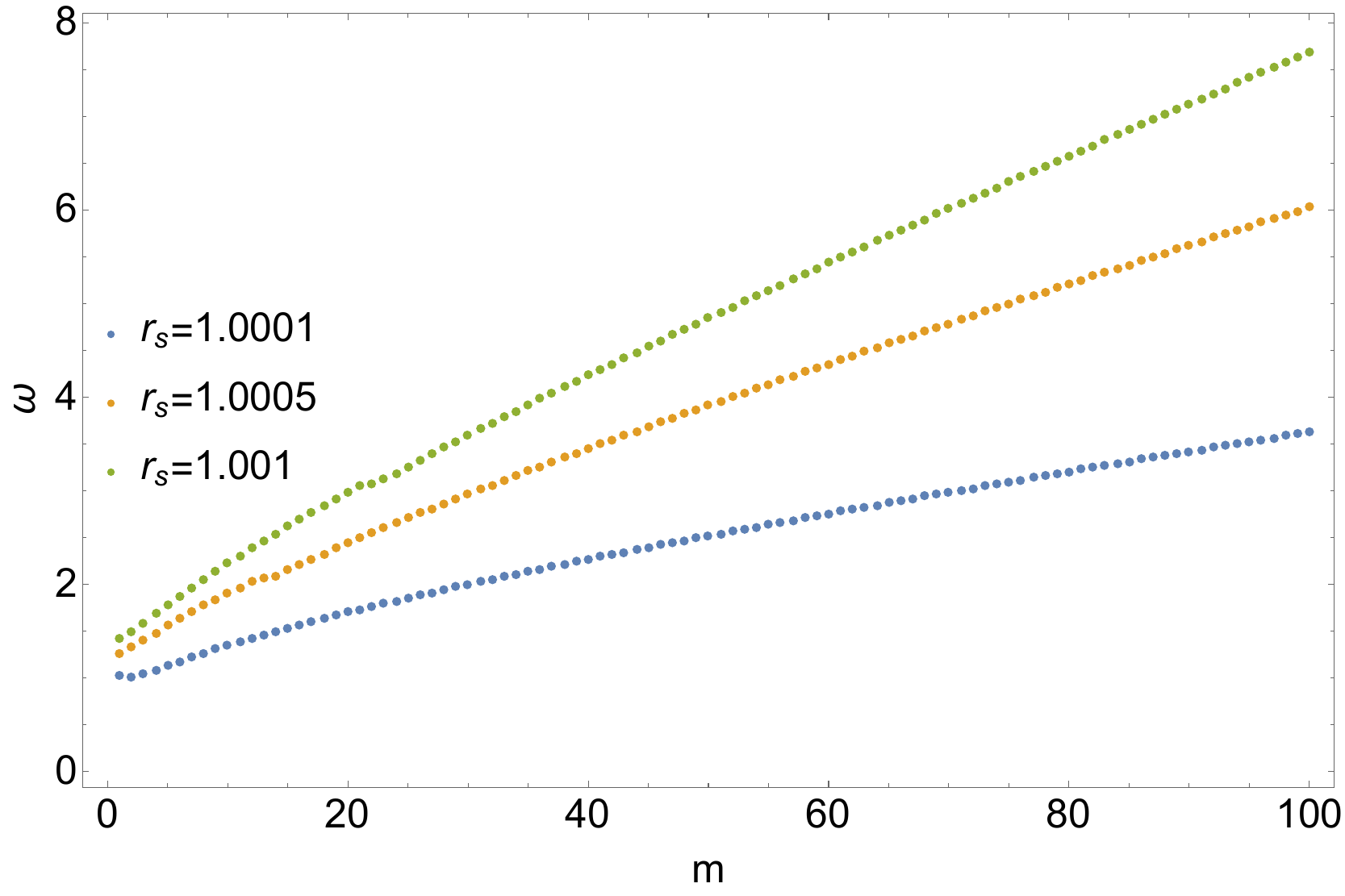}
   \caption{The figure shows the $r_s$ dependence of the normal modes along the $m$-direction for fixed $n=1$. The blue, yellow, and green curves correspond to $r_s = 1.0001$, $1.0005$, and $1.001$, respectively. Other parameters are $\mu = 1.1$ and $r_H = 1$.}
    \label{spectrum_m}
\end{figure}

These expressions can be substituted into \eqref{green1} to obtain the Green's function of the boundary theory. Here, in Figure \ref{green_plot}, the Green's function is plotted for fixed $m=1$. These two pictures clearly demonstrate that as $r_s$ is decreased, {\it i.e.}~as the shell moves closer to $r_H$, the poles in the Green function come closer to each other.
Figure \ref{spectrum_m} shows the normal modes of the system along the $m$ directions.
%

\subsubsection{Case I (when $r_s\rightarrow \infty$)}

Let us also comment on a particular limit, when $r_s\rightarrow \infty$. In this limit, we can approximate $\phi_1$ and $\phi_2^{(+)}$ as follows:
\begin{align}
    \phi_1 &\approx r^{\omega+m} (C_1 \, r^{-2 a}+C_2 \, r^{-2 b}) \ , \\
    \phi_2^{(+)} &\approx \left(1-\frac{r_H^2}{r^2} \right)^{-\frac{i \omega}{2 r_H}}  \left(\frac{r_H}{r} \right)^{\Delta_+} \ .
\end{align}
Substituting these approximations back into the denominator of equation \eqref{green1} and expanding it near $r_s \rightarrow \infty$, we find the denominator to be:
\begin{equation}
    \text{denominator}=-\frac{2 r_H \, \Gamma \left(\sqrt{\mu ^2+1}+1\right) \Gamma (m+1) \, r_s ^{\sqrt{\mu ^2+1}-1} \left(\frac{r_H}{r_s}\right)^{\sqrt{\mu ^2+1}}}{\Gamma \left(\frac{m-\omega +\sqrt{\mu ^2+1}+1}{2} \right) \Gamma \left(\frac{m+\omega +\sqrt{\mu ^2+1}+1}{2} \right)}+\ldots \ . 
\end{equation}
Setting the denominator equal to zero implies poles of the Green's function, which are at:
\begin{equation}\label{omegan}
    \omega_n = \pm (1+m+2n+\sqrt{\mu ^2+1} ) \ .
\end{equation}
A more accurate approach involves substituting equations \eqref{solin2}, \eqref{phi21} and \eqref{phi22} into equation \eqref{green1}, and then expanding that expression near $r_s \rightarrow \infty$. Interestingly, this procedure yields the same values for $\omega_n$ as given in equation \eqref{omegan}.

\subsubsection{Case II (when $r_s\rightarrow r_H$)}

\begin{figure}
\begin{subfigure}{0.47\textwidth}
    \centering
    \includegraphics[width=\textwidth]{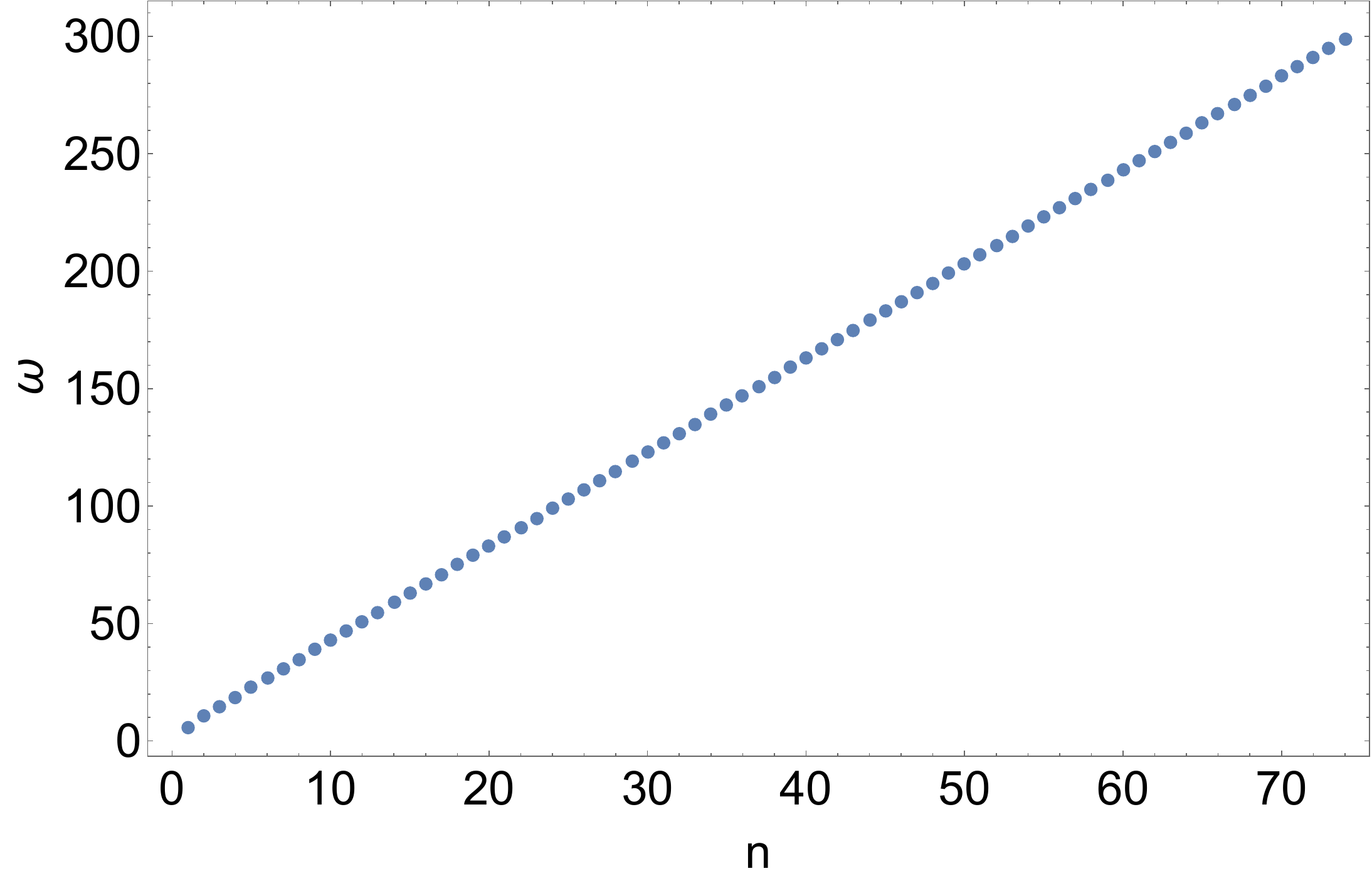}
    \end{subfigure}
    \hfill
    \begin{subfigure}{0.47\textwidth}
    \includegraphics[width=\textwidth]{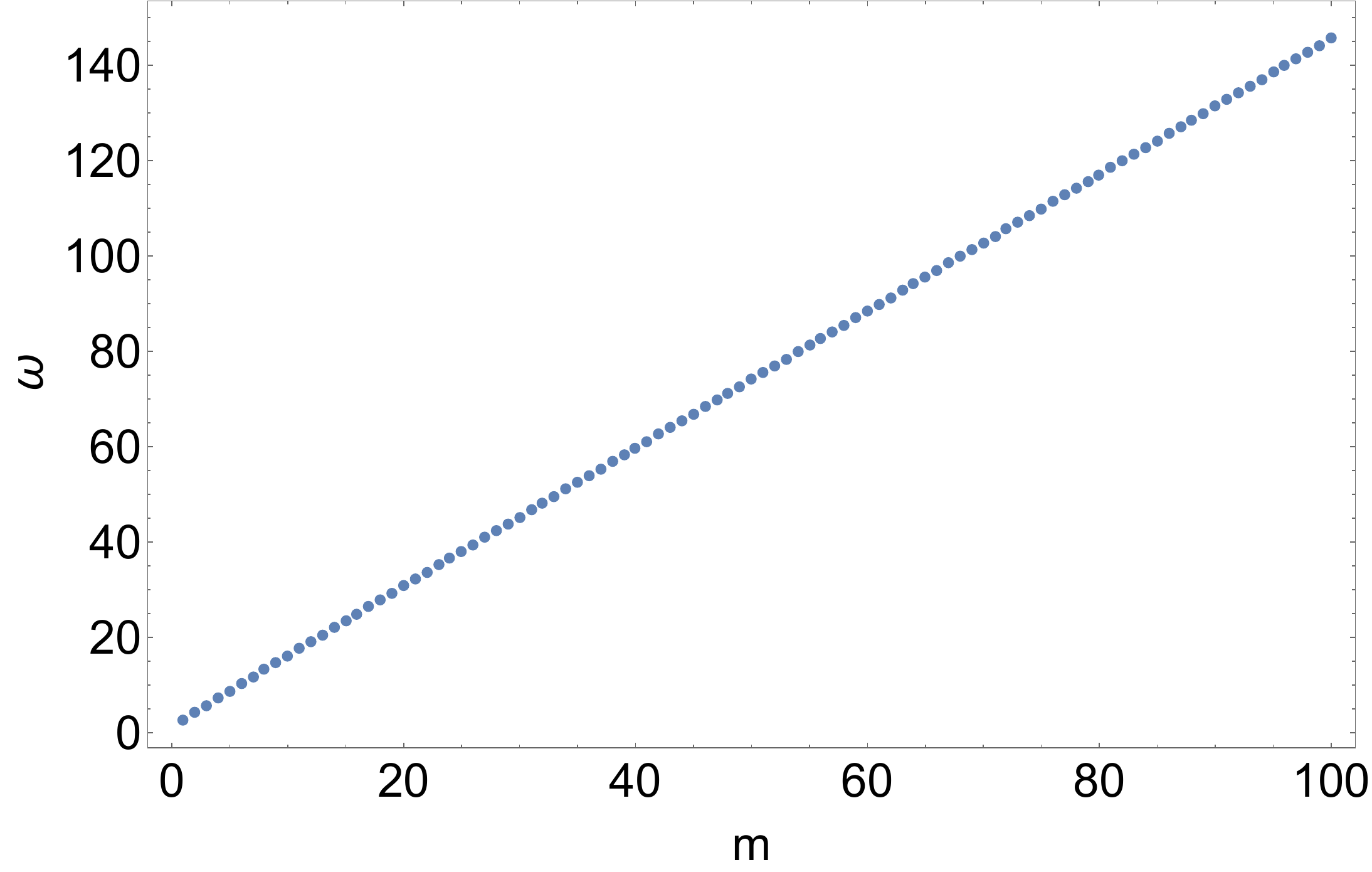}
    \end{subfigure}
    \caption{These pictures are showing normal modes corresponding to the quantization condition \eqref{quantzn2} along $n$-direction (left) and $m$-direction (right) for fixed $m=3$ and $n=1$ respectively. Other parameters are $\mu=1.1$ and $r_H=1$.}
    \label{collapsing_mn}
\end{figure}

Poles of the green function occurs when the denominator is zero, i.e.,
\begin{equation}\label{denozero}
    \phi_1 f_2 \partial_r \phi_2^{(+)} - \phi_2^{(+)}f_1 \partial_r \phi_1 =0 \  .
\end{equation}
Expanding the LHS of \eqref{denozero} around $r_s=r_H$ we get:
\begin{equation}\label{quante1}
   \Tilde{A} \, \left(\frac{r_s}{r_H}-1 \right)^{-\frac{i \omega}{2 r_H}} \left( \left( P_1+ P_2 \left(\frac{r_s}{r_H}-1 \right)^{\frac{i \omega}{r_H}} \right)+ O(r_s-r_H) \right) =0 \ ,
\end{equation}
where
\begin{align}\label{quante2}
    \Tilde{A} &=  2^{-i \frac{\omega}{2r_H} }  r_H^{m-1} \left(r_H^2+1\right)^{\omega /2} \Bigg(\frac{\left(m \,r_H^2+m+ r_H^2 \omega \right) }{r_H^2+1} \, _2F_1\left(\alpha,\beta; \gamma;-r_H^2\right) \nonumber \\ 
    & \hspace{3cm}-2 \frac{\alpha \, \beta}{\gamma} \, r_H^2 \, _2F_1\left(\alpha+1,\beta+1;\gamma+1;-r_H^2\right)\Bigg)+O(r-r_H) \ ,
\end{align}
and:
\begin{align}
    P_1 &= \frac{\Gamma(1-a-b+c) \Gamma(1-c)}{\Gamma(1-b) \Gamma(1-a)} \ , \\
    P_2 &= \frac{\Gamma(1-a-b+c) \Gamma(c-1)}{\Gamma(c-b) \Gamma(c-a)} \ , \\
    Q_1 &= \frac{\Gamma(1+a+b-c) \Gamma(1-c)}{\Gamma(1+b-c) \Gamma(1+a-c)} \ , \\
    Q_2 &= \frac{\Gamma(1+a+b-c) \Gamma(c-1)}{\Gamma(a) \Gamma(b)} \ .
\end{align}
It is noteworthy that \eqref{quante2} is nothing but $\phi_1'(r)\Big|_{r=r_H}$. Now \eqref{quante1}
implies either:
\begin{equation}\label{quantzn1}
     P_1+ P_2 \left(\frac{r_s}{r_H}-1 \right)^{\frac{i \omega}{r_H}} =0 \ ,
\end{equation}
or
\begin{equation}\label{quantzn2}
    \frac{\left(m \,r_H^2+m+ r_H^2 \omega \right) }{r_H^2+1} \, _2F_1\left(\alpha,\beta; \gamma;-r_H^2\right) -2 \frac{\alpha \, \beta}{\gamma} \, r_H^2 \, _2F_1\left(\alpha+1,\beta+1;\gamma+1;-r_H^2\right)=0 \ .
\end{equation}
The first condition in \eqref{quantzn1} implies:
\begin{equation}
    \frac{P_2}{P_1}= \frac{\Gamma(c-1) \Gamma(1-a) \Gamma(1-b)}{\Gamma(1-c) \Gamma(c-a) \Gamma(c-b)} = - \left(\frac{r_s}{r_H}-1 \right)^{-\frac{i \omega}{r_H}} \ .
\end{equation}
which can be written as:
\begin{eqnarray}\label{quant22}
     & \left(\frac{r_s}{r_H}-1 \right)^{\frac{i \omega}{r_H}} \, \frac{\Gamma(c) \Gamma(1-a) \Gamma(1-b)}{\Gamma(2-c) \Gamma(c-a) \Gamma(c-b)} \equiv e^{i \theta(\omega, m)} =1  \ , \nonumber \\
     \nonumber \\ 
     & \Rightarrow \theta(\omega_n, p)= 2n \pi \hspace{1cm}\text{where}, n\in \mathbf{Z} \ .
\end{eqnarray}
This leads to the following quantization condition,
\begin{equation}\label{quant24}
    \frac{\omega}{2 r_H} \log \left(\frac{r_s}{r_H}-1 \right)+\text{Arg}\left[\frac{\Gamma(c)}{\Gamma(c-a)\Gamma(c-b)} \right]=n \pi \ , \hspace{1.5cm}\text{with} \hspace{0.2cm} n\in \mathbf{Z} \ .
\end{equation}
This is same as our previous Dirichlet (brickwall) quantization condition. However, in the collapsing scenario, we also have \eqref{quantzn2}. Solving this equation also allows us to find the normal modes of the system. The normal modes are shown in Figure \ref{collapsing_mn}.

Let us offer a few comments at this point. As we observed explicitly, in the limit $r_s \to r_{\rm H}$, there are two in-equivalent quantization conditions: One coincides with the Dirichlet boundary condition on the scalar field and therefore yields the non-trivial and interesting and chaotic spectrum\footnote{We emphasize that here by quantum chaos we essentially mean the presence of the linear ramp of slope unity in the corresponding spectral form factor. In this sense, there is a non-trivial spectral rigidity.} that was observed in \cite{Das:2022evy, Das:2023yfj}. The other boundary condition, however, seems to yield a linear spectrum and is therefore non-chaotic.

The emergence of two boundary conditions in the $r_s \to r_{\rm H}$ limit is simple to understand. Recall that the general boundary conditions can be re-written succintly as:
\begin{eqnarray}
 \left.    \phi_1 f_2 \partial_r \phi_2^{(+)} - \phi_2^{(+)} f_1 \partial_r \phi_1 \right|_{r_s}= 0 \ .
\end{eqnarray}
In the limit $r_s \to r_{\rm H}$, $f_2 \to 0$ which yields:
\begin{eqnarray}
   \left. \phi_2^{(+)} \partial_r \phi_1 \right|_{r_s} = 0\ . \label{rsonrH}
\end{eqnarray}
This indicates two in-equivalent choices for boundary conditions: either a Dirichlet one $\phi_2^{(+)}=0$ or a Neumann one $\partial_r \phi_1 = 0$. The Dirichlet boundary condition is equivalent to the brickwall Dirichlet boundary condition and therefore the corresponding spectrum has the non-trivial features thermal and chaotic features that we have explained before. The Neumann boundary condition, on the other hand, does not yield a non-trivial level correlation and is integrable. Note further that, the Dirichlet boundary condition is imposed on the scalar {\it outside} the Schwarzschild radius while the Neumann condition is imposed {\it inside} of it. If we hypothesize that there is no interior of a quantum black hole, then we are only left with the choice of the Dirichlet boundary condition of $\phi_2^{(+)}$.\footnote{This is of course not a completely consistent statement, since the boundary conditions have been arrived at within a purely classical GR framework.}

For the above picture to hold, we need the following physical picture to emerge: Suppose we drop a shell from the boundary which collapses radially according to the classical equations of motion of GR. Once the shell approaches the corresponding Schwarzschild radius $r_{\rm H}$, classically, there is no force to stop is from further collapsing into a classical black hole which is described by a smooth event horizon and a classical interior. However, if a fuzzball-like description is to be believed, then quantum aspects are already at play and the true geometric configuration tunnels to a horizon-less fuzzball like geometry. The wavefunctional of the collapsing shell tunnels into a horizon-less fuzzball geometry near the black hole threshold, as argued in \cite{Kraus:2015zda}. This is based on the assumption that the exponential suppression of tunneling to an individual fuzzball geometry is compensated for by an exponentially large number of such configurations.\footnote{Note that in \cite{Bena:2015dpt} the Authors have estimated the quantum tunnelling probability of a collapsing shell of branes into a smooth horizonless geometry, within a supergravity description. This provide, at least, an evidence that classical gravity can break-down at the scale of the horizon.} Viewed from this perspective, the emerging Dirichlet boundary condition of the {\it outside} mode of the scalar field appears rather intriguing, albeit in a completely classical description. At this point, we will not offer a more precise statement. However, we will note that recent conjectures in \cite{Mathur:2024mtf} on the structure of fluctuations around such a microstate, which may allow for an {\it ab initio} calculation of the collapsing shell. 

Finally, note that, in principle one can also obtain a perturbative correction to the boundary condition in (\ref{rsonrH}), as $r_s$ is taken perturbatively further from the horizon. It seems reasonable to expect that there will still be two in-equivalent quantizations: one with a non-trivial level-correlation and one with less interesting spectral rigidity. At this point, however, it appears numerically unwieldy and lacking in control to systematically analyze this. We hope to address this in future.

\section{Algebraic consequences}
When the cut-off is at a finite distance away from the event horizon, the algebra of observables is of type I${_\infty}$, which follows from an equivalent quantum description of the system in terms of a collection of harmonic oscillators composing the scalar field $\phi$\cite{Soni:2023fke}. This cutoff also appears in the level spacing $\Delta \omega$ between different eigenvalues of the modular Hamiltonian, $\Delta \omega\sim-\ln\delta$. To be precise, at finite $\delta$, the scalar field can be written as\,\footnote{For simplicity, we set $t=0$ and $\phi_0=0=\phi_1$, as in \cite{Soni:2023fke}.}   
\begin{align}
    \phi=\sum_n\big[a_{r,\omega_n}e^{-\i\omega_nt}+a_{r,\omega_n}^\dagger e^{\i\omega_nt}\big]N_\omega\sin(\omega_n\rho)\,,
\end{align}
where $N_\omega$ is the normalization and
\begin{align}
    \omega_n=\frac{2\pi n}{\frac{l}{r_H}\ln\frac{r_H}{\delta}}\,.
\end{align}
As long as $\delta>0$, this provides the setting for a type I algebra. In the limit $\delta \rightarrow 0$, the exterior state is a thermofield doubled state of the form 
\begin{align}
 \label{eqn:TFDn}   |\beta\rangle_\delta=\bigotimes_{\omega_n}\sqrt{1-e^{-\beta\omega_n}}e^{-\beta\frac{\omega_n}{2}a_{l,\omega_n}^\dagger a_{r,\omega_n}^\dagger}|0_{l,\omega_n},0_{r,\omega_n}\rangle\,,
\end{align}
with an effective inverse temperature $\beta=\frac{2\pi l^2}{r_H}$.

The corresponding algebra of operators $a_{l/r,\omega_n}$ are of type I$_\infty$. The subscript $_\infty$ corresponds to the fact that the state \eqref{eqn:TFDn} comprises of a countably infinite number of energy eigenstates. This state belongs to a separable Hilbert space. Accordingly, one can define reduced density matrices and consequently, entanglement entropy is well defined. The latter can be identified as the black hole entropy with a proper partitioning of the Hilbert space. From the bulk perspective, one can, in principle, introduce a second brickwall, still outside the horizon, such that integrating out one side of that wall reproduces the desired entanglement entropy which equals the thermal entropy to a good approximation. Thus, even without having any notion of interior, one can still have a notion of entanglement entropy. The latter does not possess any geometric interpretation in the form of ER=EPR, but can still reproduce a value close to that of black hole entropy. This is a very crucial point and we will revisit this further in a bit. Before that let us quickly review the quantitative measure to understand, algebraically, the physics of continuum in the limit $\delta \rightarrow 0$ \cite{Banerjee:2024dpl}.

When the algebra is of type I,
there exists the natural notion of the trace as we are used to in quantum mechanics, except that it is not defined for every possible operator. As an example, the trace of the identity operator in the continuous setting is not defined/ diverges. In terms of formulae, this is reflected by $\lim\limits_{\epsilon\to0}\delta_D(\epsilon)\to\infty$. Note the subscript $D$ to distinguish the cutoff $\delta$ and the Dirac-Delta $\delta_D$. In general, the existence of a tracial state for an algebra $\mathcal A$ is characterized by 
\begin{align}
\label{eqn:trace}
    \omega_{\tr}(ab)=\omega_{\tr}(ba)\quad\text{for any}\quad a,b\in\mathcal{A}\,.
\end{align}
For practical purposes, this tracial state can have the following representation using a normalized state vector of the corresponding Hilbert space
\begin{align}
\omega_{\tr}^{|\phi\rangle}(\cdot)=\langle\phi|\cdot|\phi\rangle\,,
\end{align}
so that \eqref{eqn:trace} can be represented as a condition of vanishing commutators in that state vector.

Using this definition and the state 
vector \eqref{eqn:TFDn}, one can compute the trace explicitly. In this representation, the states on either sides of \eqref{eqn:trace} represent thermal expectation values in the state vector \eqref{eqn:TFDn}. In \cite{Banerjee:2024dpl} we explicitly computed the thermal expectation values of $a_{r,\omega_n}(t)a_{r,\omega_n}^\dagger(0)$ and $a_{r,\omega_n}^\dagger(0)a_{r,\omega_n}(t)$ and found that they in general differ for arbitrary $\delta$. Adopting the formalism developed in \cite{Witten:2023qsv}, we denote
\begin{align}
    F(t)&=\langle a_{r,\omega_n}(t)a_{r,\omega_n}^\dagger(0)\rangle_{\beta,\delta}=\frac{1}{Z}\tr\big(e^{-\beta H}e^{\i Ht}a_{r,\omega_n}(0)e^{-\i Ht}a_{r,\omega_n}^\dagger(0)\big) \ ,\\
    G(t)&=\langle a_{r,\omega_n}^\dagger(0)a_{r,\omega_n}(t)\rangle_{\beta,\delta}=\frac{1}{Z}\tr\big(e^{-\beta H}a_{r,\omega_n}^\dagger(0)e^{\i Ht}a_{r,\omega_n}(0)e^{-\i Ht}\big)\,.
\end{align}
These two expressions are related as
\begin{align}
    G(t)=F(t-\i\beta)\,,
\end{align}
which is the KMS condition. In terms of the Fourier modes $f(\omega)$ and $g(\omega)$ of these functions, this condition can be expressed in the familiar form
\begin{align}
    g(\omega)=e^{-\beta\omega_n}f(\omega)\,.
\end{align}
The relative scaling with the exponential tells that the two functions are not generally equal. Therefore, $F\neq G$ and their difference is non-zero. In the discrete case valid for finite $\delta$, the exponential behaves as
\begin{align}
    e^{-\beta\omega_n}\sim e^{\frac{\#}{\ln\delta}}\,,
\end{align}
where $\#$ is an order one constant.
Clearly, in the limit of $\delta\to0$, this does not vanish but tends to 1. Therefore, one can still define a tracial state, at any finite temperature. This compels the reader to conclude that naively taking $\delta \rightarrow 0$ limit, does not change the algebra and one still has an algebra of type I$_\infty$. However, there is a subtlety. It is worth recalling the computation of the holographic Green's function where we emphasized the role of energy scale. We precisely determined the cut-off energy scale (both for fixed $m$ and fixed $n$) below which the $\delta \rightarrow 0$ limit will give rise to a continuum.

This is the limit in which the accumulated discrete poles will be indistinguishable from a continuum branch cut from the perspective of a low-energy asymptotic observer. It turns out that when we take this continuum approximation into account that would also give rise to an algebraic approximation \cite{Banerjee:2024dpl, Soni:2023fke}. In this continuum limit, the trace of the algebra also assumes an approximately continuum form  $e^{-\beta\omega_n} \to e^{-\beta\omega}$ and we can no longer define a tracial state.
This manifests emergence of a type III$_1$ vN algebra in the continuum $\delta\to0$ limit which in turn justifies the effective thermal behaviour advocated through our holographic computation in this limit.

While this is a simple example establishing the effective thermalization from the perspective of vN algebra, this formalism has something deeper to offer. For this, let us reconsider our comments on  entanglement entropy mentioned at the beginning of this section. It is worth emphasizing here that the most important consequence of the emergence of type III algebra is that the TFD state \eqref{eqn:TFDn} does no longer belong to a factorized Hilbert space \cite{Witten:2021jzq}\footnote{It rather serves as the identity element of a GNS Hilbert space.}.   Furthermore, the algebra being of type III, there is no way one can define a reduced density matrix and it does not make any sense to talk about entanglement entropy. Therefore, to demystify the entanglement between the exterior and interior of the black hole, one needs to either add perturbative and non-perturbative corrections, or devise a quantum mechanical system which is ``entanglement-equivalent" of that of a black hole to a good approximation. Note this ``entanglement-equivalent" system does not have an ``interior'' but nevertheless can have a well-defined entanglement entropy between two subsystems, partitioned in a fine-tuned manner such that the value of the entanglement entropy between the two parts can be identified with that of the thermal entropy of a semiclassical black hole.
This reproduction of entanglement entropy from an ``equivalent'' type I system lacking a notion of ``interior" is, in fact, the statement of the "factorization map" first proposed in \cite{Jafferis:2019wkd}. 

Type I nature of the algebra of operators associated with the cutoff geometry can also be justified as follows. Note that the bulk solution with a cutoff outside the horizon is not a solution of semiclassical gravity, i.e. this configuration does not solve the semiclassical Einstein's equations. If it were a solution, it would correspond to operator algebra of type III. The examples of the latter correspond to configurations with end-of-the-world branes resulting in a backreacted semi-classical geometry. In our case, the cutoff scale quantifies our ignorance about the UV completion of our setup. Setting the scale $\delta$ to zero corresponds to moving to the semiclassical regime, namely the black hole solution of semiclassical Einstein's equations.\footnote{Note that, in this limit, we have to also smoothly approach to a purely ingoing boundary condition, from the Dirichlet boundary condition on the cutoff surface.} In this limiting case, the algebra of operators naturally becomes type III, as reflected through the computation of the trace of the algebra. In the example of spherical collapse, to stabilise the wall sufficiently close to the location of the event horizon, one needs UV fields, effects whereof are again hidden in the energy scale set by the location of the wall.

\section{Summary \& Discussions}

Let us briefly summarize various aspects of this article. Building on \cite{Banerjee:2024dpl}, we have explored several consequences of the condensation of poles of the Green's function, as the cut-off surface is taken towards the event horizon, thereby creating a large red-shift effect. In particular, we have demonstrated how a simple effective description of quasi-normal modes emerge, which is qualitatively rather similar to a top-down heavy-heavy-light-light correlator computation using supergravity\cite{Giusto:2023awo}, in a classical microstate geometry\cite{Bena:2016ypk}.\footnote{At this point, it is important to note that several physical aspects of the Fuzzball-geometries are currently under active exploration. Some representative such works are: \cite{Bena:2024qed, Guo:2024pvv, Ganchev:2023sth, Bena:2022fzf, Bena:2022rna, Bena:2022sge, Ganchev:2021pgs, Martinec:2020cml, Bena:2020yii, Heidmann:2019xrd}.} This geometry interpolates between global AdS$_3$ and an extremal BTZ geometry. Correspondingly, the brickwall model captures this interpolation as the location of the Dirichlet hypersurface is varied from the conformal boundary of AdS to the near horizon limit. Note that, for the quasi-normal modes to emerge, an extremal black hole geometry is enough and therefore the brickwall model actually capture more features than the classical microstate considered in \cite{Giusto:2023awo}. In this sense, the emergence of the quasi-normal modes are a ``weakly kinematic feature" of such models.

It is straightforward to check that the above universality will not occur for an arbitrary quantum mechanical system with a linear spectrum, like the harmonic oscillator. More precisely, for the above mechanism to hold, we need two basic ingredients: First, a parameter by tuning which the gap in the spectrum can be arbitrarily reduced. Secondly, the residues at the poles of the Green's function should contain the desired properties so that the jump function across the branch-cut reproduces the black hole quasi-normal modes. While the first ingredient is available for most quantum systems, {\it e.g.}~for a particle in a box (by tuning the length of the box) or a harmonic oscillator (by tuning $\hbar$), the characterization of the residued at the physical poles contain non-trivial dynamical features of the underlying system.

In particular, the non-extremal Rindler structure of the near horizon limit in the effective brickwall model can give rise to further features of quantum chaos in the corresponding spectral form factor in the one-loop scalar determinant\cite{Das:2022evy, Das:2023ulz, Das:2023yfj}. This is particularly tied to exciting non-vanishing angular momentum along the compact direction of the geometry. This non-vanishing angular momentum yields a non-perturbatively faster condensation of poles as the Dirichlet hypersurface is moved towards the event-horizon. On top of the universal kinematic aspects, this fast condensation is expected to give rise to the ramp of slope unity, resulting from a sufficient spectral rigidity. This is unexpected from a currently available top-down model in \cite{Bena:2016ypk}, since the spectrum in the probe scalar sector is linear. Thus, it is fair to state that the brickwall model captures two different dynamical aspects of the physics of thermalization for such systems.

In fact, we have made qualitative statements regarding how the non-vanishing angular momenta can give rise to a strong thermalization, as opposed to a weak thermalization in the absence of them. Our arguments are fairly generic and are based on the simple observation that the rate of pole condensation carries a hierarchy along the angular momenta modes. Note, however, that this is not precisely the notion of strong and weak thermalization of \cite{Ba_uls_2011}. It will, nonetheless, be interesting to explore how different initial states lead to thermalization in the scalar Hilbert space. We leave this for future work.

We have also demonstrated explicitly that in a classical shell-collapse model the one-loop determinant can be similarly quantized yielding a non-trivial spectral rigidity as the shell approaches the event horizon scale. In particular, exactly at the event horizon, the quantization condition reduces to the Dirichlet boundary condition that were used in\cite{Das:2022evy, Das:2023xjr, Das:2023yfj}. Thus, the collapsing shell reduces to a Dirichlet hypersurface as far as the probe scalars are concerned. This provides us with a plausible interpretation of the dual state of the brickwall model: It is a result of a global quench in the CFT. Furthermore, the classical description is naturally consistent with the Fuzzball paradigm, since the Dirichlet boundary condition on the scalar can be interpreted as having no interior of the state. It is, however, important to note that an alternate quantization condition also emerges in this limit, which is a Neumann boundary condition on the scalar from the inside. This quantization does not lead to a non-trivial spectral correlation. It will be very interesting to understand this aspect better.

Note that, real-time correlators  at weak coupling is known to have a distinct analytic structure from that at strong coupling, see {\it e.g.}~\cite{Hartnoll:2005ju}. The perturbative analyses of glueballs in ${\cal N}=4$ SYM yield branch-cuts, whereas results in the dual gravitational picture yields poles at the quasi-normal modes, at strong coupling. {\it A priori}, this is a different physical effect from what we have discussed here; it is, nonetheless, intriguing to explore a possible connection between them. We hope to address this in future. One related aspect is to first understand whether the analytic structure observed in the scalar sector generalizes to gauge fields and fermions, which is currently under investigation.

From an observational perspective, one essential feature to dintinguish black hole mimickers from true black holes is to perform controlled analyses of multipole moments, see {\it e.g.}~\cite{Heynen:2023sin}. While the brickwall model is completely stationary and isotropic, in \cite{Das:2023ulz} a ``typical curve" was considered for the brickwall which possesses a non-trivial angular profile. While this profile is chosen from an appropriate distribution in this framework, it is intriguing to analyze what universal features emerge for multipole moments as one interpolates between the chaotic and the  integrable regimes.

In this context, it is useful to note that potential Extremely Compact Objects (ECOs) are a rather intriguing possibility from an astrophysical perspective. Such BH mimickers may exist as a result of dark matter, along with the standard matter, providing enough Pauli pressure to balance the gravitational pressure at the almost horizon scale. Such ECOs are typically expected to give rise to stable photon rings, see {\it e.g.}~\cite{Cardoso:2019rvt, Cardoso:2014sna, Keir:2014oka, Cunha:2017qtt} for more detailed account of such possibilities, as well as {\it e.g.}~\cite{Danielsson:2021ykm, Danielsson:2023onu, Giri:2024cks}. It will be very interesting to understand how our framework and studies are relevant in such cases as well.

\section{Acknowledgements}

We thank Mohsen Alishahiha, Elena Caceres, Roberto Emparan, Johanna Erdmenger, Bobby Ezhuthachan, Abhijit Gadde, Jani Kastikainen, Chethan Krishnan, Gautam Mandal, Samir Mathur, Shiraz Minwalla, Onkar Parrikar, Sandip Trivedi, Nicholas P.~Warner for various conversations and comments related to this work. AK is partially supported by 
DAE-BRNS 58/14/12/2021-BRNS and CRG/2021/004539 of Govt. of India. AK further acknowledges the support of Humboldt Research Fellowship for Experienced Researchers by the Alexander von Humboldt Foundation and for the hospitality of the Theoretical
Physics III, Department of Physics and Astronomy, University of Wurzburg during the course of this work.


\appendix

\section{Scalar Field in BTZ-background}
\label{s.scalarBTZqnm}

\begin{figure}
    \centering
    \includegraphics[width=\columnwidth]{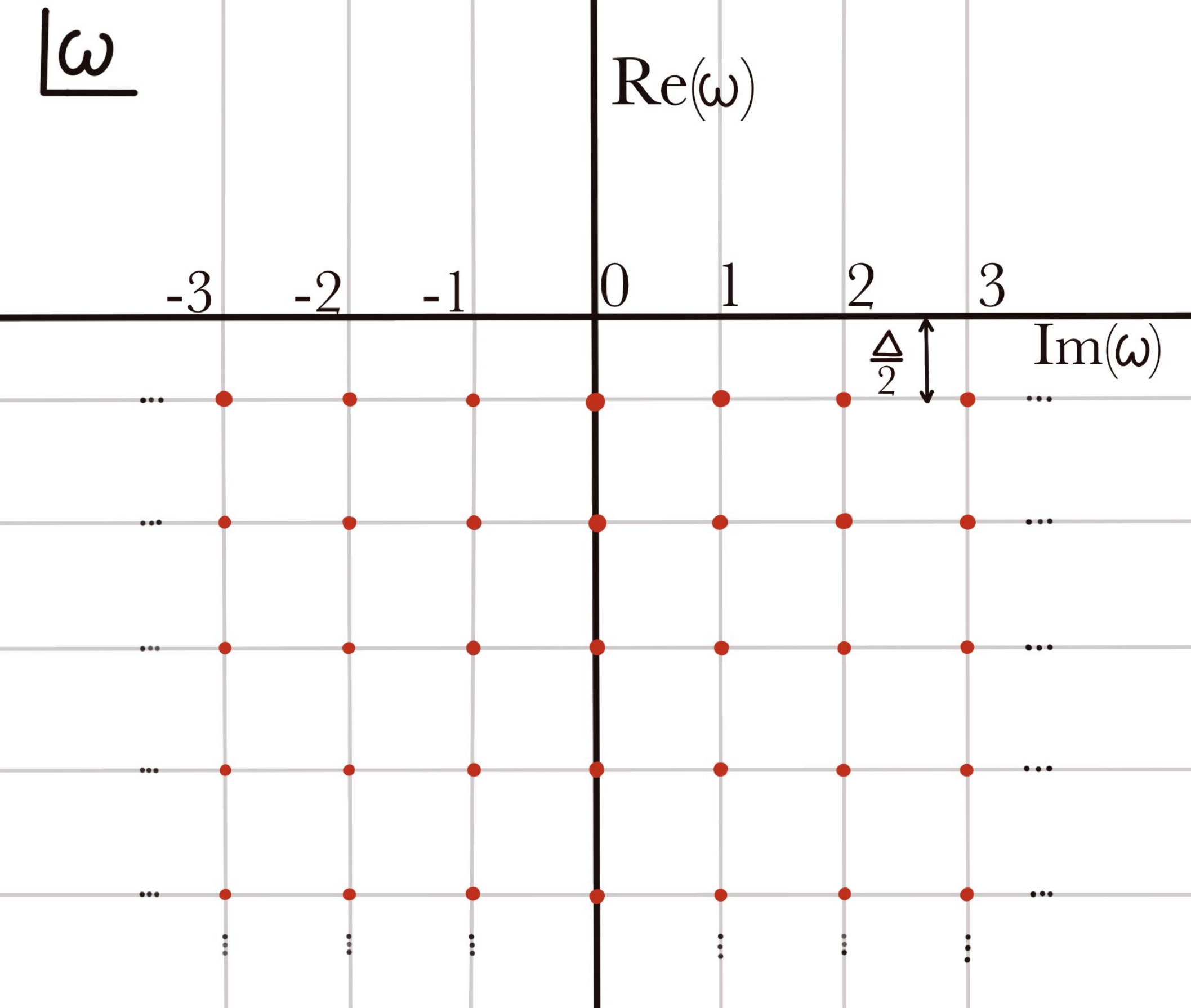}
   \caption{ Schematic diagram of the pole structure of $G_{\text{bh}}(\omega)$}
    \label{polesBH}
\end{figure}

In this appendix, we will review the well-known computation of quasi-normal modes of scalar fields in the BTZ-geometry. Recall that the BTZ-geometry is given by
\begin{eqnarray}
ds_{\rm BTZ}^2 = - \left(r^2- r_{\rm H}^2 \right) dt^2 + \frac{dr^2}{\left(r^2- r_{\rm H}^2 \right)} + r^2 d\psi^2 \ ,
\end{eqnarray}
and the Klein-Gordon equation is given by
\begin{eqnarray}
\Box\Phi = \mu^2 \Phi \ .
\end{eqnarray}
Decomposing the field $\Phi = \sum_{\omega,m} e^{-i \omega t} e^{i m \psi} \tilde{\phi}(r)$, the solution of the radial part is obtained to be:
\begin{eqnarray}
    {\tilde\phi}\left(r(z)\right) & \equiv \phi\left(z\right)  =(1-z)^{\beta}\left[ C_1\, z^{-i \alpha} {}_2F_1(a, b; c; z) \right. \nonumber \\
    & \left. + \,  C_2\, z^{i \alpha} \,{}_2F_1(1+a-c, 1+b-c; 2-c; z)\right] \ ,
\end{eqnarray} 
where
\begin{align}\label{abappen}
  &  a=\beta-\frac{i}{2r_{\rm H}}(\omega + m) \ , \ b=\beta-\frac{i}{2r_{\rm H}}(\omega - m) \ , \ c=1-2i\alpha \ , \nonumber \\
  & \text{with}  \ \ \
   \alpha=\frac{\omega}{2r_{\rm H}} \ , \ \ \beta=\frac{1}{2}(1-\sqrt{1+\mu^2}) \ ,
\end{align}
with
\begin{eqnarray}
 z = 1 - \frac{r_{\rm H}^2}{r^2} \ .
\end{eqnarray}
The standard ingoing boundary condition at the black hole horizon can therefore be implemented simply by setting $C_2=0$ in \eqref{sol1} so that
\begin{equation}
\label{eqn:sol-ingoing}
    \phi(z) = (1-z)^{\beta} z^{-i \alpha} {}_2F_1(a, b; c; z) \ .
\end{equation}
The next step would be to expand the solution \eqref{eqn:sol-ingoing} near the boundary. At the leading order, this expansion yields:
\begin{equation}
\label{eqn:expansion-boundary}
    \phi_{\text{bdry}}\sim z^{-i \alpha} \left(R_1 (1-z)^{\frac{1}{2}(1-\nu)} +R_2 (1-z)^{\frac{1}{2}(1+\nu)} \right) \ ,
\end{equation}
with $\nu =\sqrt{1+\mu^2}$ and
\begin{equation*}
    R_1=\frac{\Gamma(c)\Gamma(c-a-b)}{\Gamma(c-a)\Gamma(c-b)}, \, \,R_2=\frac{\Gamma(c)\Gamma(a+b-c)}{\Gamma(a)\Gamma(b)} \ .
\end{equation*}
Clearly, the first term is non-normalizable and second, normalizable near the boundary. If we now impose the normalizability of the bulk field at the boundary, it implies $R_1=0$, i.e. $c-a=-n$ or $c-b=-n$ where $n \in \mathbf{Z}$. These conditions give rise to the quasinormal modes
\begin{equation}
    \omega_{n,m}=\pm m- i\left(n+\frac{\Delta}{2} \right), \hspace{0.2cm} \text{with} \quad \Delta=1+\sqrt{1+\mu^2} \ .
\end{equation}
Following the prescription of \cite{Son:2002sd}, the boundary Green's function is given by the ratio between the normalizable and the non-normalizable components of the field expansion \eqref{eqn:expansion-boundary} and is given by
\begin{equation}\label{greenBH}
    G_{\text{bh}}(\omega)=\frac{R_2}{R_1}=\frac{\Gamma(a+b-c)\Gamma(c-a)\Gamma(c-b)}{\Gamma(a)\Gamma(b)\Gamma(c-a-b)} \ .
\end{equation}
The pole structure of \eqref{greenBH} is shown in Figure \ref{polesBH}. This modes have been extensively studied in literature, {\it e.g.}~\cite{Birmingham:2001pj, Hartnoll:2016apf}.

\section{Green's Function \& Analytic Properties}
\label{s.GreenAnalytic}

In this appendix, we will review basic aspects of the Green's function on which we base our analyses of its analytic properties. We will closely follow the notation and discussions on \cite{Giusto:2023awo}.

\begin{figure}
    \centering
    \includegraphics[width=\columnwidth]{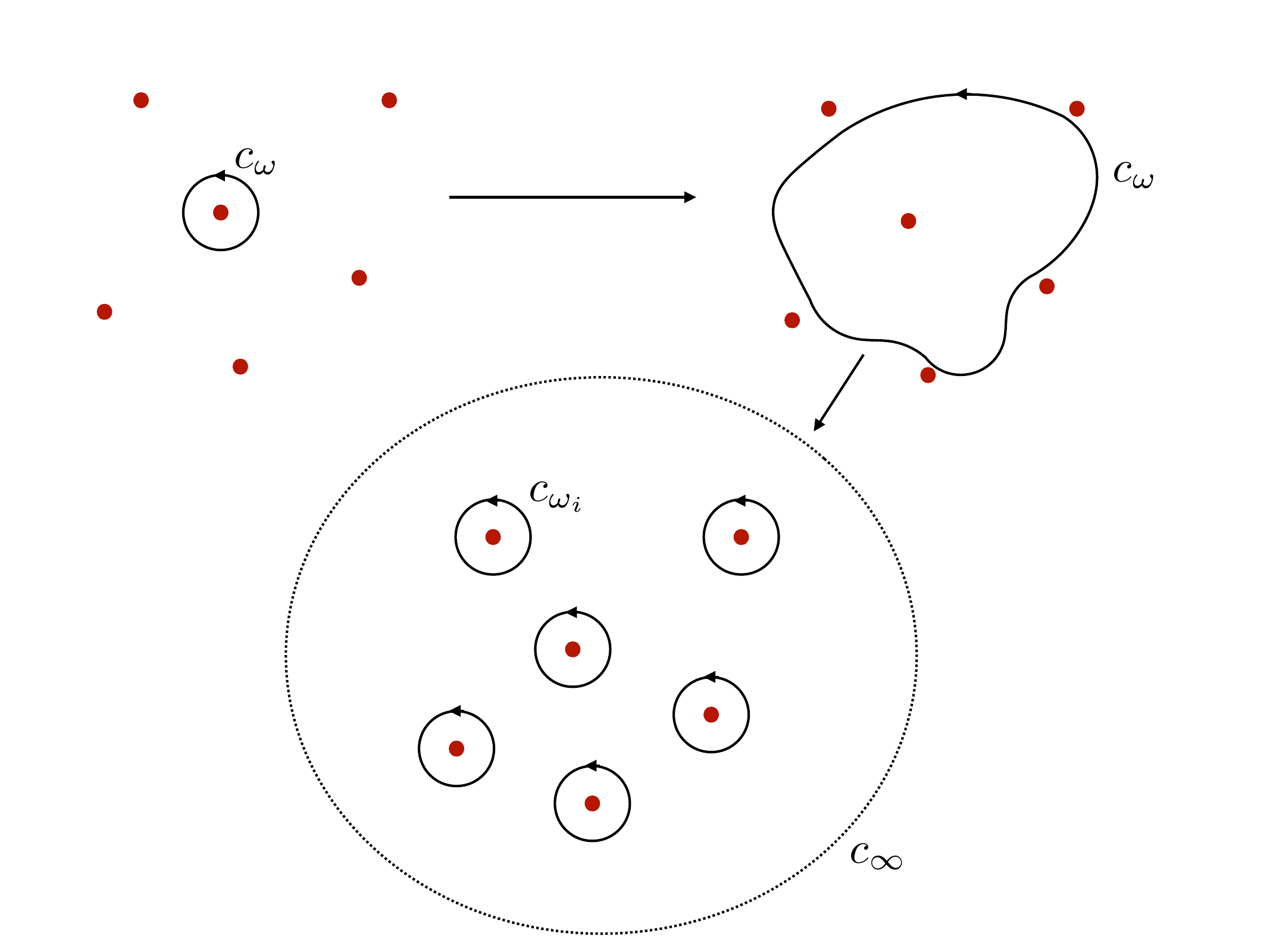}
   \caption{Representation of Green's function in the complex frequency plane.}
    \label{polescontour}
\end{figure}

Let us denote the Green's function by $G(\omega)$. Using a contour prescription, we can write:
\begin{eqnarray}
   \frac{1}{\omega^\Delta} G(\omega) = \frac{1}{2\pi i} \oint_{c_\omega} d\omega' \frac{1}{\omega'^{\Delta}} \frac{G(\omega')}{\omega'-\omega} \ , \label{Gcont}
\end{eqnarray}
where $c_\omega$ is a small contour around the pole at $\omega$. This is depicted pictorially in the top-left part of figure\ref{polescontour}. Here $\Delta$ is a positive real number. By performing a contour deformation (as shown in figure \ref{polescontour}), the RHS of (\ref{Gcont}) can now be written in terms of the sum of the residues evaluated at the poles of the function $G(\omega')$ in the complex-plane. This yields:
\begin{eqnarray}
    \frac{1}{2\pi i} \oint_{c_\omega} d\omega' \frac{1}{\omega'^{\Delta}} \frac{G(\omega')}{\omega'-\omega}  = \sum_j \frac{1}{\omega - \omega_j} {\rm Res\left( \frac{G(\omega_j)}{\omega_j^\Delta}\right)} \ . \label{Gcont2}
\end{eqnarray}
It is clear from that, because of the $\omega^\Delta$ factor in the denominator, the $c_\infty$ contour yields a vanishing contribution since the integrand falls-off rapidly enough. Here $\omega_j$ collectively denote the poles of the Green's function. 

Now equating (\ref{Gcont}) with (\ref{Gcont2}), we obtain:
\begin{eqnarray}
    G(\omega) = \sum_j \left( \frac{\omega}{\omega_j}\right)^\Delta \frac{1}{\omega - \omega_j} {\rm Res\left( G (\omega_j) \right) } \ .
\end{eqnarray}
The corresponding Feynman, retarded and Wightman correlators are given by
\begin{eqnarray}
&&  G_{\rm F} (\omega) = \sum_{\omega_j>0} \left(\frac{\omega}{\omega_j} \right)^\Delta \frac{{\rm Res}\left( G (\omega_j) \right)}{\omega - \omega_j + i \epsilon} + \sum_{\omega_j<0} \left(\frac{\omega}{\omega_j} \right)^\Delta \frac{{\rm Res}\left( G (\omega_j) \right)}{\omega - \omega_j - i \epsilon} \ , \\
&& G_{\rm R} (\omega) = \sum_j  \left(\frac{\omega}{\omega_j} \right)^\Delta \frac{{\rm Res}\left( G (\omega_j) \right)}{\omega - \omega_j + i \epsilon}  \ , \label{greenret} \\
&& G_{\rm W}(\omega) = - {\rm sign}(\omega) {\rm Im}G_{\rm R}(\omega) \ .
\end{eqnarray}
\begin{eqnarray}
    {\rm Im}\left[G_{\rm R}(\omega) \right]  \equiv \frac{1}{2i} \left[ G(\omega+i \epsilon) - G(\omega- i \epsilon) \right] \ .
\end{eqnarray}

When the poles condense, we can now approximate the discrete sum by a continuous integral:
\begin{eqnarray}
    && G(\omega) = \sum_j \left( \frac{\omega}{\omega_j}\right)^\Delta \frac{1}{\omega - \omega_j} {\rm Res\left( G (\omega_j) \right) } \to \int d\omega_j \frac{\rho_\omega(\omega_j)}{\omega - \omega_j} \ , \\
    && \rho_\omega(\omega_j) = \frac{dj}{d\omega_j} \left( \frac{\omega}{\omega_j}\right)^\Delta {\rm Res\left( G (\omega_j) \right) } \ . 
\end{eqnarray}
Now, using (\ref{greenret}), we obtain:\footnote{Using further the identity
\begin{eqnarray}
    {\rm Im}\left( \frac{1}{x+ i \epsilon}\right) = \pi \delta(x) \ . 
\end{eqnarray}
}
\begin{eqnarray}
    {\rm Im} \left[G_{\rm R} (\omega) \right] = {\rm Im} \left[ \sum_j  \left(\frac{\omega}{\omega_j} \right)^\Delta \frac{{\rm Res}\left( G (\omega_j) \right)}{\omega - \omega_j + i \epsilon} \right] & = & \sum_j \pi \delta \left( \omega - \omega_j\right) {\rm Res}\left( G (\omega_j) \right) \nonumber\\
    & \approx & \int d\omega_i \pi \delta\left( \omega - \omega_i \right) \rho_\omega \nonumber\\
    & = & \pi \rho_\omega(\omega) \ .
\end{eqnarray}
%

\section{Extremal Black Holes}
\label{s.extremalBTZ}

Let us consider a scalar field of mass $\mu$ in extremal BTZ-background descried by the metric:
\begin{equation}
   \begin{split}
        ds^2= &-f(\rho)dt^2+\frac{d\rho^2}{f(\rho)}+\rho^2\left(dy-\frac{r_H^2}{\rho^2}dt\right)^2 \ ,
   \end{split}
\end{equation}
with $f(\rho)=\frac{(\rho^2-r_H^2)^2}{\rho^2}$. The left and right-moving temperatures are given by $T_L=$zero and $T_R=\frac{r_H}{\pi}$.\\
Introducing a new radial coordinate $r$ defined by $r^2=\rho^2-r_H^2$, the metric takes a simpler form:
\begin{equation}\label{met1}
    ds^2=-f(r)dt^2+\frac{r^2}{f(r)(r^2+r_H^2)}dr^2+(r^2+r_H^2)\left(dy^2-\frac{r_H^2}{r^2+r_H^2}dt \right)^2 \ ,
\end{equation}
with $ f(r)=\frac{r^4}{r^2+r_H^2}$ and where $r=0$ is horizon and $r\rightarrow \infty$ is the conformal boundary. With the decomposition of scalar field $\Phi(t,,r,y)\sim e^{-i \omega t} e^{i m y} \phi(r)$, KG-equation takes the following form:
\begin{equation}\label{extrad}
    \begin{split}
         \phi(r) (m^2 \left(r_H^2-r^2\right)-& 2 m r_H^2 \omega -\mu ^2 r^4+\omega ^2 \left(r^2+r_H^2\right))+\\
         & r^5 \left(r \phi''(r)+3 \phi'(r)\right)=0 \ .
    \end{split}
\end{equation}
Let us define two new variables $p$ and $q$ as $p=\frac{m+\omega}{2}$ and $q=\frac{m-\omega}{2}$. In terms of $z=\frac{2 i r_H q}{r^2}$, \eqref{extrad} is a confluent hypergeometric equation:
\begin{equation}
    \phi''(z)+\left(-\frac{\mu^2}{4z^2}+\frac{i p}{2 r_H z} -\frac{1}{4}  \right)\phi(z)=0 \ ,
\end{equation}
with solution is in terms of  Whittaker functions:
\begin{equation}\label{whit1}
    \phi(z)=C_1 \, M_{\kappa,\nu}(z)+C_2 \, M_{\kappa, -\nu}(z) \ , 
\end{equation}
with $ \kappa=\frac{i p}{2 r_H}, \, \nu=\frac{\sqrt{1+\mu^2}}{2}$. Near horizon ($z\rightarrow \infty$) behaviour of Whittaker M function is the following:
\begin{equation}
     M_{\kappa,\nu}(z)  \sim  e^{\frac{z}{2}} z^{-\kappa} \frac{\Gamma(1+2\nu)}{\Gamma(\frac{1}{2}-\kappa+\nu)} +  e^{-\frac{z}{2}} z^{\kappa} (-1)^{\frac{3}{2}+\kappa-\nu}  \frac{\Gamma(1+2 \nu)}{\Gamma(\frac{1}{2}+\kappa+\nu)} \ .
\end{equation}
Using this, \eqref{whit1} can be written as:
\begin{equation}
    \begin{split}
        & \phi_{\text{hor}}(z) \sim  e^{\frac{z}{2}} z^{-\kappa} \left( C_1 \frac{\Gamma(1+2\nu)}{\Gamma(\frac{1}{2}-\kappa+\nu)} +  C_2 \frac{\Gamma(1-2\nu)}{\Gamma(\frac{1}{2}-\kappa-\nu)}  \right) +  e^{-\frac{z}{2}}  \\ 
        &  z^{\kappa} \left( C_1 \, (-1)^{\frac{3}{2}+\kappa-\nu} \frac{\Gamma(1+2\nu)}{\Gamma(\frac{1}{2}+\kappa+\nu)} +C_2 \, (-1)^{\frac{3}{2}+\kappa+\nu} \frac{\Gamma(1-2\nu)}{\Gamma(\frac{1}{2}+\kappa-\nu)}  \right) \ .
    \end{split}
\end{equation}
where the first term is outgoing and second term is ingoing. Ingoing boundary condition implies that the first term is zero which fixes the ratio between $C_1$ and $C_2$. This yields:
\begin{equation}
    \frac{C_1}{C_2} = -\frac{\Gamma(1-2\nu) \Gamma(\frac{1}{2}-\kappa+\nu)}{\Gamma(1+2\nu)\Gamma(\frac{1}{2}-\kappa-\nu)} \ .
\end{equation}
Near boundary ($z\rightarrow 0$) expansion of \eqref{whit1} is given by
\begin{equation}\label{extboundary}
    \begin{split}
         \phi_{\text{bdry}}(z) & \sim C_1 z^{\frac{1}{2}+\nu} +C_2 z^{\frac{1}{2}=\nu} \\
          & = C_1 \, (2i r_H q)^{\frac{1}{2}+\nu} r^{-1-2\nu} + C_2 \,  (2 i r_H q)^{\frac{1}{2}-\nu} r^{-1+2\nu} \ . 
    \end{split}
\end{equation}
Green's function is given by the ratio of normalizable and non-normalizable part, {\it i.e.}
\begin{equation}\label{greenextremal}
    \begin{split}
        G_{\text{ext}} & = (2i r_H q)^{2\nu} \frac{C_1}{C_2}\\
        & = -(2i r_H q)^{2\nu} \frac{\Gamma(1-2\nu) \Gamma(\frac{1}{2}-\kappa+\nu)}{\Gamma(1+2\nu) \Gamma(\frac{1}{2}-\kappa-\nu)} \\
        &= -(2i r_H q)^{\Delta-1} \frac{\Gamma(2-\Delta) \Gamma(\frac{\Delta}{2}-\frac{i p}{2r_H})}{\Gamma(\Delta) \Gamma(1-\frac{\Delta}{2}-\frac{ip}{2r_H})} \ .
    \end{split}
\end{equation}
Poles of this \eqref{greenextremal} are the quasinormal modes and is given by the following:
\begin{equation}
    \begin{split}
        \frac{\Delta}{2}-\frac{i p}{2r_H} = -n  \ , \hspace{2cm} \\
        \Rightarrow \omega_{n,m} =-m-2i r_H (2n+\Delta) \ ,
    \end{split}
\end{equation}
where $n\in \mathbf{Z}$ and $\Delta=1+\sqrt{1+\mu^2}$.

\subsection{Brickwall in Extremal BTZ}
\label{s.extremalBTZbrick}

In this Appendix, we review the key quantitative aspects of the Brickwall model in an extremal BTZ geometry. Many of these results were already obtained in \cite{Das:2023xjr} and we review them for the sake of completeness. 

Instead of ingoing boundary condition, we impose Dirichlet boundary condition that $\phi_{\text{hor}}(z=z_0)=0$ which fixes the ratio between $C_1$ and $C_2$ to be:
\begin{equation}\label{ratioext}
    \frac{C_1}{C_2}=-\frac{   \Gamma(1-2\nu)   }{   \Gamma(1+2\nu) }   \frac{  \left( \frac{e^{\frac{z_0}{2}}z_0^{-\kappa}}{\Gamma(\frac{1}{2}-\kappa-\nu)}  +\frac{e^{-\frac{z_0}{2}}z_0^{\kappa}(-1)^{\frac{3}{2}+\kappa+\nu}}{\Gamma(\frac{1}{2}+\kappa-\nu)}  \right)     }{  \left( \frac{e^{\frac{z_0}{2}}z_0^{-\kappa}}{\Gamma(\frac{1}{2}-\kappa+\nu)}  +\frac{e^{-\frac{z_0}{2}}z_0^{\kappa}(-1)^{\frac{3}{2}+\kappa-\nu}}{\Gamma(\frac{1}{2}+\kappa+\nu)}  \right)  } \ . 
\end{equation}
Now from the boundary expansion \eqref{extboundary}, Green's function is given by the following,
\begin{equation}\label{extremalgreen3}
    \begin{split}
         G &=(i r_H q)^{2\nu} \frac{C_1}{C_2} \\
         &= -(i r_H q)^{2\nu}  \, \frac{   \Gamma(1-2\nu)   }{   \Gamma(1+2\nu) }   \frac{  \left( \frac{e^{\frac{z_0}{2}}z_0^{-\kappa}}{\Gamma(\frac{1}{2}-\kappa-\nu)}  +\frac{e^{-\frac{z_0}{2}}z_0^{\kappa}(-1)^{\frac{3}{2}+\kappa+\nu}}{\Gamma(\frac{1}{2}+\kappa-\nu)}  \right)     }{  \left( \frac{e^{\frac{z_0}{2}}z_0^{-\kappa}}{\Gamma(\frac{1}{2}-\kappa+\nu)}  +\frac{e^{-\frac{z_0}{2}}z_0^{\kappa}(-1)^{\frac{3}{2}+\kappa-\nu}}{\Gamma(\frac{1}{2}+\kappa+\nu)}  \right)  } \\
         &=    \frac{  G_{\text{ext}} - (i r_H q)^{2\nu}  \frac{\Gamma(1-2\nu)}{\Gamma(1+2\nu)} \frac{e^{-z_0} z_0^{2\kappa} (-1)^{\frac{3}{2}+\kappa+\nu} \Gamma(\frac{1}{2}-\kappa+\nu) }{\Gamma{\frac{1}{2}+\kappa-\nu}}  }{ 1+  e^{-z_0} z_0^{2\kappa} (-1)^{\frac{3}{2}+\kappa-\nu}  \frac{\Gamma{\frac{1}{2}-\kappa+\nu}}{\Gamma{\frac{1}{2}+\kappa+\nu}}    } \ . 
    \end{split}
\end{equation}
Where $G_{\text{ext}}$ is written in \eqref{greenextremal}. The equation \eqref{extremalgreen3} has poles when the denominator is zero which gives the normal modes $\omega_n$.
\begin{equation}\label{normalext}
    \begin{split}
        1+  e^{-z_0} z_0^{2\kappa} (-1)^{\frac{3}{2}+\kappa-\nu}  \frac{\Gamma{\frac{1}{2}-\kappa+\nu}}{\Gamma{\frac{1}{2}+\kappa+\nu}}   =0  \ , \\
        \Rightarrow e^{-z_0} z_0^{2\kappa} = - \frac{\Gamma{\frac{1}{2}+\kappa+\nu}}{\Gamma{\frac{1}{2}-\kappa+\nu}}  (-1)^{   -\frac{3}{2}-\kappa+\nu} \ . 
    \end{split}
\end{equation}
We can compute the corresponding residues  \eqref{extremalgreen3} at these poles $\omega_n$:
\begin{equation}
     \frac{ \left( G_{\text{ext}} - (i r_H q)^{2\nu}  \frac{\Gamma(1-2\nu)}{\Gamma(1+2\nu)} \frac{e^{-z_0} z_0^{2\kappa} (-1)^{\frac{3}{2}+\kappa+\nu} \Gamma(\frac{1}{2}-\kappa+\nu) }{\Gamma{\frac{1}{2}+\kappa-\nu}}\right) \bigg\rvert_{\omega=\omega_n}  }{ \partial_{\omega}\left( 1+  e^{-z_0} z_0^{2\kappa} (-1)^{\frac{3}{2}+\kappa-\nu}  \frac{\Gamma{\frac{1}{2}-\kappa+\nu}}{\Gamma{\frac{1}{2}+\kappa+\nu}} \right) \bigg\rvert_{\omega=\omega_n}   }  \ . 
\end{equation}
Using \eqref{normalext} we can write the numerator as,
\begin{equation}
    \begin{split}
         & G_{\text{ext}}(\omega_n)+  (i r_H q)^{2\nu} \, (-1)^{2\nu} \, \frac{\Gamma(1-2\nu)}{\Gamma(1+2\nu)}  \frac{\Gamma{\frac{1}{2}+\kappa+\nu}}{\Gamma{\frac{1}{2}+\kappa-\nu}} \bigg\rvert_{\omega=\omega_n} \\
         &= G_{\text{ext}}(\omega_n)+  (-i r_H q)^{2\nu} \, \frac{\Gamma(1-2\nu)}{\Gamma(1+2\nu)}  \frac{\Gamma{\frac{1}{2}+\kappa+\nu}}{\Gamma{\frac{1}{2}+\kappa-\nu}} \bigg\rvert_{\omega=\omega_n} \\
         &= G_{\text{ext}} -G_{\text{ext}}^* \bigg\rvert_{\omega=\omega_n} \\
         &= 2i \, \text{Im}  G_{\text{ext}} \bigg\rvert_{\omega=\omega_n}  \ . \label{GextGextstarext}
    \end{split}
\end{equation}
The denominator is given by
\begin{equation}
    { \partial_{\omega}\left( 1+  e^{-z_0} z_0^{2\kappa} (-1)^{\frac{3}{2}+\kappa-\nu}  \frac{\Gamma{\frac{1}{2}-\kappa+\nu}}{\Gamma{\frac{1}{2}+\kappa+\nu}} \right) \bigg\rvert_{\omega=\omega_n}   }= \partial_{\omega} e^{i \theta(\omega, m)} \bigg\rvert_{\omega=\omega_n} \ . 
\end{equation}
So we can finally write:
\begin{equation}
    \text{Res}(G, \omega_n)= 2 i \frac{\text{Im} G_{\text{ext}}(\omega, m)\bigg\rvert_{\omega=\omega_n} }{\partial_{\omega} e^{i \theta(\omega, m)} \bigg\rvert_{\omega=\omega_n}} \ . \label{resGext}
\end{equation}
When the stretched horizon comes very close to the event horizon, a similar phenomenon of pole condensation occurs, as discussed in Section 3. In this regime, the poles on the real line can be approximated as a branch cut, allowing us to replace the discrete sum over $q_n$ with an integral over a continuous variable $q$. Using a similar approach as outlined in Section 3, the position space Green’s function can be expressed as:
\begin{align}
 C(u, v)= \theta(u) \, \mathcal{N} \int_{\mathbf{R}^2} dp \, dq \, e^{ipv-iqu}   \text{Im}G_{\text{ext}}(p, q) \ . \label{Cuvext}
\end{align}
where,
\begin{align}
    \text{Im}G_{\text{ext}}(p, q) =-\frac{1}{2i}\left((2i r_H q)^{\Delta-1} \frac{\Gamma(2-\Delta) \Gamma(\frac{\Delta}{2}-\frac{i p}{2r_H})}{\Gamma(\Delta) \Gamma(1-\frac{\Delta}{2}-\frac{ip}{2r_H})} - (-2i r_H q)^{\Delta-1} \frac{\Gamma(2-\Delta) \Gamma(\frac{\Delta}{2}+\frac{i p}{2r_H})}{\Gamma(\Delta) \Gamma(1-\frac{\Delta}{2}+\frac{ip}{2r_H})} \right).
\end{align}
Where the first term has poles at lower half $p$-plane at $p=-i r_H(\Delta+2n)$ and the second term has poles in the upper half plane. If we choose $v<0$ then the poles in the lower half plane only contribute and as a result only the first term contributes while integrating over $p$ in \eqref{Cuvext}, which gives \footnote{In the following steps, we absorb various constant factors into the definition of $\mathcal{N}$ at various stages of the calculation.},
\begin{align}
     C(u, v)&= \theta(u) \theta(-v)  \, \mathcal{N} \int_{\mathbf{R^2}} dp \, dq \, e^{ipv-iqu}   (2i r_H q)^{\Delta-1} \frac{\Gamma(2-\Delta) \Gamma(\frac{\Delta}{2}-\frac{i p}{2r_H})}{\Gamma(\Delta) \Gamma(1-\frac{\Delta}{2}-\frac{ip}{2r_H})}  \nonumber \\
     &= \theta(u) \theta(-v)  \, \mathcal{N} \int_{\mathbf{R}}  dq \, e^{-iqu} (2i r_H q)^{\Delta-1} \frac{\Gamma(2-\Delta)}{\Gamma(\Delta)} \sum_{n} \frac{e^{vr_H (\Delta+2n)}\frac{(-1)^n}{n!}}{\Gamma(1-n-\Delta)} \nonumber \\
     &= \theta(u) \theta(-v)  \, \mathcal{N} \int_{\mathbf{R}}  dq \, e^{-iqu} (2i r_H q)^{\Delta-1} \frac{\Gamma(2-\Delta)}{\Gamma(\Delta)\Gamma(1-\Delta)} (\sinh{r_H v})^{-\Delta} \nonumber \\
     &= \theta(u) \theta(-v)  \, \mathcal{N} \int_{\mathbf{R}}  dq \, e^{-iqu} r_H^{\Delta-1}(q)^{\Delta-1} (\sinh{r_H v})^{-\Delta} \nonumber \\
     &=\mathcal{N} \, \theta(u) \theta(-v)  \, \left(\frac{\pi}{\beta_R} \right)^{\Delta-1} u^{-\Delta} \left(\sinh{ \frac{\pi v}{\beta_R}} \right)^{-\Delta} \ . \nonumber \\
\end{align}
It is now straightforward to read off the temperature, $T_R$, from the above position-space correlator. Therefore, in the extremal case, the branch-cut approximation yields a thermal correlator with the right-moving temperature.

\section{Analytic WKB-results}

It is evident from our discussions that obtaining an exact expression for normal modes is challenging, if possible at all. However, it is possible to obtain analytic regimes using the WKB-approximation. Note that the specific WKB-method has already been used in \cite{Das:2023xjr} and we review it here for the sake of completeness.

The WKB-method involves determining the bound states of the effective potential. To proceed, let us express the radial equation in terms of the $r$-coordinate, setting $r_{\rm H}=1$, given by:
\begin{eqnarray}
\left( r^2 -1\right)^2 \phi''(r) + 2r (r^2-1) \phi'(r) + \left(\omega^2 - (r^2 -1) \left( \frac{1}{r^2} \left( m^2 + \frac{1}{4} \right) + \frac{3}{4} \right)  \right) \phi(r) = 0 \ , \nonumber\\
\end{eqnarray}
which can be written in  Schr\"{o}dinger equation form (see Appendix of \cite{Das:2023xjr}),
\begin{eqnarray}
\frac{d^2 \Psi}{dr^2} - V(r) \Psi(r) = 0 \ .
\end{eqnarray}
with
\begin{eqnarray} \label{pot_noJ}
&& V(r) = \frac{r^2 \left(4 m^2-4 \omega ^2-6\right)-4 m^2+3 r^4-1}{4 r^2
   \left(r^2-1\right)^2} \ ,    
\end{eqnarray}
A general structure of $V(r)$ is shown in figure. This effective potential is $\infty$ at $r=r_0$ (position of the brick wall) and has a turning point at:
\begin{equation}
     r_c^2 = \frac{1}{3} \left(-2 m^2+2 \sqrt{m^4-2 m^2 \omega ^2+\omega ^4+3
   \omega ^2+3}+2 \omega ^2+3\right) \ .
\end{equation}
The corresponding WKB-formula with two turning points is given by
\begin{equation}
    \int_{r_0}^{r_c} |V(r)|^{\frac{1}{2}} dr=\frac{3\pi}{4}+ 2 n \pi \ . 
\end{equation}
where $n$ is the principal quantum number. Fortunately, for \eqref{pot_noJ}, this integration can be performed, resulting in the following:
\begin{eqnarray}\label{int_WKB}
&& \int_{r_0}^{r_c} |V(r)|^{\frac{1}{2}} dr = \frac{1}{8} \left(-2 \sqrt{3} \tan ^{-1}(a_1)+2 \sqrt{\omega^2+1} \log \left(\frac{b_1+1}{b_1 - 1}\right) - \sqrt{4 m^2+1} \log \left(\frac{c_1+1}{c_1-1}\right)-\sqrt{3} \pi \right) \nonumber\\
&& a_1=\frac{-2 m^2-3 r_0^2+2 \omega ^2+3}{\sqrt{3} \sqrt{-4 m^2 \left(r_0^2-1\right)-3 r_0^4+r^2 \left(4 \omega^2 + 6 \right)+1}} \ , \\
&& b_1 = \frac{m^2 \left(-\left(r_0^2 - 1\right) \right) + \left(r_0^2+1\right) \omega ^2+2}{\sqrt{\omega ^2+1} \sqrt{-4 m^2 \left(r_0^2 - 1\right)-3 r_0^4 + r_0^2 \left(4 \omega^2+6\right)+1}} \ , \\
&& c_1 = \frac{-2 m^2 \left(r_0^2-2\right) + r_0^2 \left(2 \omega^2+3\right)+1}{\sqrt{4 m^2+1} \sqrt{-4 m^2 \left(r_0^2 - 1\right) - 3r_0^4 + r_0^2 \left(4 \omega^2+6\right)+1}} \ .       
\end{eqnarray}
When the cut-off surface is an $\epsilon$ (coordinate) distance away from the horizon ($r_0=r_H+\epsilon$) we can do a small $\epsilon$ expansion of \eqref{int_WKB} which in the limit $\omega \gg 1$ , $m\gg 1$ such that $\omega (m) \ll m$ gives:
\begin{equation}
    \frac{3\pi}{4}+ 2 n \pi  = \frac{1}{4} \left(\omega  \log \left(\frac{4 \omega ^4}{m^4
   \epsilon ^2}\right)-\sqrt{3} \tan ^{-1}\left(\frac{\sqrt{3}
   \omega }{m^2}\right)\right) \ . 
\end{equation}
The first some gives dominating contribution at $\epsilon \rightarrow 0$ limit, which yields:
\begin{eqnarray}
\omega(m) = \frac{\pi  (8 n+3)}{4 W\left(\frac{8 \sqrt{2} \pi  n+3 \sqrt{2} \pi }{4 m \sqrt{\epsilon }}\right)} \ , \label{omegaexact}
\end{eqnarray}
where $W$ is the product log function.

\section{The Boulware, stretched horizon and Hartle-Hawking states}
\label{s.boulwarehartlehawking}
The general theory of relativity states that in a curved space there is no unambiguous definition of a vacuum. The inherent diffeomorphism invariance enables to choose different canonical time coordinates for different observers. The various ground states are connected to each other by a Bogoliubov transformation.\\
The Boulware groud state is essential to provide a model of a black hole mimicker. Therefore, this vacuum is introduced below, and furthermore compared to the Hartle-Hawking state, which is the ground state of a black hole configuration. Both ground states are explained in a (1+1)-dimensional example, using the metric 
\begin{align}
    ds^2=-h(r)dt^2+\frac{dr^2}{h(r)},\label{eq.2dmetric}
\end{align}
where the horizon is located at $r_0$, implying $h(r_0)=0$. The energy density $\rho=-T^t_t$ and the pressure $P=T^r_r$ are entirely determined by demanding the conservation of energy, and further by evaluating the trace anomaly
\begin{align}
    D_\mu T^\mu_\nu=0,\,\,\,T^\mu_\mu=\frac{R}{24\pi},
\end{align}
respectively \cite{Mukohyama:1998rf}. The off-diagonal elements of the stress-energy tensor vanish. The Richi scalar for the given metric is given by $R=-h^{\prime\prime}(r)$. Using the previous definitions one obtains
\begin{align}
    P(r)=\frac{-\kappa^2(r)+c}{24\pi h(r)},\label{eq.defp}
\end{align}
where partial integration is used. The integration constant $c$ is determined by the choice of the ground state. \\
In a spacetime containing a stationary eternal black hole, the Hartle-Hawking ground state is the appropriate ground state. This state is identified by considering the annihilation operator $a_{\rm Kruskal}|0\rangle_{HH}=0$ which is related to positive frequency modes in the Kruskal-coordinates. This state seems empty of particles for a free-falling observer, and further the corresponding stress-energy is bounded at the black hole horizon. Therefore, using equation \ref{eq.defp} one obtains for the pressure and the energy density
\begin{align}
     P_{HH}=\frac{\kappa_0^2-\kappa^2(r)}{24\pi h(r)},\,\,\rho_{HH}=P_{HH}+\frac{h^{\prime\prime}(r)}{24\pi},
\end{align}
respectively \cite{Mukohyama:1998rf}.\\
In contrast, considering a compact spherically symmetric object embedded in a Schwarzschild manifold the Boulware ground state provides a bound for the present energy densities \cite{Marecki:2006wc}. It is a zero temperature state and in an asymptotically flat space the Boulware ground state approaches the Minkowski vacuum at infinity \cite{Zaslavskii:2003is}. The Boulware ground state is annihilated by the annihilation operator associated with the Killing time parameter $a_{\rm Killing}|0\rangle_{B}=0$. For this configuration, pressure and density are obtained to be given by
\begin{align}
    P_{B}=-\frac{\kappa^2(r)}{24\pi h(r)},\,\,\rho_B=P_B+\frac{h^{\prime\prime}(r)}{24\pi},
\end{align}
where equation \ref{eq.defp} is used.\\
In order to continue, the density and the pressure for both configurations are compared. Using the redshifted Hawking temperature one can calculate
\begin{align}
    \Delta \rho=\Delta P= \frac{T^2(r)}{6\pi}.
\end{align}
Therefore, the stress-energy inherent in the black hole configuration can be decomposed into
\begin{align}(T_{HH})^\mu_\nu=(T_B)^\mu_\nu+(T_{th})^\mu_\nu,
\end{align}
where $(T_{th})^\mu_\nu$ refers to thermal contributions. Thus, the stress-energy of a black hole is reproduced by thermal quantum fields having the Boulware ground state \cite{Mukohyama:1998rf}.
%
%
%
%
%
\subsection{The stretched Horizon}
%
The stretched horizon model provides a way of understanding the information paradox in terms of quantum fields located just outside of this surface. The primary idea here is to impose a Dirichlet boundary condition to the quantum fields and thus prevent the bulk degrees of freedom from reaching the horizon. Generally, such a Dirichlet boundary condition can break the rotational symmetry around the event horizon, however, for simplicity, we will consider only symmetric cases. The radius of this spherical surface is slightly larger than the gravitational radius of the mass configuration $r_s=r_0+\varepsilon$, $\varepsilon>0$. The mode functions of the quantum fields are vanishing at the stretched horizon \cite{Burman:2023kko}. Therefore, observables stay finite, which is similar to the cut-off regularization in quantum field theory in flat space. This can be considered to be a mathematical model which provides a way to remove divergences and yields regularized quantum expectation values. Furthermore, this results in an area-worth entropy on the stretched horizon which mimics the entropy of a black hole. In a two dimensional example of a stretched horizon, pressure and density are determined by 
\begin{align}
    P_s=\frac{c^2-\kappa^2 (r)}{24\pi h(r)}, \rho_s=P_s+\frac{h^{\prime\prime}(r)}{24\pi} \ ,
\end{align}
where the metric \ref{eq.2dmetric} is used \cite{Mukohyama:1998rf}. Here, $\kappa (r)=1/2h^\prime(r)$ defines the surface gravity, while $c^2$ is a regulation parameter, that the observables are finite at the stretched horizon. This can be compared to the results for a black hole configuration with the Hartle-Hawking state:
\begin{align}
     P_{HH}=\frac{\kappa_0^2-\kappa^2(r)}{24\pi h(r)},\,\,\rho_{HH}=P_{HH}+\frac{h^{\prime\prime}(r)}{24\pi} \ ,
\end{align}
where $\kappa_0=\kappa(r_0)$ is the surface gravity at the horizon. The difference of the pressure and energy densities is given by
\begin{align}
    \Delta P=\Delta\rho=\frac{\kappa_0^2-c^2}{24\pi h(r)} \ .
\end{align}

To estimate the contribution relative to the position of the stretched horizon, we can rewrite the difference as:
\begin{align}
    \Delta P=\Delta\rho\approx \frac{\pi}{6}T^2(r_0+\varepsilon)-\frac{c^2}{24\pi}\left(\frac{1}{h(r_0+\varepsilon)}+\mathcal{O}(\varepsilon)\right) \\=\frac{\pi}{6}T^2(r_0+\varepsilon)-\left(T^2_c(r_0+\varepsilon)+\mathcal{O}(\varepsilon)\right)<\frac{\pi}{6}T^2(r_0+\varepsilon) \ ,\label{eq.expand}
\end{align}
where $\varepsilon >0$ is small. The first therm in equation \ref{eq.expand} refers to the local temperature which is derived in analogue to section \ref{s.boulwarehartlehawking} for the brick-wall setup. The second term is a zeroth-order correction term, which is derived by expanding the metric in terms of $\varepsilon$ and by using $h(r_0)/h(r_s)=h(r_o)/h(r_0+\varepsilon)\approx 1+\mathcal{O}(\varepsilon)$. Choosing $c^2=0$ results in the expression discussed in section \ref{s.boulwarehartlehawking}, where it is shown that the topped-up Boulware state can reproduce the stress-energy present in a Hartle-Hawking configuration.

One can, nonetheless, consider this correction term as a reduction of radiation which is necessary for mimicking a Hartle-Hawking configuration. This can be understood as follows: The ground state energy in the stretched horizon model is higher in comparison to the energy corresponding to the Boulware state, in which the stretched horizon is placed on top of the event horizon. Thus, the stretched horizon state is exited relative to the Boulware state.

In section \ref{s.structureofgreen} the condensation of the poles of the Green's function in the horizon limit ($\varepsilon\to 0$) corresponds to the emergence of an effective branch cut in the correlater. It is subsequently shown that the retarded correlater of a scalar field propagating in a BTZ background with an ingoing boundary condition is represented by the Green's function of the same scalar field, where a Dirichlet boundary is imposed on the stretched horizon, in the limit when $\varepsilon$ is sufficiently small. The thermal contributions that emerge in this limit are discussed in section \ref{s.positionspace} in detail.

\section{The Brick-Wall model}
\label{s.bwm}
The brick-wall model describes a spherically symmetric starlike object with a reflecting surface. The radius of the surface is slightly larger than would be horizon of a black hole configuration, and thus can be considered a cut-off parameter for thermal quantum fields propagating in the background geometry \cite{tHooft:1984kcu}. In the thermal equilibrium, at infinity the temperature of the quantum fields is given by $T_\infty$. The background geometry is described by the spherically symmetric metric
\begin{align}
    ds^2=-h(r)dt^2+\frac{dr^2}{h(r)}+r^2d\Omega^2, \label{eq.bwmetric}
\end{align}
where $d\Omega^2=d\phi^2+\sin(\phi)^2d\theta^2$ is the metric on the two-sphere. For a black hole configuration the horizon is located at $h(r_0)=0$, which is assumed to be a simple zero. The temperature is shifted according to Tolman's law and at the would be horizon of a black hole approaches the Hawking value. Nevertheless, it is important to notice that the cut-off is outside off the horizon $r_0<r_1$ \cite{Mukohyama:1998rf}.\\
The thermodynamic properties of this configuration can be derived using statistical mechanics. For an on-shell particle gas the density and pressure are given by
\begin{align}
    P=\frac{1}{3}\rho=\frac{4\pi \sigma}{(2\pi)^3}\int dpY_{(\rho/P)}e^{-E(p)/T(r)},\label{eq.rhpp}
\end{align}
with the Boltzmann weights $\exp(- E(p)/T(r))$. For density and pressure applies $Y_\rho=p^2E$, $Y_P=vp^3$, respectively. $E$, $p$ are the particle energy and momentum, whereas $v$ denotes the velocity. Deviations which appear by considering the Bose-Einstein or Fermi-Dirac distribution are absorbed into $\sigma$. In the ultra relativistic limit ($E>m$, $E=p$, $v=1$) one obtains
\begin{align}
    \rho=3P\approx \frac{3\sigma}{\pi^2}T^4(r),
\end{align}
with the local temperature $T(r)=T_\infty/\sqrt{h(r)}$. The entropy density is given by 
\begin{align}
    s(r)= \frac{(\rho+P)}{T(r)},
\end{align}
and furthermore one can calculate the wall entropy
\begin{align}
    S_w =\frac{4\sigma}{\pi^2}4\pi r_1^2T^3_\infty\int^{r_1+\delta}_{r_1}dr h(r)^{-2}
    \approx \frac{\sigma}{90\pi\alpha^2}\left(\frac{T_\infty}{\kappa_0/2\pi}\right)^3\frac{A}{4}.
\end{align}
where the proper line element is used. Here, $A$ is the surface area, and further $\alpha$ is the proper distance of the wall to the surface. Moreover, the wall extensions are in the order of magnitude $\varepsilon\ll\delta\ll r_1$. In analogy, the contribution of the quantum fields to the wall mass is given by
\begin{align}
    M_{th,w}=4\pi\int^{r_1+\delta}_{r_1}drr^2\rho(r)\approx \frac{\sigma}{480\pi\alpha^2}\left(\frac{T_\infty}{\kappa_0/2\pi}\right)^3AT_\infty.
\end{align}
For these derivations it is assumed that $h(r)\approx 2\kappa_0(r-r_0)$ in the neighborhood of the wall. The free energy can be calculated by using thermodynamics
\begin{align}
    F_w=M_{th,w}-T_\infty S_w=\frac{-\sigma}{1440\pi\alpha^2}\left(\frac{T_\infty}{\kappa/2\pi}\right)^3AT_\infty.
\end{align}
Furthermore, one can calculate 
\begin{align}
    S_{w}=-\frac{\partial F_{w}}{\partial T_\infty}=\frac{-\sigma A}{360\pi \alpha^2}\left(\frac{T_\infty}{T_H}\right)^3.
\end{align}
In order to reproduce the Bekenstein-Hawking value, on-shell, the cut-off is determined to be given by $\alpha=\sqrt{\sigma/90\pi}$.\\
Moreover, using the Gibbs-Duhem relation yields
\begin{align}
    F=M-T_\infty S=\frac{3A}{16}T_H.
\end{align}
In section \ref{s.boulwarehartlehawking} it was shown that the stress-energy of a black hole configuration can be reproduced by the stress-energy of thermal quantum fields which have the Boulware ground state. On-shell, the negative energy density which is due to the Boulware ground state almost cancels the thermal energy density and results in the Hartle-Hawking value
\begin{align}
(T_{HH})^\mu_\nu=(T_{th,T_\infty=T_H})^\mu_\nu+(T_B)^\mu_\nu.
\end{align}
For a very thin layer surrounding the wall, the stress-energy almost vanishes. Evaluating
the stress-energy near the horizon, one finds that it is bounded. An analogue approach for the mass of the wall yields
\begin{align}
    M_{w}=M_{th,w}+ M_{B,w}=\frac{3}{16}AT_H\left(\left(\frac{T_\infty}{T_H}\right)^4-1\right),
\end{align}
which vanishes on-shell. Thus, there is no gravitational back-reaction between the background manifold and the wall mass for $T_\infty =T_H$.\\
In order to justify the results from the particle description, the thermodynamic properties are reproduced using quantum filed theory. The action of a real scalar field in the container around the star is given by
\begin{align}
    S[\phi]=-\frac{1}{2}\int d^4x\sqrt{-g}(g^{\mu\nu}\partial_\mu\phi\partial_\nu\phi +m^2_\phi\phi^2),
\end{align}
where again the metric equation \ref{eq.bwmetric} is used. Moreover, for the radial part the radius is within the boundaries $r\in [r_1,L]$ defined. The field equation is obtained by varying the action and yields
\begin{align}
    \phi(r,x^i)=\sum_{nlm}\frac{1}{\sqrt{2\omega_{nl}}}\left(a_{nlm}\varphi_{nl}(r)\mathcal{K}_{lm}(x^i)e^{-i\omega_{nl}t}+a^\dagger_{nlm}\varphi_{nl}(r)\mathcal{K}_{lm}(x^i)e^{i\omega_{nl}t}\right),
\end{align}
where a Kaluza–Klein reduction is performed. Furthermore, the fields are expanded in terms of the mode functions $\varphi_{nl}$, and $\mathcal{K}_{lm}$ denotes the spherical harmonics. The mode functions as well as the spherical harmonics are normalized and fulfill the completes relation, providing a complete basis. On the spherical container boundaries, which are located outside the star, Dirichlet boundary conditions are imposed $\varphi_{nl}(r_1)=\varphi_{nl}(L)=0$. The conjugated moment is given by 
\begin{align}
     \pi(r,x^i)=-i\frac{r^2\sqrt{\Omega(x^i)}}{h(r)}\sum_{nlm}\sqrt{\frac{\omega_{nl}}{2}}\left\lbrack a_{nlm}\varphi_{nl}(r)\mathcal{K}_{lm}(x^i)e^{-i\omega_{nl}t}-a^\dagger_{nlm}\varphi_{nl}(r)\mathcal{K}_{lm}(x^i)e^{i\omega_{nl}t}\right\rbrack,
\end{align}
which is derived using the appropriate definition for a curved space. The quantization is performed by the equal-time commutator relation with respect to the killing time parameter $t$. Thus, the appropriate ground state is the Boulware state \cite{Mukohyama:1998rf}. The normal ordered Hamiltonian can be calculated by performing a Legendre transformation of the Lagrangian, and thus yields
\begin{align}
    :H:=\sum_{nlm}\omega_{nl}a^\dagger_{nlm}a_{nlm},
\end{align}
where $a^\dagger_{nlm}$ and $a_{nlm}$ are the creation and annihilation operator, obtained by the quantisation pattern, respectively. Thus, the Hilbert space can be constructed upon the Boulware ground state by applying the creation operator
\begin{align}
    |\{N_{nlm}\}\rangle=\prod_{nlm}\frac{1}{\sqrt{N_{nlm}!}}\left(a^\dagger_{nlm}\right)^{N_{nlm}}|0\rangle_B.
\end{align}
Thus, the eigenvalue equation is given by
\begin{align}
    :H:|\{N_{nlm}\}\rangle =\left(\sum_{nlm}\omega_{nl}N_{nlm}\right)|\{N_{nlm}\}\rangle,
\end{align}
where the normal ordered Hamiltonian is used.\\
Following statistical mechanics, the free energy is given by 
\begin{align}
    e^{-F_w/T_\infty}=\mathrm{Tr}\lbrack e^{- :H:/T_\infty}\rbrack =\prod_{nlm}\frac{1}{1-e^{- \omega_{nl}/T_\infty}},
\end{align}
in terms of the Boltzmann weights. The mode functions are written as $\varphi_{nl}(r)=\psi_{nl}(r)\exp(-ikr)$ in order to use the WKB-approximation. Therefore, the free energy can be calculated to be given by
\begin{align}
    F_w(r)\approx 4\pi T(r)\int^\infty_0dp\frac{p^2}{(2\pi)^3}\ln{\left(1-e^{-\omega/T(r)}\right)},\label{eq.freee}
\end{align}
where the condition
\begin{align}
    \left|\frac{\partial_r(r^2h(r))}{r^2h(r)}\right|\ll \frac{T_\infty}{h(r)},
\end{align}
must be satisfied \cite{Mukohyama:1998rf}. The free energy given in equation \ref{eq.freee} can be associated to the cumulant generating function, and thus by performing derivatives, one can calculate control parameters. Using this method one obtains for the density and pressure
\begin{align}
    \rho(r)&=\frac{\partial}{\partial\beta(r)}\beta(r)\Tilde{F}(r)\approx\frac{4\pi}{(2\pi)^3}\int^\infty_0dpp^2Ee^{-E/T(r)},\\
    P(r)&=-\Tilde{F}(r)\approx\frac{4\pi}{3(2\pi)^3}\int^\infty_0dpvp^3e^{-E/T(r)},
\end{align}
and further 
\begin{align}
    s(r)=\beta^2(r)\frac{\partial}{\partial\beta(r)}\Tilde{F}(r)=\beta(r)(\rho(r)+P(r)).
\end{align}
These expressions coincide with the approach which is used for the particle description equation \ref{eq.rhpp}. In analogy, one obtains for the total energy and the entropy
\begin{align}
     U=4\pi\int^L_{r_1}drr^2\rho(r),\,\,\,
    S=4\pi \int^L_{r_1}dr\frac{r^2}{f(r)}s(r).
\end{align}
The mass contribution of the thermal quantum fields is given by
\begin{align}
    \langle M \rangle =4\pi \mathrm{Tr}\left\lbrack e^{\beta_\infty (F-:H:)}\int^L_{r_1}drr^2{T^{(ren)}}^t_t\right\rbrack,
\end{align}
where the renormalized stress-energy tensor is used which can be calculated starting from the action. Furthermore, one can calculated
\begin{align}
     M_B =-4\pi\int^L_{r_1}drr^2{}_B\langle 0|{T^{(ren)}}^t_t|0\rangle_B,
\end{align}
which corresponds to the vacuum energy of the thermal quantum fields having the Boulware ground state. Thus, one can show that the gravitational mass of the configuration can be decomposed into
\begin{align}
     \langle M \rangle &= U + M_B
    = M_{th} + M_B,
\end{align}
where $M_{th}$ is the mass from the thermal quantum fields. The mass contribution $M_B$ is due to the quantum fields having the Boulware ground state \cite{Mukohyama:1998rf}. On-shell the mass contribution of the brick-wall to the gravitational mass vanishes which is necessary for a consistent model. The configuration of having a Boulware ground state with added radiation is called a topped-up Boulware state. The Bekenstein-Hawking entropy is reproduced by adjusting the cut-off parameter $\alpha$.

Given the above, one can now calculate an $n$-point correlation function constructed from a string of local operators, {\it e.g.}~$\Phi_1 \ldots \Phi_n$, in any state in the Boulware Fock space.\footnote{Recently, in \cite{Krishnan:2023jqn}, such correlation functions have been calculated explicitly, with a special focus on two-point functions in the bulk geometry. Subsequently, it was argued that, as the Dirichlet hypersurface moves closer to the horizon, a smooth horizon correlator emerges. The smooth horizon correlator corresponds to the Hartle-Hawking vacuum.} Particularly, the stress-tensor expectation value can be obtained by computing a Wightman correlator $\langle | \Phi(x) \Phi(y) | \rangle$ and taking its derivatives with respect to $x$ and $y$ in the limit $x \to y$ limit. While an explicit computation of the stress-tensor expectation value is possible, there is a quick argument to reach a qualitative answer, see {\it e.g.}~\cite{Mathur:2024mvo} for a recent discussion on this. 

The Hartle-Hawking vacuum corresponds to a black hole in equilibrium with its radiation and therefore there is no net energy flux in or out at the horizon. Consequently, the stress-tensor expectation value vanishes in this case. On the other hand, in a near-horizon limit, we obtain a Rindler patch in which the stress-tensor expectation value matches with that of the Minkowski vacuum. Therefore it vanishes. Thus, local expectation value of the stress-tensor in both Hartle-Hawking state and in the Rindler patch match.

The Hartle-Hawking vacuum corresponds to a black hole in thermal equilibrium with its radiation and correspondingly the quantum state is regular at the past and the future event horizons. This corresponds to an eternal black hole. The Unruh vacuum, on the other hand, corresponds to a black hole at a given temperature, while its surroundings have a vanishing temperature. This corresponds to a black hole formed due to a collapse process and is regular at the future event horizon but is not regular at the past event horizon.\footnote{The Unruh vacuum eventually becomes the Hartle-Hawking vacuum if we impose reflecting boundary condition at the asymptotics so that the radiation does not escape the geometry and eventually thermalizes with the black hole.} The Boulware vacuum, locally, corresponds to the Rindler vacuum. The Minkowski vacuum appears as a thermal one in the Rindler patch. Since the stress-tensor expectation value vanishes in the Minkowski vacuum, the Rindler expectation value is also thermal with a negative contribution. Therefore, this yields a non-vanishing stress-tensor in the Boulware vacuum. We have reviewed this in details above. 
%
%
%
%
\section{Correlators in terms of geodesic length}
%
A thermal two point function should decay in time, while a two point function in an arbitrary excited pure state is expected to display a thermal behaviour for a particular time-regime beyond the dissipation scale. However, it is expected to rise back up at a time-scale which will have the information of the gap in the spectrum. This behaviour is dubbed as the Echo. Since the Dirichlet wall is, by construction, a pure state. Therefore, a two-point function in this geometry will not decay for arbitrarily long time. Nonetheless, as the Dirichlet wall approaches the event horizon, the time-scale till the decay behaviour continues will be become larger and larger. In this appendix, we provide elementary estimates on the Echo behaviour. Unlike the Fuzzball geometries, which have been shown to exhibit a series of Echoes\cite{Bena:2019azk}, we will observe only a single Echo and a subsequent plateau in the two-point function.

It is well known that in the limit of large conformal dimension of the boundary operator, equal-time correlators can be approximated by $e^{-\Delta L}$, where $L$ is the normalized geodesic length. Here we will compute this by computing the geodesic length in a brickwall type model.
\begin{figure}[H]
    \centering
    \includegraphics[width=.47\textwidth]{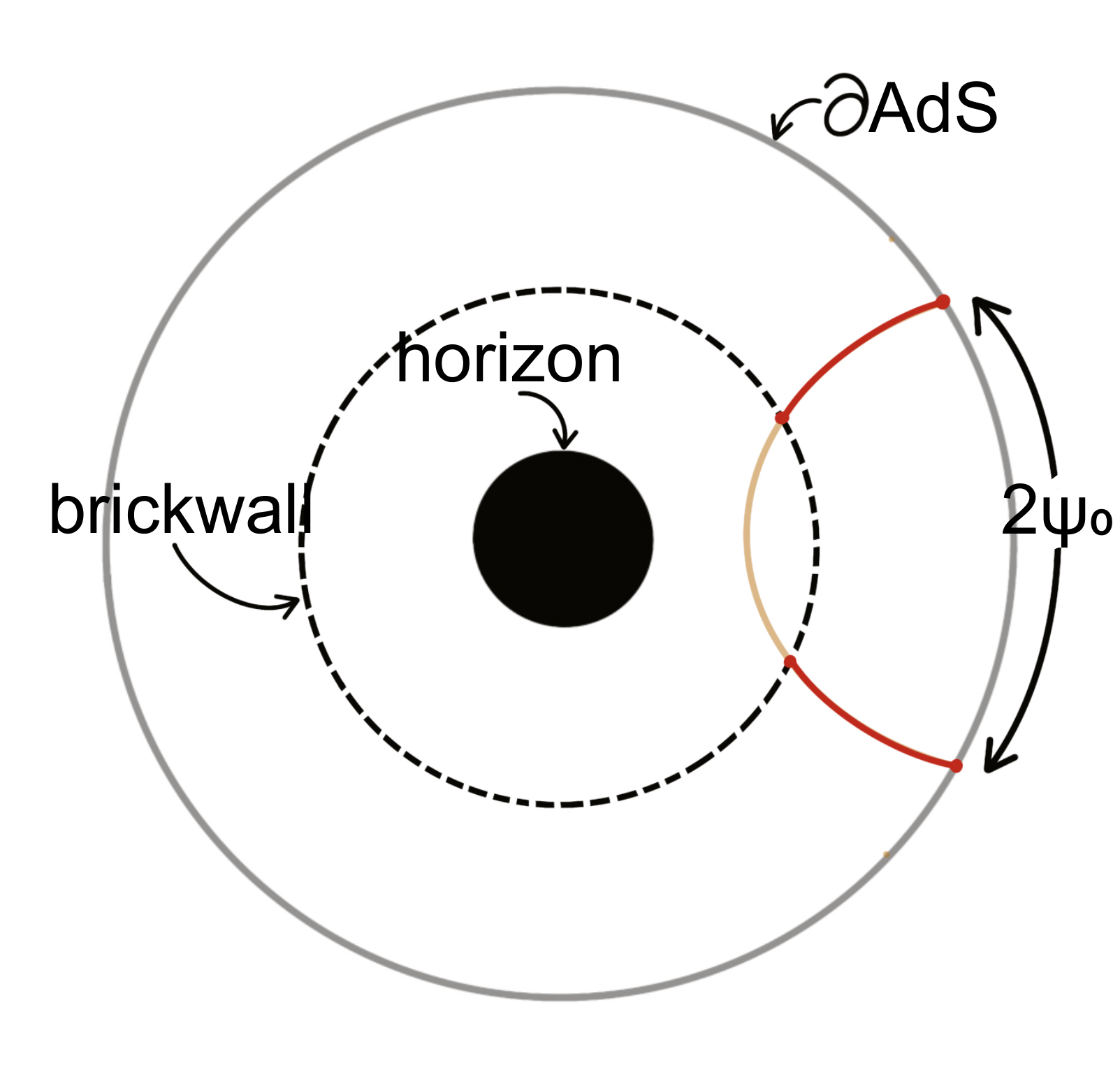}
   \caption{A schematic diagram illustrating geodesics with (red) and without (yellow) the brick wall. In the presence of the wall, the dotted line connecting the two red curves is excluded from the total length computation, as it does not represent a geodesic.}
    \label{toypic}
\end{figure}
For simplicity, we consider the BTZ metric given by,
\begin{equation}\label{btzmet}
    ds^2=-(r^2-r_H^2)dt^2+\frac{dr^2}{r^2-r_H^2}+r^2 d\psi^2 .
\end{equation}

\begin{figure}[H]
    \centering
    \includegraphics[width=.47\textwidth]{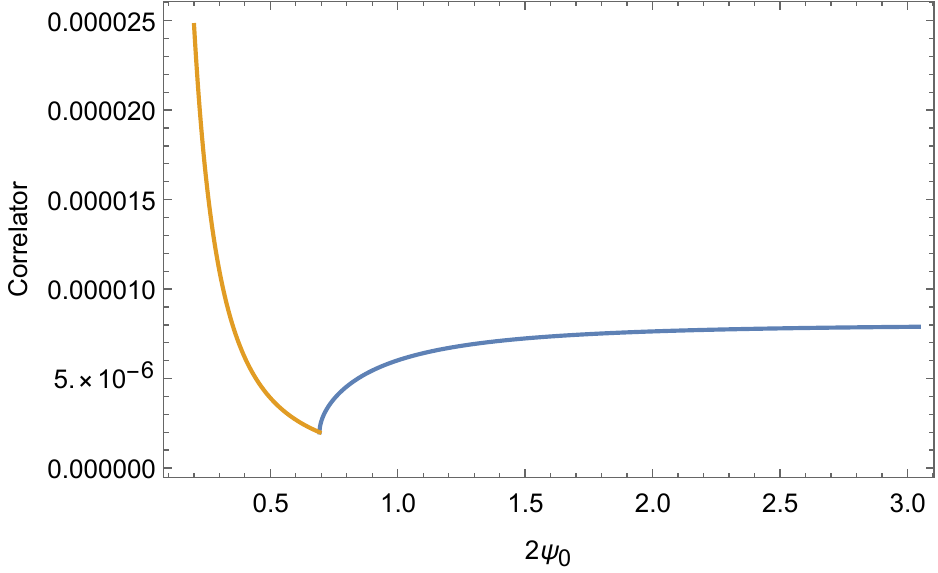}
   \caption{Figure illustrating the change in the correlator as the distance between the insertion points increases. Unlike the exponential decay (yellow curve), the correlator increases and remains non-zero beyond a certain point (blue curve) which depends the position of the brick wall.}
    \label{lengthcor}
\end{figure}

Let's focus on the constant $t$-slice of \eqref{btzmet}, which is,
\begin{equation}
    ds^2\big|_{t=const}=\frac{dr^2}{r^2-r_H^2}+r^2 d\psi^2.
\end{equation}
The geodesic equation in this geometry is the following,
\begin{equation}
    \frac{\dot{r}^2}{r^2-r_H^2}+r^2 \dot{\psi}^2=1,
\end{equation}
where dot represents derivative with respect to proper length, $\lambda$. We can express $dr/d\psi$ as,
\begin{equation}\label{geo2}
    \frac{dr}{d\psi}=\frac{r^2}{l} \sqrt{\left(1-\frac{l^2}{r^2}\right)(r^2-r_H^2)},
\end{equation}
where $l=r^2\dot{\psi}=$constant. The turning points of the geodesic are located at $r_*=l, r_H$. We will focus on the non-trivial turning point at $r_*=l$. \\ By integrating \eqref{geo2}, we can can determine $2\psi_0$, which represents the separation between the operators under consideration, as depicted in Figure \ref{toypic}.
\begin{equation}
     \int_{0}^{\psi_0} \, d\psi= \int_{r_*=l}^{\Lambda} \frac{l}{r^2} \frac{dr}{\sqrt{\left(1-\frac{l^2}{r^2}\right)(r^2-r_H^2)}}
\end{equation}
where $\Lambda$ is some UV cut-off. Which implies,
\begin{equation}\label{subsys}
    \psi_0=\frac{1}{r_H}\tanh ^{-1}\left(\frac{r_H}{l}  \sqrt{\frac{\Lambda^2-l^2}{\Lambda^2-r_H^2}}\right).
\end{equation}
Similarly, for a general $r$,
\begin{align}
    \psi(r) &= \int_{r_*=l}^{r} \frac{l}{r^2} \frac{dr}{\sqrt{\left(1-\frac{l^2}{r^2}\right)(r^2-r_H^2)}}\nonumber  \\
    &= \frac{1}{r_H}\tanh ^{-1}\left(\frac{r_H}{l}  \sqrt{\frac{r^2-l^2}{r^2-r_H^2}}\right).
\end{align}
The corresponding geodesic length is given by,
\begin{equation}
    L=2\int d\lambda=2\int\frac{dr}{\sqrt{\left(1-\frac{l^2}{r^2}\right)(r^2-r_H^2)}}.
\end{equation}
In the absence of any brickwall, it simplifies to,
\begin{align}
    L_{BH} &=2\int_{r_*}^{\Lambda}  \frac{dr}{\sqrt{\left(1-\frac{l^2}{r^2}\right)(r^2-r_H^2)}} \nonumber \\
    &= 2 \sinh^{-1} \left(\frac{\sqrt{\Lambda^2-l^2}}{\sqrt{l^2-r_H^2}}   \right)
\end{align}
In the presence of a stretched horizon, the situation is modified depending on the position of the wall $r_0$, resulting in two cases:
\begin{enumerate}
    \item When $r_0<l$ (corresponding to the yellow part in Figure \ref{lengthcor}): The geodesic length is
\begin{equation}
    L_1=2\int_{r_*=l}^{\Lambda} \frac{dr}{\sqrt{\left(1-\frac{l^2}{r^2}\right)(r^2-r_H^2)}}= 2\sinh^{-1}\left(\frac{\sqrt{\Lambda^2-l^2}}{\sqrt{l^2-r_H^2}}   \right)
\end{equation}
which is identical to $L_{BH}$ since the geodesic does not encounter the brickwall. The corresponding subsystem size is given by twice of \eqref{subsys}.

\item When $r_0\geq l$: The geodesic length of interest,
\begin{align}
    L_2 &=2\int_{r_0}^{\Lambda} \frac{dr}{\sqrt{\left(1-\frac{l^2}{r^2}\right)(r^2-r_H^2)}}\nonumber \\
    &= 2\sinh^{-1}\left(\frac{\sqrt{\Lambda^2-l^2}}{\sqrt{l^2-r_H^2}}   \right)-2\sinh^{-1}\left(\frac{\sqrt{r_0^2-l^2}}{\sqrt{l^2-r_H^2}}   \right).
\end{align}
In this scenario, we consider only the red segments of the geodesic (refer to Figure \ref{toypic}), which are not continuous. The circular part near the brickwall is excluded since it is not a geodesic.
\end{enumerate}
Figure \ref{lengthcor} illustrates the dependence of the correlation function (derived from the geodesic length) as the separation between the insertion points increases.

\bibliography{Bibliography}

\providecommand{\href}[2]{#2}\begingroup\raggedright\begin{thebibliography}{10}

\bibitem{Bekenstein1972}
J.D.~Bekenstein, \emph{{Black holes and the second law}},
  \href{https://doi.org/10.1007/BF02757029}{\emph{Lettere al Nuovo Cimento
  (1971-1985)} {\bfseries 4} (1972) 737}.

\bibitem{Hawking:1975vcx}
S.W.~Hawking, \emph{{Particle Creation by Black Holes}},
  \href{https://doi.org/10.1007/BF02345020}{\emph{Commun. Math. Phys.}
  {\bfseries 43} (1975) 199}.

\bibitem{Almheiri:2012rt}
A.~Almheiri, D.~Marolf, J.~Polchinski and J.~Sully, \emph{{Black Holes:
  Complementarity or Firewalls?}},
  \href{https://doi.org/10.1007/JHEP02(2013)062}{\emph{JHEP} {\bfseries 02}
  (2013) 062} [\href{https://arxiv.org/abs/1207.3123}{{\ttfamily 1207.3123}}].

\bibitem{Almheiri:2013hfa}
A.~Almheiri, D.~Marolf, J.~Polchinski, D.~Stanford and J.~Sully, \emph{{An
  Apologia for Firewalls}},
  \href{https://doi.org/10.1007/JHEP09(2013)018}{\emph{JHEP} {\bfseries 09}
  (2013) 018} [\href{https://arxiv.org/abs/1304.6483}{{\ttfamily 1304.6483}}].

\bibitem{Almheiri:2019hni}
A.~Almheiri, R.~Mahajan, J.~Maldacena and Y.~Zhao, \emph{{The Page curve of
  Hawking radiation from semiclassical geometry}},
  \href{https://doi.org/10.1007/JHEP03(2020)149}{\emph{JHEP} {\bfseries 03}
  (2020) 149} [\href{https://arxiv.org/abs/1908.10996}{{\ttfamily
  1908.10996}}].

\bibitem{Almheiri:2020cfm}
A.~Almheiri, T.~Hartman, J.~Maldacena, E.~Shaghoulian and A.~Tajdini,
  \emph{{The entropy of Hawking radiation}},
  \href{https://doi.org/10.1103/RevModPhys.93.035002}{\emph{Rev. Mod. Phys.}
  {\bfseries 93} (2021) 035002}
  [\href{https://arxiv.org/abs/2006.06872}{{\ttfamily 2006.06872}}].

\bibitem{Lewkowycz:2013nqa}
A.~Lewkowycz and J.~Maldacena, \emph{{Generalized gravitational entropy}},
  \href{https://doi.org/10.1007/JHEP08(2013)090}{\emph{JHEP} {\bfseries 08}
  (2013) 090} [\href{https://arxiv.org/abs/1304.4926}{{\ttfamily 1304.4926}}].

\bibitem{Penington:2019npb}
G.~Penington, \emph{{Entanglement Wedge Reconstruction and the Information
  Paradox}}, \href{https://doi.org/10.1007/JHEP09(2020)002}{\emph{JHEP}
  {\bfseries 09} (2020) 002}
  [\href{https://arxiv.org/abs/1905.08255}{{\ttfamily 1905.08255}}].

\bibitem{Saad:2019lba}
P.~Saad, S.H.~Shenker and D.~Stanford, \emph{{JT gravity as a matrix
  integral}},  \href{https://arxiv.org/abs/1903.11115}{{\ttfamily 1903.11115}}.

\bibitem{Penington:2019kki}
G.~Penington, S.H.~Shenker, D.~Stanford and Z.~Yang, \emph{{Replica wormholes
  and the black hole interior}},
  \href{https://doi.org/10.1007/JHEP03(2022)205}{\emph{JHEP} {\bfseries 03}
  (2022) 205} [\href{https://arxiv.org/abs/1911.11977}{{\ttfamily
  1911.11977}}].

\bibitem{Saad:2021rcu}
P.~Saad, S.H.~Shenker, D.~Stanford and S.~Yao, \emph{{Wormholes without
  averaging}},  \href{https://arxiv.org/abs/2103.16754}{{\ttfamily
  2103.16754}}.

\bibitem{Bousso:2022ntt}
R.~Bousso, X.~Dong, N.~Engelhardt, T.~Faulkner, T.~Hartman, S.H.~Shenker
  et~al., \emph{{Snowmass White Paper: Quantum Aspects of Black Holes and the
  Emergence of Spacetime}},  \href{https://arxiv.org/abs/2201.03096}{{\ttfamily
  2201.03096}}.

\bibitem{Harlow:2022qsq}
D.~Harlow et~al., \emph{{TF1 Snowmass Report: Quantum gravity, string theory,
  and black holes}},  \href{https://arxiv.org/abs/2210.01737}{{\ttfamily
  2210.01737}}.

\bibitem{Bena:2022ldq}
I.~Bena, E.J.~Martinec, S.D.~Mathur and N.P.~Warner, \emph{{Snowmass White
  Paper: Micro- and Macro-Structure of Black Holes}},
  \href{https://arxiv.org/abs/2203.04981}{{\ttfamily 2203.04981}}.

\bibitem{Bena:2022rna}
I.~Bena, E.J.~Martinec, S.D.~Mathur and N.P.~Warner, \emph{{Fuzzballs and
  Microstate Geometries: Black-Hole Structure in String Theory}},
  \href{https://arxiv.org/abs/2204.13113}{{\ttfamily 2204.13113}}.

\bibitem{Buchdahl:1959zz}
H.A.~Buchdahl, \emph{{General Relativistic Fluid Spheres}},
  \href{https://doi.org/10.1103/PhysRev.116.1027}{\emph{Phys. Rev.} {\bfseries
  116} (1959) 1027}.

\bibitem{Gibbons:2013tqa}
G.W.~Gibbons and N.P.~Warner, \emph{{Global structure of five-dimensional
  fuzzballs}},
  \href{https://doi.org/10.1088/0264-9381/31/2/025016}{\emph{Class. Quant.
  Grav.} {\bfseries 31} (2014) 025016}
  [\href{https://arxiv.org/abs/1305.0957}{{\ttfamily 1305.0957}}].

\bibitem{Mathur:2016ffb}
S.D.~Mathur, \emph{{What prevents gravitational collapse in string theory?}},
  \href{https://doi.org/10.1142/S0218271816440181}{\emph{Int. J. Mod. Phys. D}
  {\bfseries 25} (2016) 1644018}
  [\href{https://arxiv.org/abs/1609.05222}{{\ttfamily 1609.05222}}].

\bibitem{tHooft:1984kcu}
G.~'t~Hooft, \emph{{On the Quantum Structure of a Black Hole}},
  \href{https://doi.org/10.1016/0550-3213(85)90418-3}{\emph{Nucl. Phys. B}
  {\bfseries 256} (1985) 727}.

\bibitem{Das:2022evy}
S.~Das, C.~Krishnan, A.P.~Kumar and A.~Kundu, \emph{{Synthetic fuzzballs: a
  linear ramp from black hole normal modes}},
  \href{https://doi.org/10.1007/JHEP01(2023)153}{\emph{JHEP} {\bfseries 01}
  (2023) 153} [\href{https://arxiv.org/abs/2208.14744}{{\ttfamily
  2208.14744}}].

\bibitem{Das:2023ulz}
S.~Das, S.K.~Garg, C.~Krishnan and A.~Kundu, \emph{{Fuzzballs and random
  matrices}}, \href{https://doi.org/10.1007/JHEP10(2023)031}{\emph{JHEP}
  {\bfseries 10} (2023) 031}
  [\href{https://arxiv.org/abs/2301.11780}{{\ttfamily 2301.11780}}].

\bibitem{Das:2023xjr}
S.~Das and A.~Kundu, \emph{{Brickwall in rotating BTZ: a dip-ramp-plateau
  story}}, \href{https://doi.org/10.1007/JHEP02(2024)049}{\emph{JHEP}
  {\bfseries 02} (2024) 049}
  [\href{https://arxiv.org/abs/2310.06438}{{\ttfamily 2310.06438}}].

\bibitem{Das:2023yfj}
S.~Das, S.K.~Garg, C.~Krishnan and A.~Kundu, \emph{{What is the Simplest Linear
  Ramp?}}, \href{https://doi.org/10.1007/JHEP01(2024)172}{\emph{JHEP}
  {\bfseries 01} (2024) 172}
  [\href{https://arxiv.org/abs/2308.11704}{{\ttfamily 2308.11704}}].

\bibitem{Krishnan:2023jqn}
C.~Krishnan and P.S.~Pathak, \emph{{Normal modes of the stretched horizon: a
  bulk mechanism for black hole microstate level spacing}},
  \href{https://doi.org/10.1007/JHEP03(2024)162}{\emph{JHEP} {\bfseries 03}
  (2024) 162} [\href{https://arxiv.org/abs/2312.14109}{{\ttfamily
  2312.14109}}].

\bibitem{Burman:2023kko}
V.~Burman, S.~Das and C.~Krishnan, \emph{{A smooth horizon without a smooth
  horizon}}, \href{https://doi.org/10.1007/JHEP03(2024)014}{\emph{JHEP}
  {\bfseries 2024} (2024) 014}
  [\href{https://arxiv.org/abs/2312.14108}{{\ttfamily 2312.14108}}].

\bibitem{Banerjee:2024dpl}
S.~Banerjee, S.~Das, M.~Dorband and A.~Kundu, \emph{{Brickwall, normal modes,
  and emerging thermality}},
  \href{https://doi.org/10.1103/PhysRevD.109.126020}{\emph{Phys. Rev. D}
  {\bfseries 109} (2024) 126020}
  [\href{https://arxiv.org/abs/2401.01417}{{\ttfamily 2401.01417}}].

\bibitem{Krishnan:2024kzf}
C.~Krishnan and R.~Mondol, \emph{{Young Black Holes Have Smooth Horizons: A
  Swampland Argument}},  \href{https://arxiv.org/abs/2407.11952}{{\ttfamily
  2407.11952}}.

\bibitem{Burman:2024egy}
V.~Burman and C.~Krishnan, \emph{{A Bottom-Up Approach to Black Hole
  Microstates}},  \href{https://arxiv.org/abs/2409.05850}{{\ttfamily
  2409.05850}}.

\bibitem{Krishnan:2024sle}
C.~Krishnan and P.S.~Pathak, \emph{{Holomorphic Factorization at the Quantum
  Horizon}},  \href{https://arxiv.org/abs/2410.00732}{{\ttfamily 2410.00732}}.

\bibitem{Papadodimas:2012aq}
K.~Papadodimas and S.~Raju, \emph{{An Infalling Observer in AdS/CFT}},
  \href{https://doi.org/10.1007/JHEP10(2013)212}{\emph{JHEP} {\bfseries 10}
  (2013) 212} [\href{https://arxiv.org/abs/1211.6767}{{\ttfamily 1211.6767}}].

\bibitem{Papadodimas:2013jku}
K.~Papadodimas and S.~Raju, \emph{{State-Dependent Bulk-Boundary Maps and Black
  Hole Complementarity}},
  \href{https://doi.org/10.1103/PhysRevD.89.086010}{\emph{Phys. Rev. D}
  {\bfseries 89} (2014) 086010}
  [\href{https://arxiv.org/abs/1310.6335}{{\ttfamily 1310.6335}}].

\bibitem{Papadodimas:2013wnh}
K.~Papadodimas and S.~Raju, \emph{{Black Hole Interior in the Holographic
  Correspondence and the Information Paradox}},
  \href{https://doi.org/10.1103/PhysRevLett.112.051301}{\emph{Phys. Rev. Lett.}
  {\bfseries 112} (2014) 051301}
  [\href{https://arxiv.org/abs/1310.6334}{{\ttfamily 1310.6334}}].

\bibitem{Papadodimas:2015jra}
K.~Papadodimas and S.~Raju, \emph{{Remarks on the necessity and implications of
  state-dependence in the black hole interior}},
  \href{https://doi.org/10.1103/PhysRevD.93.084049}{\emph{Phys. Rev. D}
  {\bfseries 93} (2016) 084049}
  [\href{https://arxiv.org/abs/1503.08825}{{\ttfamily 1503.08825}}].

\bibitem{Papadodimas:2015xma}
K.~Papadodimas and S.~Raju, \emph{{Local Operators in the Eternal Black Hole}},
  \href{https://doi.org/10.1103/PhysRevLett.115.211601}{\emph{Phys. Rev. Lett.}
  {\bfseries 115} (2015) 211601}
  [\href{https://arxiv.org/abs/1502.06692}{{\ttfamily 1502.06692}}].

\bibitem{Raju:2020smc}
S.~Raju, \emph{{Lessons from the information paradox}},
  \href{https://doi.org/10.1016/j.physrep.2021.10.001}{\emph{Phys. Rept.}
  {\bfseries 943} (2022) 1} [\href{https://arxiv.org/abs/2012.05770}{{\ttfamily
  2012.05770}}].

\bibitem{Geng:2021hlu}
H.~Geng, A.~Karch, C.~Perez-Pardavila, S.~Raju, L.~Randall, M.~Riojas et~al.,
  \emph{{Inconsistency of islands in theories with long-range gravity}},
  \href{https://doi.org/10.1007/JHEP01(2022)182}{\emph{JHEP} {\bfseries 01}
  (2022) 182} [\href{https://arxiv.org/abs/2107.03390}{{\ttfamily
  2107.03390}}].

\bibitem{Raju:2021lwh}
S.~Raju, \emph{{Failure of the split property in gravity and the information
  paradox}}, \href{https://doi.org/10.1088/1361-6382/ac482b}{\emph{Class.
  Quant. Grav.} {\bfseries 39} (2022) 064002}
  [\href{https://arxiv.org/abs/2110.05470}{{\ttfamily 2110.05470}}].

\bibitem{Raju:2018xue}
S.~Raju and P.~Shrivastava, \emph{{Critique of the fuzzball program}},
  \href{https://doi.org/10.1103/PhysRevD.99.066009}{\emph{Phys. Rev. D}
  {\bfseries 99} (2019) 066009}
  [\href{https://arxiv.org/abs/1804.10616}{{\ttfamily 1804.10616}}].

\bibitem{Calabrese:2009qy}
P.~Calabrese and J.~Cardy, \emph{{Entanglement entropy and conformal field
  theory}}, \href{https://doi.org/10.1088/1751-8113/42/50/504005}{\emph{J.
  Phys. A} {\bfseries 42} (2009) 504005}
  [\href{https://arxiv.org/abs/0905.4013}{{\ttfamily 0905.4013}}].

\bibitem{Cardy:2014rqa}
J.~Cardy, \emph{{Thermalization and Revivals after a Quantum Quench in
  Conformal Field Theory}},
  \href{https://doi.org/10.1103/PhysRevLett.112.220401}{\emph{Phys. Rev. Lett.}
  {\bfseries 112} (2014) 220401}
  [\href{https://arxiv.org/abs/1403.3040}{{\ttfamily 1403.3040}}].

\bibitem{Calabrese:2016xau}
P.~Calabrese and J.~Cardy, \emph{{Quantum quenches in 1 + 1 dimensional
  conformal field theories}},
  \href{https://doi.org/10.1088/1742-5468/2016/06/064003}{\emph{J. Stat. Mech.}
  {\bfseries 1606} (2016) 064003}
  [\href{https://arxiv.org/abs/1603.02889}{{\ttfamily 1603.02889}}].

\bibitem{Das:2021qsd}
S.~Das, B.~Ezhuthachan, A.~Kundu, S.~Porey and B.~Roy, \emph{{Critical
  quenches, OTOCs and early-time chaos}},
  \href{https://doi.org/10.1007/JHEP07(2022)046}{\emph{JHEP} {\bfseries 07}
  (2022) 046} [\href{https://arxiv.org/abs/2108.12884}{{\ttfamily
  2108.12884}}].

\bibitem{Akal:2021foz}
I.~Akal, Y.~Kusuki, N.~Shiba, T.~Takayanagi and Z.~Wei, \emph{{Holographic
  moving mirrors}},
  \href{https://doi.org/10.1088/1361-6382/ac2c1b}{\emph{Class. Quant. Grav.}
  {\bfseries 38} (2021) 224001}
  [\href{https://arxiv.org/abs/2106.11179}{{\ttfamily 2106.11179}}].

\bibitem{Akal:2022qei}
I.~Akal, T.~Kawamoto, S.-M.~Ruan, T.~Takayanagi and Z.~Wei, \emph{{Zoo of
  holographic moving mirrors}},
  \href{https://doi.org/10.1007/JHEP08(2022)296}{\emph{JHEP} {\bfseries 08}
  (2022) 296} [\href{https://arxiv.org/abs/2205.02663}{{\ttfamily
  2205.02663}}].

\bibitem{Cotler:2022weg}
J.~Cotler and A.~Strominger, \emph{{The Universe as a Quantum Encoder}},
  \href{https://arxiv.org/abs/2201.11658}{{\ttfamily 2201.11658}}.

\bibitem{Biswas:2024mlq}
P.~Biswas, B.~Ezhuthachan, A.~Kundu and B.~Roy, \emph{{Moving Mirrors, OTOCs
  and Scrambling}},  \href{https://arxiv.org/abs/2406.05772}{{\ttfamily
  2406.05772}}.

\bibitem{Mayerson:2023wck}
D.R.~Mayerson and B.~Vercnocke, \emph{{Observational Opportunities for the
  Fuzzball Program}},  \href{https://arxiv.org/abs/2306.01565}{{\ttfamily
  2306.01565}}.

\bibitem{Banerjee:2023uto}
S.~Banerjee, U.~Danielsson and M.~Zemsch, \emph{{The dark bubbleography}},
  \href{https://doi.org/10.1007/JHEP02(2024)102}{\emph{JHEP} {\bfseries 02}
  (2024) 102} [\href{https://arxiv.org/abs/2311.16242}{{\ttfamily
  2311.16242}}].

\bibitem{Banerjee:2021qei}
S.~Banerjee, U.~Danielsson and S.~Giri, \emph{{Dark bubbles and black holes}},
  \href{https://doi.org/10.1007/JHEP09(2021)158}{\emph{JHEP} {\bfseries 09}
  (2021) 158} [\href{https://arxiv.org/abs/2102.02164}{{\ttfamily
  2102.02164}}].

\bibitem{Giri:2024cks}
S.~Giri, U.~Danielsson, L.~Lehner and F.~Pretorius, \emph{{Exploring Black Hole
  Mimickers: Electromagnetic and Gravitational Signatures of AdS Black
  Shells}},  \href{https://arxiv.org/abs/2405.08062}{{\ttfamily 2405.08062}}.

\bibitem{Danielsson:2023onu}
U.~Danielsson and S.~Giri, \emph{{Horizonless black hole mimickers with spin}},
  \href{https://doi.org/10.1103/PhysRevD.109.024038}{\emph{Phys. Rev. D}
  {\bfseries 109} (2024) 024038}
  [\href{https://arxiv.org/abs/2310.12148}{{\ttfamily 2310.12148}}].

\bibitem{Mukohyama:1998rf}
S.~Mukohyama and W.~Israel, \emph{{Black holes, brick walls and the Boulware
  state}}, \href{https://doi.org/10.1103/PhysRevD.58.104005}{\emph{Phys. Rev.
  D} {\bfseries 58} (1998) 104005}
  [\href{https://arxiv.org/abs/gr-qc/9806012}{{\ttfamily gr-qc/9806012}}].

\bibitem{Denef:2009kn}
F.~Denef, S.A.~Hartnoll and S.~Sachdev, \emph{{Black hole determinants and
  quasinormal modes}},
  \href{https://doi.org/10.1088/0264-9381/27/12/125001}{\emph{Class. Quant.
  Grav.} {\bfseries 27} (2010) 125001}
  [\href{https://arxiv.org/abs/0908.2657}{{\ttfamily 0908.2657}}].

\bibitem{Banados:1992wn}
M.~Banados, C.~Teitelboim and J.~Zanelli, \emph{{The Black hole in
  three-dimensional space-time}},
  \href{https://doi.org/10.1103/PhysRevLett.69.1849}{\emph{Phys. Rev. Lett.}
  {\bfseries 69} (1992) 1849}
  [\href{https://arxiv.org/abs/hep-th/9204099}{{\ttfamily hep-th/9204099}}].

\bibitem{Birrell:1982ix}
N.D.~Birrell and P.C.W.~Davies, \emph{{Quantum Fields in Curved Space}},
  Cambridge Monographs on Mathematical Physics, Cambridge Univ. Press,
  Cambridge, UK (2, 1984),
  \href{https://doi.org/10.1017/CBO9780511622632}{10.1017/CBO9780511622632}.

\bibitem{Leutheusser:2021frk}
S.A.W.~Leutheusser, \emph{{Emergent Times in Holographic Duality}},
  \href{https://doi.org/10.1103/PhysRevD.108.086020}{\emph{Phys. Rev. D}
  {\bfseries 108} (2023) 086020}
  [\href{https://arxiv.org/abs/2112.12156}{{\ttfamily 2112.12156}}].

\bibitem{Emparan:2023ypa}
R.~Emparan and J.M.~Magan, \emph{{Tearing down spacetime with quantum
  disentanglement}}, \href{https://doi.org/10.1007/JHEP03(2024)078}{\emph{JHEP}
  {\bfseries 03} (2024) 078}
  [\href{https://arxiv.org/abs/2312.06803}{{\ttfamily 2312.06803}}].

\bibitem{Mathur:2024mvo}
S.D.~Mathur and M.~Mehta, \emph{{The universal thermodynamic properties of
  Extremely Compact Objects}},
  \href{https://arxiv.org/abs/2402.13166}{{\ttfamily 2402.13166}}.

\bibitem{Mathur:2024mtf}
S.D.~Mathur, \emph{{The secret structure of the gravitational vacuum}},
  \href{https://arxiv.org/abs/2405.08945}{{\ttfamily 2405.08945}}.

\bibitem{Maldacena:2001kr}
J.M.~Maldacena, \emph{{Eternal black holes in anti-de Sitter}},
  \href{https://doi.org/10.1088/1126-6708/2003/04/021}{\emph{JHEP} {\bfseries
  04} (2003) 021} [\href{https://arxiv.org/abs/hep-th/0106112}{{\ttfamily
  hep-th/0106112}}].

\bibitem{Giusto:2023awo}
S.~Giusto, C.~Iossa and R.~Russo, \emph{{The black hole behind the cut}},
  \href{https://doi.org/10.1007/JHEP10(2023)050}{\emph{JHEP} {\bfseries 10}
  (2023) 050} [\href{https://arxiv.org/abs/2306.15305}{{\ttfamily
  2306.15305}}].

\bibitem{Bena:2019azk}
I.~Bena, P.~Heidmann, R.~Monten and N.P.~Warner, \emph{{Thermal Decay without
  Information Loss in Horizonless Microstate Geometries}},
  \href{https://doi.org/10.21468/SciPostPhys.7.5.063}{\emph{SciPost Phys.}
  {\bfseries 7} (2019) 063} [\href{https://arxiv.org/abs/1905.05194}{{\ttfamily
  1905.05194}}].

\bibitem{Ikeda:2021uvc}
T.~Ikeda, M.~Bianchi, D.~Consoli, A.~Grillo, J.F.~Morales, P.~Pani et~al.,
  \emph{{Black-hole microstate spectroscopy: Ringdown, quasinormal modes, and
  echoes}}, \href{https://doi.org/10.1103/PhysRevD.104.066021}{\emph{Phys. Rev.
  D} {\bfseries 104} (2021) 066021}
  [\href{https://arxiv.org/abs/2103.10960}{{\ttfamily 2103.10960}}].

\bibitem{Ba_uls_2011}
M.C.~Bañuls, J.I.~Cirac and M.B.~Hastings, \emph{Strong and weak
  thermalization of infinite nonintegrable quantum systems},
  \href{https://doi.org/10.1103/physrevlett.106.050405}{\emph{Physical Review
  Letters} {\bfseries 106} (2011) }.

\bibitem{Barbon:2003aq}
J.L.F.~Barbon and E.~Rabinovici, \emph{{Very long time scales and black hole
  thermal equilibrium}},
  \href{https://doi.org/10.1088/1126-6708/2003/11/047}{\emph{JHEP} {\bfseries
  11} (2003) 047} [\href{https://arxiv.org/abs/hep-th/0308063}{{\ttfamily
  hep-th/0308063}}].

\bibitem{Barbon:2014rma}
J.L.F.~Barbon and E.~Rabinovici, \emph{{Geometry And Quantum Noise}},
  \href{https://doi.org/10.1002/prop.201400044}{\emph{Fortsch. Phys.}
  {\bfseries 62} (2014) 626} [\href{https://arxiv.org/abs/1404.7085}{{\ttfamily
  1404.7085}}].

\bibitem{Cardoso:2004hs}
V.~Cardoso and O.J.C.~Dias, \emph{{Small Kerr-anti-de Sitter black holes are
  unstable}}, \href{https://doi.org/10.1103/PhysRevD.70.084011}{\emph{Phys.
  Rev. D} {\bfseries 70} (2004) 084011}
  [\href{https://arxiv.org/abs/hep-th/0405006}{{\ttfamily hep-th/0405006}}].

\bibitem{Cardoso:2004nk}
V.~Cardoso, O.J.C.~Dias, J.P.S.~Lemos and S.~Yoshida, \emph{{The Black hole
  bomb and superradiant instabilities}},
  \href{https://doi.org/10.1103/PhysRevD.70.049903}{\emph{Phys. Rev. D}
  {\bfseries 70} (2004) 044039}
  [\href{https://arxiv.org/abs/hep-th/0404096}{{\ttfamily hep-th/0404096}}].

\bibitem{Danielsson:1999zt}
U.H.~Danielsson, E.~Keski-Vakkuri and M.~Kruczenski, \emph{{Spherically
  collapsing matter in AdS, holography, and shellons}},
  \href{https://doi.org/10.1016/S0550-3213(99)00511-8}{\emph{Nucl. Phys. B}
  {\bfseries 563} (1999) 279}
  [\href{https://arxiv.org/abs/hep-th/9905227}{{\ttfamily hep-th/9905227}}].

\bibitem{Danielsson:1999fa}
U.H.~Danielsson, E.~Keski-Vakkuri and M.~Kruczenski, \emph{{Black hole
  formation in AdS and thermalization on the boundary}},
  \href{https://doi.org/10.1088/1126-6708/2000/02/039}{\emph{JHEP} {\bfseries
  02} (2000) 039} [\href{https://arxiv.org/abs/hep-th/9912209}{{\ttfamily
  hep-th/9912209}}].

\bibitem{Son:2002sd}
D.T.~Son and A.O.~Starinets, \emph{{Minkowski space correlators in AdS / CFT
  correspondence: Recipe and applications}},
  \href{https://doi.org/10.1088/1126-6708/2002/09/042}{\emph{JHEP} {\bfseries
  09} (2002) 042} [\href{https://arxiv.org/abs/hep-th/0205051}{{\ttfamily
  hep-th/0205051}}].

\bibitem{Kraus:2015zda}
P.~Kraus and S.D.~Mathur, \emph{{Nature abhors a horizon}},
  \href{https://doi.org/10.1142/S0218271815430038}{\emph{Int. J. Mod. Phys. D}
  {\bfseries 24} (2015) 1543003}
  [\href{https://arxiv.org/abs/1505.05078}{{\ttfamily 1505.05078}}].

\bibitem{Bena:2015dpt}
I.~Bena, D.R.~Mayerson, A.~Puhm and B.~Vercnocke, \emph{{Tunneling into
  Microstate Geometries: Quantum Effects Stop Gravitational Collapse}},
  \href{https://doi.org/10.1007/JHEP07(2016)031}{\emph{JHEP} {\bfseries 07}
  (2016) 031} [\href{https://arxiv.org/abs/1512.05376}{{\ttfamily
  1512.05376}}].

\bibitem{Soni:2023fke}
R.M.~Soni, \emph{{A Type $I$ Approximation of the Crossed Product}},
  \href{https://arxiv.org/abs/2307.12481}{{\ttfamily 2307.12481}}.

\bibitem{Witten:2023qsv}
E.~Witten, \emph{{Algebras, Regions, and Observers}},
  \href{https://arxiv.org/abs/2303.02837}{{\ttfamily 2303.02837}}.

\bibitem{Witten:2021jzq}
E.~Witten, \emph{{Why Does Quantum Field Theory In Curved Spacetime Make Sense?
  And What Happens To The Algebra of Observables In The Thermodynamic Limit?}},
   \href{https://arxiv.org/abs/2112.11614}{{\ttfamily 2112.11614}}.

\bibitem{Jafferis:2019wkd}
D.L.~Jafferis and D.K.~Kolchmeyer, \emph{{Entanglement Entropy in
  Jackiw-Teitelboim Gravity}},
  \href{https://arxiv.org/abs/1911.10663}{{\ttfamily 1911.10663}}.

\bibitem{Bena:2016ypk}
I.~Bena, S.~Giusto, E.J.~Martinec, R.~Russo, M.~Shigemori, D.~Turton et~al.,
  \emph{{Smooth horizonless geometries deep inside the black-hole regime}},
  \href{https://doi.org/10.1103/PhysRevLett.117.201601}{\emph{Phys. Rev. Lett.}
  {\bfseries 117} (2016) 201601}
  [\href{https://arxiv.org/abs/1607.03908}{{\ttfamily 1607.03908}}].

\bibitem{Bena:2024qed}
I.~Bena, R.~Dulac, A.~Houppe, D.~Toulikas and N.P.~Warner, \emph{{Waves on
  Mazes}},  \href{https://arxiv.org/abs/2404.14477}{{\ttfamily 2404.14477}}.

\bibitem{Guo:2024pvv}
B.~Guo, S.D.~Hampton and N.P.~Warner, \emph{{Inscribing geodesic circles on the
  face of the superstratum}},
  \href{https://doi.org/10.1007/JHEP05(2024)224}{\emph{JHEP} {\bfseries 05}
  (2024) 224} [\href{https://arxiv.org/abs/2401.17366}{{\ttfamily
  2401.17366}}].

\bibitem{Ganchev:2023sth}
B.~Ganchev, S.~Giusto, A.~Houppe, R.~Russo and N.P.~Warner,
  \emph{{Microstrata}},
  \href{https://doi.org/10.1007/JHEP10(2023)163}{\emph{JHEP} {\bfseries 10}
  (2023) 163} [\href{https://arxiv.org/abs/2307.13021}{{\ttfamily
  2307.13021}}].

\bibitem{Bena:2022fzf}
I.~Bena, N.~\v{C}eplak, S.D.~Hampton, A.~Houppe, D.~Toulikas and N.P.~Warner,
  \emph{{Themelia: the irreducible microstructure of black holes}},
  \href{https://arxiv.org/abs/2212.06158}{{\ttfamily 2212.06158}}.

\bibitem{Bena:2022sge}
I.~Bena, N.~Ceplak, S.~Hampton, Y.~Li, D.~Toulikas and N.P.~Warner,
  \emph{{Resolving black-hole microstructure with new momentum carriers}},
  \href{https://doi.org/10.1007/JHEP10(2022)033}{\emph{JHEP} {\bfseries 10}
  (2022) 033} [\href{https://arxiv.org/abs/2202.08844}{{\ttfamily
  2202.08844}}].

\bibitem{Ganchev:2021pgs}
B.~Ganchev, A.~Houppe and N.P.~Warner, \emph{{Q-balls meet fuzzballs: non-BPS
  microstate geometries}},
  \href{https://doi.org/10.1007/JHEP11(2021)028}{\emph{JHEP} {\bfseries 11}
  (2021) 028} [\href{https://arxiv.org/abs/2107.09677}{{\ttfamily
  2107.09677}}].

\bibitem{Martinec:2020cml}
E.J.~Martinec and N.P.~Warner, \emph{{The Harder They Fall, the Bigger They
  Become: Tidal Trapping of Strings by Microstate Geometries}},
  \href{https://doi.org/10.1007/JHEP04(2021)259}{\emph{JHEP} {\bfseries 04}
  (2021) 259} [\href{https://arxiv.org/abs/2009.07847}{{\ttfamily
  2009.07847}}].

\bibitem{Bena:2020yii}
I.~Bena, F.~Eperon, P.~Heidmann and N.P.~Warner, \emph{{The Great Escape:
  Tunneling out of Microstate Geometries}},
  \href{https://doi.org/10.1007/JHEP04(2021)112}{\emph{JHEP} {\bfseries 04}
  (2021) 112} [\href{https://arxiv.org/abs/2005.11323}{{\ttfamily
  2005.11323}}].

\bibitem{Heidmann:2019xrd}
P.~Heidmann, D.R.~Mayerson, R.~Walker and N.P.~Warner, \emph{{Holomorphic Waves
  of Black Hole Microstructure}},
  \href{https://doi.org/10.1007/JHEP02(2020)192}{\emph{JHEP} {\bfseries 02}
  (2020) 192} [\href{https://arxiv.org/abs/1910.10714}{{\ttfamily
  1910.10714}}].

\bibitem{Hartnoll:2005ju}
S.A.~Hartnoll and S.P.~Kumar, \emph{{AdS black holes and thermal Yang-Mills
  correlators}},
  \href{https://doi.org/10.1088/1126-6708/2005/12/036}{\emph{JHEP} {\bfseries
  12} (2005) 036} [\href{https://arxiv.org/abs/hep-th/0508092}{{\ttfamily
  hep-th/0508092}}].

\bibitem{Heynen:2023sin}
J.~Heynen and D.R.~Mayerson, \emph{{Gravitational Multipoles in Five
  Dimensions}},  \href{https://arxiv.org/abs/2312.04352}{{\ttfamily
  2312.04352}}.

\bibitem{Cardoso:2019rvt}
V.~Cardoso and P.~Pani, \emph{{Testing the nature of dark compact objects: a
  status report}},
  \href{https://doi.org/10.1007/s41114-019-0020-4}{\emph{Living Rev. Rel.}
  {\bfseries 22} (2019) 4} [\href{https://arxiv.org/abs/1904.05363}{{\ttfamily
  1904.05363}}].

\bibitem{Cardoso:2014sna}
V.~Cardoso, L.C.B.~Crispino, C.F.B.~Macedo, H.~Okawa and P.~Pani, \emph{{Light
  rings as observational evidence for event horizons: long-lived modes,
  ergoregions and nonlinear instabilities of ultracompact objects}},
  \href{https://doi.org/10.1103/PhysRevD.90.044069}{\emph{Phys. Rev. D}
  {\bfseries 90} (2014) 044069}
  [\href{https://arxiv.org/abs/1406.5510}{{\ttfamily 1406.5510}}].

\bibitem{Keir:2014oka}
J.~Keir, \emph{{Slowly decaying waves on spherically symmetric spacetimes and
  ultracompact neutron stars}},
  \href{https://doi.org/10.1088/0264-9381/33/13/135009}{\emph{Class. Quant.
  Grav.} {\bfseries 33} (2016) 135009}
  [\href{https://arxiv.org/abs/1404.7036}{{\ttfamily 1404.7036}}].

\bibitem{Cunha:2017qtt}
P.V.P.~Cunha, E.~Berti and C.A.R.~Herdeiro, \emph{{Light-Ring Stability for
  Ultracompact Objects}},
  \href{https://doi.org/10.1103/PhysRevLett.119.251102}{\emph{Phys. Rev. Lett.}
  {\bfseries 119} (2017) 251102}
  [\href{https://arxiv.org/abs/1708.04211}{{\ttfamily 1708.04211}}].

\bibitem{Danielsson:2021ykm}
U.~Danielsson, L.~Lehner and F.~Pretorius, \emph{{Dynamics and observational
  signatures of shell-like black hole mimickers}},
  \href{https://doi.org/10.1103/PhysRevD.104.124011}{\emph{Phys. Rev. D}
  {\bfseries 104} (2021) 124011}
  [\href{https://arxiv.org/abs/2109.09814}{{\ttfamily 2109.09814}}].

\bibitem{Birmingham:2001pj}
D.~Birmingham, I.~Sachs and S.N.~Solodukhin, \emph{{Conformal field theory
  interpretation of black hole quasinormal modes}},
  \href{https://doi.org/10.1103/PhysRevLett.88.151301}{\emph{Phys. Rev. Lett.}
  {\bfseries 88} (2002) 151301}
  [\href{https://arxiv.org/abs/hep-th/0112055}{{\ttfamily hep-th/0112055}}].

\bibitem{Hartnoll:2016apf}
S.A.~Hartnoll, A.~Lucas and S.~Sachdev, \emph{{Holographic quantum matter}},
  \href{https://arxiv.org/abs/1612.07324}{{\ttfamily 1612.07324}}.

\bibitem{Marecki:2006wc}
P.~Marecki, \emph{{On quantum effects in the vicinity of would-be horizons}},
  in \emph{{11th Marcel Grossmann Meeting on General Relativity}},
  pp.~1517--1519, 12, 2006,
  \href{https://doi.org/10.1142/9789812834300_0183}{DOI}
  [\href{https://arxiv.org/abs/gr-qc/0612178}{{\ttfamily gr-qc/0612178}}].

\bibitem{Zaslavskii:2003is}
O.B.~Zaslavskii, \emph{{Boulware state and semiclassical thermodynamics of
  black holes in a cavity}},
  \href{https://doi.org/10.1103/PhysRevD.68.127502}{\emph{Phys. Rev. D}
  {\bfseries 68} (2003) 127502}
  [\href{https://arxiv.org/abs/gr-qc/0310090}{{\ttfamily gr-qc/0310090}}].

\end{thebibliography}\endgroup
\bibliographystyle{JHEP}
\end{document}